%% file: thesis.tex
\documentclass[12pt,titlepage]{book} 
\usepackage{macros}
\input{header}


\begin{document} 

%\selectlanguage{english}

%\title{The static quark potential and scaling behavior of 
%${\rm SU(3)}$ lattice Yang-Mills theory}
%\maketitle

\input frontpage.tex
\cleardoublepage
\input abstract.tex
%\input zusammen.tex
%\cleardoublepage
\lhead[\fancyplain{}{\bfseries\thepage}]%
        {\fancyplain{}{\bfseries\rightmark}}
        \rhead[\fancyplain{}{\bfseries\leftmark}]%
        {\fancyplain{}{\bfseries\thepage}}
        \chead{}
        \lfoot{}
        \cfoot{}
        \rfoot{}
        \renewcommand{\sectionmark}[1]{\markright{\thesection\ #1}}
        \renewcommand{\chaptermark}[1]%
        {\markboth{\chaptername\ \thechapter\ #1}{}}
\selectlanguage{english}
\pagenumbering{roman}
\cleardoublepage
\tableofcontents
\listoffigures
\listoftables
\cleardoublepage
\newpage
\pagenumbering{arabic}
\input intro_new.tex
\cleardoublepage
\lhead[\fancyplain{}{\bfseries\thepage}]%
        {\fancyplain{}{\bfseries\rightmark}}
        \rhead[\fancyplain{}{\bfseries\leftmark}]%
        {\fancyplain{}{\bfseries\thepage}}
        \chead{}
        \lfoot{}
        \cfoot{}
        \rfoot{}
        \renewcommand{\sectionmark}[1]{\markright{\thesection\ #1}}
        \renewcommand{\chaptermark}[1]%
        {\markboth{\chaptername\ \thechapter\ #1}{}}
% chapter 2: lattice gauge fields
\input lattice.tex
% chapter 3: the static quark potential
\input potential.tex
\input tcr0.tex

\input glueball.tex

\newpage
\input concl.tex

\cleardoublepage
\lhead[\fancyplain{}{\bfseries\thepage}]%
        {\fancyplain{}{\bfseries\rightmark}}
        \rhead[\fancyplain{}{\bfseries\leftmark}]%
        {\fancyplain{}{\bfseries\thepage}}
        \chead{}
        \lfoot{}
        \cfoot{}
        \rfoot{}
        \renewcommand{\sectionmark}[1]{\markright{\thesection\ #1}}
        \renewcommand{\chaptermark}[1]%
        {\markboth{\chaptername\ \thechapter\ #1}{}}
\begin{appendix}
  \addcontentsline{toc}{chapter}{Appendix}
  \renewcommand{\chaptermark}[1]%
  {\markboth{\appendixname\ \thechapter\ #1}{}}

\input app_a.tex

\input app_trans.tex
\input app1.tex

%\input{app2.tex} 
 
\end{appendix}
%\bibliography{lattice}        %or whatever your .bib file is
%   \bibliographystyle{h-elsevier}   %if you use h-elsevier.bst

\input thesis.bbl
\cleardoublepage
\input akno.tex
%\cleardoublepage
%\input erklar.tex
%\cleardoublepage
%\input lebenslauf.tex
\end{document}

%% file: header.tex
\usepackage[latin1]{inputenc}
\usepackage[T1]{fontenc}
\usepackage[german,english]{babel}

\usepackage{amsmath}
\usepackage{vmargin}

\usepackage{ae,aecompl}
\usepackage{amssymb}
\usepackage{textcomp}
\usepackage{type1cm}
\usepackage{bm}
\usepackage[multiple]{footmisc}
\usepackage{multicol}
\usepackage{algorithmic}
\usepackage{algorithm}
\usepackage{ifthen}
\usepackage{layout}
\usepackage{acronym}

\usepackage{fancyhdr}
\usepackage{amssymb}
\usepackage{cite}
\usepackage{latexsym}
\usepackage{epic}
\usepackage{graphics}
\usepackage{epsfig}
\usepackage{rotating}

\usepackage{multirow}
\usepackage{hhline}
\usepackage{subfigure}
\usepackage{calc}
\usepackage{curves}
\usepackage{slashed}
\usepackage[bf,footnotesize]{caption}
\usepackage{babel}

\usepackage{ae,aecompl}

\usepackage{textcomp}
\usepackage{latexsym}

\usepackage{amssymb}
\usepackage{amsfonts}

\usepackage{type1cm}

\usepackage{bm}

\usepackage{amsmath,amsthm}
\allowdisplaybreaks[4]

\usepackage[multiple]{footmisc}
\usepackage{multicol}
\usepackage{algorithmic}
\usepackage{algorithm}
\usepackage{ifthen}

\usepackage{graphicx}
\usepackage{epsfig}
\usepackage{index} % changes \addcontentsline incompatible to hyperref!
\usepackage{slashed}
\usepackage{layout}

\usepackage{fixltx2e}[1999/12/01]
\usepackage{array}
\usepackage{blkarray}
\usepackage{acronym}

\usepackage{minitoc}

\usepackage{amsthm} % must be loaded after amsmath!

\theoremstyle{definition}

\theoremstyle{plain}

\theoremstyle{remark}

 \makeatletter
 \let\oldendpart=\@endpart
 \def\@endpart{\vskip 40\p@}
 \makeatother

\makeatletter
\setpapersize{A4}

\renewcommand{\chaptermark}[1]%
{\markboth{\chaptername\ \thechapter\ #1}{}}
\renewcommand{\sectionmark}[1]%
{\markright{\thesection\ #1}}

%% file: frontpage.tex
\begin{titlepage}
\begin{flushright}
DESY-THESIS-2003-016 \\
June 2003
\end{flushright}

\vskip 1 cm
\begin{center}
{\large Silvia Necco}
\end{center}

\vskip 1 cm
\begin{center}
 {\bf {\Large The static quark potential and scaling behavior of ${\rm SU(3)}$ lattice Yang-Mills theory}}
\end{center}

\vskip 1.0cm
\begin{center}
{\large Ph.D. Thesis}\\
{\large Humboldt Universit\"at zu Berlin}
\end{center}
\vskip 1.0cm
\begin{center}
\vskip 0.8cm
DESY, Platanenallee 6, D-15738 Zeuthen, Germany
\vskip 0.8cm
 necco@ifh.de
\end{center}
\end{titlepage}

%% file: abstract.tex
\chapter*{Abstract}
         \pagestyle{empty}
        \fancyhead{}
        \fancyhead[LE,RO]{\bfseries\thepage}
        \fancyhead[LO]{\bfseries Abstract}
        \fancyhead[RE]{\bfseries Abstract}

The potential between a static quark and antiquark in pure $\SUthree$ Yang-Mills theory is evaluated non-perturbatively through computations on the lattice in the region from short to intermediate distances ($0.05 \fm\leq r \leq 0.8\fm$). 
Renormalized dimensionless quantity are extrapolated to the continuum limit, confirming the theoretical expectation that the leading lattice artifacts are quadratic in the lattice spacing.\\
In the high energy regime the results are compared with the parameter-free prediction of perturbation theory obtained by solving the Renormalization Group equation at two and three loops. 
The choice of the renormalization scheme to define a running coupling turns out to be important for the accuracy of the perturbative prediction:
by obtaining the running coupling through the force, perturbation theory 
is applicable up to $\alpha\sim 0.3$, while from the static potential only up to $\alpha\sim 0.15$. In the region where perturbation is supposed to be reliable, 
no large unexpected non-perturbative term is observed: on the contrary, one finds a good agreement between perturbation theory and our non-perturbative data.\\
For large quark-antiquark separations our results are compared with the predictions of a bosonic effective string theory, finding a surprising good agreement already for distances $\gtrsim 0.5\fm$.\\
In the second part of this work, universality and scaling behavior of different formulations of Yang-Mills theory on the lattice are discussed.\\
In particular, the Iwasaki and DBW2 action are investigated, which were 
obtained by following renormalization group (RG) arguments. 
The length scale $\rnod\sim 0.5\fm$ is evaluated at several lattice spacings and the scaling of the critical deconfinement temperature $T_{c}\rnod$ is analyzed and confronted with the results obtained with the usual Wilson plaquette action. Since they agree in the continuum limit, the universality is confirmed. We remark that the quantity to use to set the scale has to be chosen with care in order to avoid large systematic uncertainties and $\rnod$ turns out to be appropriate. For the critical temperature the data obtained with RG actions show reduced lattice artifacts, above all with the Iwasaki action.\\
Finally the mass of the glueballs $0^{++}$ and $2^{++}$ is evaluated by 
considering the quantities $m_{0^{++}}\rnod$ and  $m_{2^{++}}\rnod$; however for those observables no clear conclusion about the scaling behavior can be drawn.
Particular attention is dedicated to the violation of physical positivity which occur in these actions and the consequences in the extraction of physical quantities from Euclidean correlation functions.\\
\\

\noindent
{\bf Keywords}:\\
Lattice gauge theory, static quark potential, renormalization group, improved actions

%% file: intro_new.tex
\chapter*{Introduction}
        \label{introduction}
         \addcontentsline{toc}{chapter}{Introduction}
        \pagestyle{fancy}
        \fancyhead{}
        \fancyhead[LE,RO]{\bfseries\thepage}
        \fancyhead[LO]{\bfseries Introduction}
        \fancyhead[RE]{\bfseries Introduction}

The strong, weak and electromagnetic interactions are up to now described successfully within the Standard Model of elementary particles.
This theory is based on a \emph{local gauge principle} \cite{Weyl:1929fm}, with the gauge group
\begin{equation}
G_{loc}=\SUthree_{c}\times \SUtwo_{L} \times {\rm U(1)}_{Y},
\end{equation}
and it is essentially determined once the matter fields and their transformation laws under $G_{loc}$ are specified.\\
The degrees of freedom are respectively the color for $\SUthree$, weak isospin for $\SUtwo$ and weak hypercharge for ${\rm U(1)}$.\\
In particular, quantum chromodynamics (QCD) associated to the color group $\SUthree$ is the currently accepted framework to describe  strong interactions. The matter components are the quarks, which are described by spinor fields carrying three color indices and appearing in six flavors (up, down, charm, strange, top and bottom). The gluons are the vector bosons that mediate the interactions and, due to the fact that the gauge group $\SUthree$ is non-Abelian, self-interactions exist, contrary to the situation in electromagnetism.\\
The formulation of QCD was the outcome of several years of investigations and discussions on experimental results.\\
The goal was to formulate a theory which is able to reproduce the two main features of strong interactions, asymptotic freedom and confinement, which are related to their properties in the high and low energy regimes, respectively.\\
The first step was the observation that the hadrons, which are the particles that participate in strong interactions, fall into multiplets that should reflect underlying internal symmetries. To express this fact in a concrete way, it was formulated the hypothesis that hadrons are composed of more elementary constituents with certain basic symmetries, which were called quarks \cite{Gell-Mann:1964nj,Zweig:1964jf}.\\

Asymptotic freedom was observed in deep inelastic scattering:
at high energies the form factors in the cross sections loose their dependence on certain mass parameters; this scale invariance (also called \emph{scaling}) was interpreted to mean that the constituents act as if they were free particles at extremely high energies. The most intuitive explanation of the scaling relation came from Feynman's \emph{parton} model \cite{Bjorken:1969ja,Feynman:1969ej,Feynman:1970fm}, where the proton was assumed to consist of point-like constituents. Later the partons were identified with the quarks on the basis of their quantum numbers.\\
Finally in 1973 Fritzsch, Gell-Mann and Leutwyler \cite{Fritzsch:1973pi} proposed QCD as the non-Abelian gauge theory built from the $\SUthree$ gauge group associated with the color symmetry.\\
The most direct experimental evidence that quarks must have an additional quantum number, the color, came from the measurements of the total cross section for the annihilation of electron-positron pairs in colliding beam experiments (SLAC, PETRA).\\
In the same year, it was shown
\cite{'tHooft:1985ir,Gross:1973id,Politzer:1973fx} that non-Abelian gauge
theory exhibits asymptotic freedom.\\
The appropriate framework to discuss this property is the renormalization group equation.\\
The insertion of quantum corrections leads to divergences in
the evaluation of physical observables; these can be eliminated
via renormalizing the theory by adding counterterms in the Lagrangian in a proper way.
The consequence of the
renormalization of a quantum field theory is that the coupling constant
that appears in the Lagrangian becomes a \emph{running coupling}.  
The so-called $\beta$-function quantifies the dependence of the coupling on the energy scale.
Through the perturbative study of the $\beta$-function it was shown 
that at very high
energies (or equivalently at small distances) the coupling becomes small, so
that the quarks behave as free particles. In this regime one then
expects that perturbative methods furnish reliable predictions for physical
observables. \\

On the other hand, quarks have never been detected in isolation, but only as
constituents of hadrons. From the theoretical point of view, this should
correspond to the fact that all physical states are singlets with respect to
the color group. In order to check whether this feature is contained in QCD, 
one can not apply perturbative methods, since the coupling is expected to be
large at scales corresponding to the size of hadrons. It is however widely
believed that this behavior is a consequence of quantum chromodynamics,
although up to now no proof exists.\\
One of the strongest evidence for this to be true comes from lattice field 
theory. In 1974 Wilson proposed \cite{wilson} a formulation of a gauge field theory
on a discretized Euclidean space-time. 
In this framework, by using a \emph{strong coupling expansion} it was possible to demonstrate that at sufficiently strong couplings, pure $\SUthree$ lattice gauge theory exhibits color confinement: in the limit of large distances, the energy needed to separate a pair of quark-antiquark grows proportionally to the separation.\\
Since then lattice field theory has been widely studied and became a powerful tool to investigate properties of strong interactions.\\
The natural framework to quantize the lattice theory is the path integral formalism;
in this formulation the system takes the form of a classical four-dimensional
statistical model. 
The matter fields are treated as classical stochastic variables assigned to
the points of the lattice, while the gauge fields are associated to the
links. 
In this analogy, the Euclidean action of the quantum field theory corresponds to the classical Hamiltonian of the statistical system and the mass of the lightest particle corresponds to the inverse of the correlation length.\\
The lattice furnishes
a regularization of the theory by providing an ultraviolet cutoff
proportional to the inverse lattice spacing and this is actually the only known non-perturbative regularization of quantum field theory.\\
The crucial question is whether the theory formulated on the lattice is well defined in the limit when the lattice spacing $a$ is sent to zero. The physical continuum limit corresponds to holding the values of physical quantities fixed while letting $a\rightarrow 0$; the renormalization group equation describes how the parameters behave by changing the scale of the theory (in this case the lattice spacing). It turns out that the continuum limit is realized when the bare gauge coupling $g_{0}$ is sent to zero. In the language of statistical mechanics 
this corresponds to a second order phase transition, where the correlation length in lattice units diverges. The basic concepts and definition of lattice gauge theory are reported in chapter \ref{c_lattice}.\\
An important advantage of the lattice formulation is that the path integrals which correspond to expectation values of physical observables can be computed numerically via Monte Carlo simulations. The idea is to generate samples consisting of a large number of field configurations according to the Boltzmann distribution and to evaluate the observables as sample average.\\
Since the first works by Creutz \cite{pot:creutz} on $\SUtwo$
this method was successfully applied 
to lattice pure gauge theories and in recent years to full QCD with two and three flavors of
dynamical fermions; in this period, increasing precision and reliability due to theoretical improvement of the techniques has been accompanied by an enhanced computing power.\\

Due to the non-perturbative nature of the formulation, lattice field theory provides a very powerful tool to be adopted for the investigation of low energy features of QCD; the final goal is to test the Standard Model by evaluating from first principles physical quantities which one may compare to the experimental results.\\
Within this challenging task, lattice gauge theories can give important indications regarding basic questions related to strong interactions.\\
For example, it is expected that at high energies perturbation theory gives
reliable estimates of physical quantities, but how high the energy must be
is still subject of debate. The first evident problem of the perturbative expansion for QCD is that it is described by an asymptotic series: this means that after a certain order the perturbative expansion looses its predictive power.\\
Apart from this, a second problem is that the accuracy of the prediction is highly dependent on
which renormalization scheme is adopted to define the observables.
It turns out that different ways to
subtract the divergences lead to different definitions of the running coupling
in terms of which the perturbative series can behave quite differently.\\
The question whether perturbation theory is really applicable can be
relevant in many practical cases; for example, the study of the perturbative
evolution of deep inelastic scattering structure functions starting at a
renormalization point below $1 \GeV$ was strongly debated. Another popular
example is the experimental measurement of the strong coupling $\alpha_{s}$ itself \cite{Bethke:2000ai},
which needs perturbation theory as theoretical input. The main phenomenological test of the applicability of perturbation theory is to check whether the determinations from different processes are compatible. From the experiments
it is indeed difficult to study a process systematically as function of the energy.\\
What lattice gauge theory can do in this context is to furnish a non
perturbative evaluation
of the running coupling in order to compare the results with the predictions
of perturbation theory and to establish at which energy one is allowed to
apply it.\\
 Here one has to overcome a stringent limitation: the lattice spacing
$a$ can not be made arbitrarily small and on the other hand the physical
extent of the system must be large ($\gtrsim 1.5\fm$) in order to approximate the
condition of infinite volume. With this limitation, one currently reaches
$a^{-1}\sim 2\GeV$ for $N_{f}=2$ flavors of dynamical quarks. A possible alternative has been proposed in \cite{alpha:sigma},
where one considers a finite size
effect as the physical observable which defines a renormalized
coupling. This method has been applied successfully in the framework of the
Schr\"odinger functional scheme for the evaluation of the running coupling for $N_{f}=0$ \cite{mbar:pap1,alpha:SU2,alpha:SU3} and $N_{f}=2$ \cite{alpha:lett,DellaMorte:2002vm}.\\
An important result was also the computation of the so-called Lambda-parameter in the theory with $\nf=0$ \cite{mbar:pap1}: this quantity plays an important phenomenological r\^{o}le since it fixes the energy scale of QCD.\\
 For this particular scheme the scale dependence of the running coupling has been found to be in good agreement with perturbation theory for $\alpha<0.3$.
These studies showed also that the perturbative scaling regime and the low-energy domain of the theory are smoothly connected, with no complicated transition region.\\

It is important to check these properties in many independent ways by comparing several quantities computed non-perturbatively with their perturbative prediction. An interesting quantity for this purpose is the static quark potential,
which defines the energy of gauge fields in presence of two static color sources separated by a distance $r$ and has been studied very intensively in the last years.\\ 
However, although many non-perturbative computations of this observables have been performed, there was up to now no convincing investigation in the short distance regime. 
With this goal we decided to evaluate the static quark potential in the $N_{f}=0$ case (for our purposes this means the pure Yang-Mills theory), for which it is possible to simulate large lattices; this allowed the investigation of the small distance region and the continuum extrapolation of the results obtained at finite lattice spacing.\\
At high energies, corresponding to small distances between the static quark-antiquark pair, the asymptotic freedom justifies the application of perturbative methods. After subtracting a constant part, the perturbative potential can be written in the form\\
\begin{equation}
V(r)=-\frac{4}{3}\frac{\alphavbar(\mu)}{r},\quad \mu=\frac{1}{r},
\end{equation}
which defines the running coupling in the so-called $\bar{V}$ scheme.\\
This definition is restricted to perturbation theory since from the non-perturbative point of view the subtraction of the self-energy is not well defined.\\
Alternatively, one can define the coupling in the $q\bar{q}$ scheme through the force 
\begin{equation}
F(r)=\frac{d V}{d r}=\frac{4}{3}\frac{\alphaqqbar(\mu)}{r^2},\quad \mu=\frac{1}{r}.
\end{equation}
In perturbation theory, the potential is known at two loops \cite{Fischler:1977yf,Appelquist:1977tw,Appelquist:1978es}; however, the 
reliability of perturbation theory has been doubted at distances as short as $0.1\fm$, and the presence of large non-perturbative terms has been argued \cite{pot:bali99,Akhoury:1998by}.\\
We started our investigation by constructing in the continuum a perturbative prediction for $\alphavbar(\mu)$ and $\alphaqqbar(\mu)$.
By using the (non-perturbative) knowledge of the Lambda-parameter in units of the low energy scale $\rnod\sim 0.5\fm$ \cite{pot:r0}, it is possible to extract a parameter-free prediction for the running coupling via the renormalization group equation (RGE). The only ingredients that one needs are, for a given renormalization scheme, the corresponding Lambda-parameter and the coefficients of the $\beta$-functions. The Lambda-parameters in different schemes are connected exacly by one-loop relations and the $\beta$-function coefficients can be taken from the literature.\\
With these ingredients we evaluated the 2-and 3-loops RG perturbative predictions for $\alphavbar(\mu)$ and $\alphaqqbar(\mu)$.\\
We want to stress that we are dealing with a single-scale problem and hence we expect that this is the best perturbative prediction, since there are no free parameters related to the energy scale to be fixed.\\
By analyzing the difference between the 2-and 3-loops behavior one can argue that for the $\bar{V}$ scheme the perturbative expansion appears to be applicable only up to $\alpha\sim 0.15$, while for the $q\bar{q}$ scheme the series is well behaved until $\alpha\sim 0.3$. This analysis gives already important indications on the quality of the perturbative prediction within these two different schemes. To clarify whether unexpectedly large 
non-perturbative terms are present in these observables was the strongest motivation of our computation.\\
First of all, we tried to give a reasonable phenomenological definition of a ``large non-perturbative term''.
We then evaluated the potential and the force on the lattice in the region $0.05\fm\leq r\leq 0.8\fm$.\\ 
The first step was to set the scale by specifying a dimensionful low energy observable in terms of which all other dimensionful quantities can be expressed.
This observable must be chosen very carefully and must be evaluable with good precision.\\
 In particular the length scale $\rnod\sim 0.5\fm$ \cite{pot:r0}, which is defined through the force, turned out to be a good choice. Due to the fact that we are interested in the short distance region, we decided to introduce a smaller reference scale $r_{c}\sim 0.26\fm$ also.\\
Once renormalized quantities have been constructed, we performed the continuum extrapolation of the potential and the force; in this procedure
we were also able to test the theoretical expectation on the leading lattice spacing errors to $F(r)$.\\
Finally we compared our continuum results with the perturbative predictions and we discussed the domain of applicability of perturbation theory for the different renormalization schemes.\\

As already mentioned, while perturbation theory describes the small distance regime, the long range properties are of non-perturbative nature.
An intuitive picture of the confinement mechanism is given by an effective bosonic string theory \cite{nambu}; a tube of color electric flux is formed between the quark and the antiquark and in the limit of large separation $r$ one expects that the flux tube behaves like a string with fixed ends.\\
With this assumption, the static quark potential has the asymptotic expansion for large $r$ \cite{Luscher:1980fr,Luscher:1981ac}
\begin{equation}\label{stringa}
V(r)=const.+\sigma r +\frac{\gamma}{r}+{\rm O}\left(\frac{1}{r^2}\right),
\end{equation}
where $\gamma$ is an universal factor depending only on the dimensions of the system and $\sigma$ defines the string tension.\\
This expression furnishes another parameter-free prediction of the potential and also in this case the 
 non-perturbative results obtained via lattice calculations can be useful to validate this picture.\\ 
The problem here is that it is very difficult to evaluate the static quark potential at large distances with a good statistical precision; special numerical techniques are required and this was beyond the purpose of our work.\\
We compared however our results with \eq{stringa} 
 also including a higher order term in the effective theory, which has been estimated in \cite{Luscher:2002qv}.\\
Chapter \ref{chapt_potential} is devoted to the static quark potential, the details of the lattice evaluation and the comparison with the predictions of perturbation theory and string model.\\

The second part of this work is dedicated to the study of the properties of
different formulations of the gauge action on the lattice.\\
The simplest way to formulate a lattice gauge action is the one introduced by
Wilson: one sums over plaquette terms, which are gauge invariant products of
link variables along the smallest closed loop that one can build on a cubic
lattice.\\ 
Within this framework one can show that the leading discretization errors that
one introduces are of order ${\rm O}(a^2)$.\\
Due to the practical limitations which prevent to study
systems with very small lattice spacings and the impossibility in many cases
to perform a continuum extrapolation of the physical observables, it would be
highly desirable to work with a formulation in which the lattice artefacts are
reduced.
This problem is even more drastic in the fermionic action, where the leading
discretization errors are linear in the lattice spacing.\\
For this reason in the last years there were big efforts in the formulation of
both gauge and fermionic \emph{improved} actions.
The basic idea is to add appropriate combinations of \emph{irrelevant} operators to the lattice action and to tune their coefficients such that the discretization errors are reduced.
Using the language of statistical mechanics, the concept of \emph{universality} ensures that the influence of irrelevant operators on the critical behavior is negligible and hence one recovers the desired physical continuum limit.
At finite lattice spacing, the irrelevant operators governe the discretization errors, also called \emph{scaling violations} \footnote{Note that in this context \emph{scaling} is used with a different meaning than before and it is related to the behavior of physical quantities by approaching the continuum limit.}, associated to renormalized dimensionless quantities.\\
Several approaches were studied with this purpose. A popular one is based on
the Symanzik picture \cite{impr:Sym1} and has the goal to eliminate the leading lattice
artefacts proportional to $a$ for the fermionic action and to $a^2$ for the
gauge action.\\
This procedure has been applied to
the gauge sector yielding the so-called L\"uscher-Weisz action \cite{Luscher:1985zq} and to the
fermionic action by defining the so called Sheikholeslami-Wohlert (clover)
action \cite{impr:SW}.\\
A different strategy is based on Renormalization Group (RG) considerations;
a popular example is the Fixed Point (FP) action \cite{Hasenfratz:1994sp}, formulated both for gauge and fermionic field.\\
Other examples are the Iwasaki \cite{Iwasaki:1983ck} and DBW2 \cite{deForcrand:1997bx,Takaishi:1996xj} gauge actions;
the philosophy here is to perform blocking transformations to study the
evolution of the couplings in a restricted parameter space, with the final
goal to define a theory which lies close to the \emph{renormalized
  trajectory}, where no lattice artefacts are expected to be present.\\
In particular, these actions are restricted to a two-parameter space and they were constructed by adding rectangular $(1\times 2)$ loops with a certain coefficient to the usual plaquette action.\\
The Iwasaki action has been used for instance by the CP-PACS collaboration
\cite{AliKhan:2001tx} for an advanced computation of light hadron
spectrum and RG improved gauge actions have been considered to be particularly suited for 
next simulations on Ginsparg-Wilson/domain wall fermions.\\
Before one starts to use extensively these alternative actions, it is
important to investigate their properties, starting by checking the fundamental one which is the universality.\\ 
Moreover, the scaling behavior has to be tested for a possibly large number of observables in order to establish how efficient is the improvement.\\
We decided in particular to study the critical temperature for deconfinement, $T_{c}$, for DBW2 and Iwasaki action (chapter \ref{improved}); by normalizing the quantity through the string tension $T_{c}/\sqrt{\sigma}$, it was observed \cite{Okamoto:1999hi} that even in the continuum limit the result obtained with Iwasaki action deviates from what was evaluated with the Wilson plaquette action.\\
A possible reason for this discrepancy could be a violation of 
universality, but we suspected that a
more natural explanation is that the string tension is difficult to determine, and it is preferable
to use the scale $\rnod$ to reliably set the scale in pure Yang-Mills theory.\\
In order to check whether this is the case, we evaluated $\rnod$ at the values of the critical couplings that can be found in the literature and built the renormalized quantity $T_{c}\rnod$ for the Iwasaki and DBW2 action.\\
In our evaluation we remarked that for these actions, and more generally for actions containing additional loop terms, the physical positivity is violated, as already pointed out in \cite{Luscher:1984is}. This fact spoils the applicability of the variational method, which in many cases is necessary to extract the physical observables. In appendix \ref{app_trans} the properties of the transfer matrix for improved actions are discussed and the presence of unphysical states is studied by investigating the location of the poles in the propagator at the lowest order of perturbation theory.\\

Finally we considered the glueball masses as additional quantities to test the properties of RG gauge actions (chapter \ref{chapter_glueball}).\\
Glueballs represent one of the most fascinating predictions of the gauge sector of QCD;
the mass of the lightest ($0^{++}$) glueball is moreover particularly promising to study the scaling behavior since several calculations with the Wilson action
\cite{Bali:1993fb,Michael:1989jr,deForcrand:1985rs,Chen:1994uw} 
showed large lattice artefacts. We evaluated the $0^{++}$ and $2^{++}$ glueball masses for the Iwasaki and DBW2 actions for several values of lattice spacings by using an appropriate basis of operators from space-like Wilson loops up to length eight.
Then we made use of our evaluation of $\rnod$ to 
form $m_{0^{++}}\rnod$ and $m_{2^{++}}\rnod$.\\
%We emphasize that our results are affected by large statistical and systematic errors and further investigations are needed. \\
There are several open questions that could be addressed and that have to be discussed in order to find an optimal combination of gauge and fermionic actions to obtain reliable results in future simulations of QCD. Some of them will be discussed at the end of chapter \ref{chapter_glueball}.

%%% Local Variables: 
%%% mode: latex
%%% TeX-master: t
%%% End: 

%% file: lattice.tex
\chapter{Lattice gauge fields}\label{c_lattice}
Although  QCD is the best candidate for a theory of the strong interactions, the quantitative information that can be extracted is still limited. It turns out that perturbation theory fails to reproduce many of the essential low-energy  properties. The lattice formulation is one of the most elegant and powerful non-perturbative methods, and was originally introduced to investigate the confinement features.
Its power comes mainly from the possibility to evaluate physical quantities ``exactly'' via numerical simulations, yielding a method to test whether QCD provides the correct framework for describing strong interactions. The QCD Lagrangian has six unknown input parameters: the coupling constant $\alpha_{s}$ and the masses of the up, down, strange, charm and bottom quarks. \footnote{The top quark is considerably heavier than the other quarks ($m_{t}=174.3\pm 5.1\GeV$ \cite{Hagiwara:2002fs}), so that its bound states decay too fast to be detected. For this reason the top quark is not included here.}
In the lattice regularization, once these free parameters have been fixed for example in terms of six measured hadron masses, then the (Euclidean) properties of the other particles made up of these quarks and gluons have to agree with experiments. Thus the lattice formulation allows the evaluation of physical observables \emph{from first principles}.
However, there are practical limitations, mainly due to the fact that the accessible lattice volumes and resolutions are
restricted by the available (finite) computer performance and memory.\\
In this chapter we will concentrate on the gauge sector of QCD and describe different lattice formulations of $\SUn$ gauge theory that will be used later for the extraction of physical observables.

%%%%%%%%%%%%%%%%%%%%%%%%%%%%%%%%%%%%%%%%%%%%%%%%%%%%%%%%%%%%%%%%%%%%%%%%%%%%%%%
\section{Wilson action}
The starting point for a lattice formulation of quantum field theory is the path integral formalism in Euclidean space-time \cite{Dyson:1949ha}, \cite{Wick:1954eu}. In this context it is possible to establish a useful analogy between Euclidean quantum field theory on a lattice and statistical mechanics, which is also the basis for numerical simulations.
In the following we will consider pure $\SUn$ gauge theories on a 4 dimensional hypercubic lattice
\begin{equation}\label{lattice}
 \Lambda=a\mathbb{Z}^{4}=\{x|x_{\mu}/a\in\mathbb{Z}\},\quad\mu=0,1,2,3,
\end{equation}
where $a$ is the lattice spacing. The lattice introduces an ultraviolet cutoff and provides the only known consistent non-perturbative regularization of non-Abelian gauge theories. The Fourier transformation of functions on the lattice is periodic in momentum space with periodicity $2\pi/a$, therefore all momenta can be restricted to the first Brillouin zone
\begin{equation}\label{brillouin_zone}
\mathcal{B}=\left\{p|-\frac{\pi}{a}<p_{\mu}\leq \frac{\pi}{a}\right\},
\end{equation}
and the momentum cutoff is $\pi/a$.\\

An $\SUn$ lattice gauge field is an assignment of a matrix $U(x,\mu)\in\SUn$ to every lattice bond with endpoints $x$ and $(x+a\hat{\mu})$, where $\hat{\mu}$ denotes the unit vector in the positive $\mu$ direction. Under a gauge transformation $V(x)$ the link variables transform as
\begin{equation}\label{gauge_transf}
U(x,\mu)\quad\rightarrow U'(x,\mu)=\quad V(x)U(x,\mu)V^{-1}(x+a\hat{\mu}),\quad\forall\quad V(x)\in \SUn.
\end{equation}
A particular gauge-invariant object one can construct on the lattice is the trace of the product of link variables along a closed curve. These loops can be of arbitrary size and shape, and can be taken to lie in any representation of $\SUn$. The simplest example is the plaquette $W_{\mu\nu}^{1\times 1}$, a $(1\times 1)$ loop
\begin{equation}\label{plaquette}
W_{\mu\nu}^{1\times 1}(x)=\Tr\left\{U(x,\mu)U(x+a\hat{\mu},\nu)U^{-1}(x+a\hat{\nu},\mu)U^{-1}(x,\nu)\right\}.
\end{equation}
The action which has been proposed by Wilson \cite{wilson} for pure lattice gauge theory is defined in terms of these plaquette variables
\begin{equation}\label{wilson_action}
S=\beta\sum_{x}\sum_{\mu<\nu}\left\{1-\frac{1}{N}\Re W_{\mu\nu}^{1\times 1}(x)\right\}.
\end{equation}
Let $A_{\mu}(x)=-ig_{0}A_{\mu}^{b}(x)T_{b}$, where $T_{b}$ are the generators of the group, be a Lie algebra valued vector field defined on the lattice and let $U(x,\mu)=e^{-aA_{\mu}(x)}$ be a smooth field; \eq{wilson_action} takes the form
\begin{equation}\label{wilson_action_lim}
S=-\frac{\beta}{4N}\sum_{x}a^{4}\Tr F_{\mu\nu}(x)F^{\mu\nu}(x)+{\rm O}(a^{6}),
\end{equation}
where 
$$
F_{\mu\nu}(x)=\Delta_{\mu}^{f}A_{\nu}(x)-\Delta_{\nu}^{f}A_{\mu}(x)+[A_{\mu}(x),A_{\nu}(x)]
$$ 
and $\Delta_{\mu}^{f}f(x)=\frac{1}{a}(f(x+a\hat{\mu})-f(x))$ is the lattice forward derivative. 
The leading term of \eq{wilson_action_lim} for small $a$ coincides with the Yang-Mills Euclidean action
\begin{equation}\label{ym_cont}
S_{YM}=-\frac{1}{2g_{0}^2}\int d^{4}x\Tr F_{\mu\nu}F_{\mu\nu}=\frac{1}{4}\int d^{4}x F^{a}_{\mu\nu}F^{a}_{\mu\nu},
\end{equation} 
if we set $\beta=\frac{2N}{g_{0}^{2}}$ and identify $g_{0}$ with the bare coupling constant of the lattice theory; thus the Wilson action gives the desired classical continuum limit. \\
After having defined the field variables and the action, the next step to
quantization is to specify the functional integral. Given an observable $\mathcal{O}[U]$ which is in general a gauge invariant function of the link variables, its expectation value is given by
\begin{equation}\label{expect_value}
\langle\mathcal{O}\rangle=\frac{1}{Z}\int D[U]\mathcal{O}[U]e^{-S[U]},
\end{equation}
where
\begin{equation}\label{partition_func}
Z=\int D[U]e^{-S[U]}
\end{equation}
and $D[U]=\prod_{x,\mu}dU(x,\mu)$ is the invariant group measure or \emph{Haar measure}. In analogy to statistical mechanics, $Z$ is called \emph{partition function}. The high dimensional integral in  \eq{expect_value} can be evaluated by means of a (stochastic) Monte-Carlo method as an average over an ensemble of $n$ representative gauge configurations 
\begin{equation}
\langle\mathcal{O}\rangle=\frac{1}{n}\sum_{i=1}^{n}\mathcal{O}[U_{i}]+\Delta\mathcal{O}\left(\frac{1}{\sqrt{n}}\right),
\end{equation}
where $\Delta\mathcal{O}$ is the statistical error;
the method represents an \emph{exact} approach in the sense that the latter can in principle be made arbitrarily small by increasing the sample size $n$.

%%%%%%%%%%%%%%%%%%%%%%%%%%%%%%%%%%%%%%%%%%%%%%%%%%%%%%%%%%%%%%%%%%%%%%%%%%%
\section{Continuum limit and improvement}
Once one has evaluated a certain observable in lattice gauge theories, it remains the issue to understand how this is related to the continuum physical world. 
In order to obtain the theory in the continuum, the lattice spacing has to be sent to zero; at the same time, the cut-off goes to infinity and one has to construct renormalized physical quantities
which remain finite in the continuum limit.
In practice one has to compute the 
quantities of interest at different values of the lattice spacing $a$ and to extrapolate
the results to $a=0$. An obvious limitation in this procedure is the fact that it is not 
possible to perform numerical simulations at arbitrarily small lattice spacings. In current calculations,
one reaches $a\sim 0.05\fm$ for pure gauge simulations and the quenched approximation, and $a\sim 0.1\fm$ for full QCD.\\
Then the \emph{improvement} of the lattice formulation turns out to be a very important topic, and 
in the last years there were a lot of efforts in this direction, both for the gauge and the 
fermionic action. The first requirement for an improvement program is a detailed theoretical
understanding of the approach to the continuum limit, then extensive numerical studies are 
needed to confirm or disprove the expected behavior. The properties that a "good" improved 
action has to satisfy are first of all the reduction of the lattice artifacts, the better 
restauration of rotational and internal symmetries; in addition,
for the study of topological 
properties, it would be desirable to work with a lattice action where small distance
dislocations are excluded and only physical instantons are present.
In the following sections several improvement procedures will be briefly explained. 

\subsection{The Symanzik program and the L\"uscher-Weisz action} 
In 1983 Symanzik \cite{impr:Sym1} suggested that the lattice theory can be approximated by a low-energy continuum effective theory,
and the associated action can be written as
\begin{equation}\label{effective_action}
S_{\rm{eff}}=\int d^{4}x\left\{\mathcal{L}_{0}(x)+a\mathcal{L}_{1}(x)+a^{2}
\mathcal{L}_{2}(x)+...\right\},
\end{equation}
where $\mathcal{L}_{0}$ denotes the continuum Lagrangian and $\mathcal{L}_{k}$, with
$k\geq 1$, are linear combinations of local operators of dimension $4+k$. From the list of all possible
such fields only those which are invariant under the symmetries of the lattice theory have to be included.

In the effective continuum theory, renormalized lattice fields are represented through effective fields of the form
\begin{equation}\label{effective_field}
\phi_{\rm{eff}}=\phi_{0}+a\phi_{1}+a^{2}\phi_{2}+...
\end{equation}
%The fields $\phi_{0},\phi_{1},...$ which can occour here must have the appropriate dimension and transform under the
%lattice symmetries in the same way as the lattice fields that they are representing.\\

\begin{figure}[ht]
\begin{center}
\includegraphics[width=6cm]{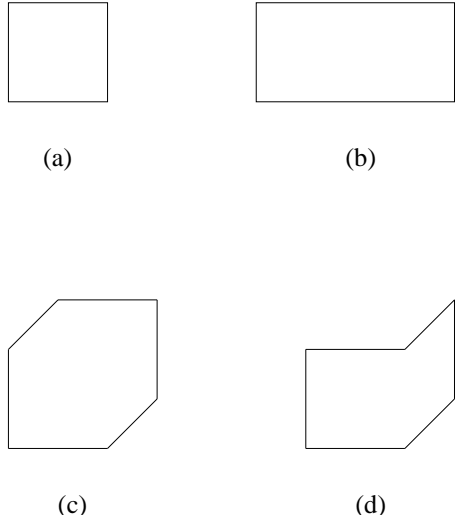}
\end{center}\vspace{-0.8cm}
\caption[Paths contributing to the L\"uscher-Weisz action]{\footnotesize{\label{symanz} List of paths which contribute to the action \eq{lw_action}.}}      
\end{figure}

The so called Symanzik improvement programme consists of removing lattice artifacts organized 
as a power series expansion in $a$, by adding irrelevant terms to both the action and the 
operators.
 For pure gauge theory the corrections begin at $O(a^2)$, since there are no dimension 5 operators.
It turns out that there are only 3 gauge-invariant dimension 6 operators
and, if the Symanzik approach is correct, these should be sufficient to cancel the $O(a^2)$ effects in the Wilson action.
The improved action for $\SUthree$ then takes the form
\begin{equation}\label{lw_action}
S=\beta\sum_{i=0}^{3}\sum_{\mathcal{C}\in T_{i}}c_{i}\left\{1-\frac{1}{3}\Re W_{\mathcal{C}}\right\},
\end{equation}
where $\mathcal{C}$ are oriented paths belonging to sets $T_{i},i=0,1,2,3$ of structurally equivalent curves, which are indicated in \fig{symanz} respectively by (a), (b), (c) and (d)\\
$T_{0}$= set of curves enclosing one plaquette;\\
$T_{1}$= set of planar curves with perimeter $6a$ enclosing two plaquettes;\\
$T_{2},T_{3}$= set of non-planar curves with perimeter $6a$.\\

%%%%%%%%%%%%%%%%%%%%%%%%%%%%%%%%%%%%%%%%%%%%%%%%%%%%%%%%%%%%%%%%%%%%%%%%%%%%%%%%%%%%%%%%%%%%%%%%%%
%%%%%%%%%%%%%%%%%%%%%%%%%%%%%%%%%%%%%%%%%%%%%%%%%%%%%%%%%%%%%%%%%%%%%%%%%%%%%%%%%%%%%%%%%%%%%%%%%%

Note that the first term corresponds to the usual plaquette action and the number of the other classes matches the number of 
independent operators of dimension 6.\\
The coefficients $c_{i}$ have to be fixed such that on-shell quantities do not have ${\rm O}(a^2)$ corrections;
at the classical level, they can be easily determined yielding the following results \cite{Weisz:1983zw},\cite{Weisz:1984bn}
\begin{eqnarray}\label{tree_level}
c_{0}&=&\frac{5}{3} =1-8c_{1},\\
c_{1}&=&-\frac{1}{12},\\
c_{2}&=&0,\\
c_{3}&=&0.
\end{eqnarray}
In quantum field theory the coefficients are subject to radiative corrections; the coefficients $c_{i}$ have to be determined non-perturbatively to fully eliminate the ${\rm O}(a^2)$ artefacts, and this has not been achieved yet.
The 1-loop improved action goes under the name L\"uscher-Weisz action \cite{Luscher:1985zq} and has coefficients
\begin{eqnarray}
c_{0}(g_{0}^2)&=&\frac{5}{3}+0.2370g_{0}^{2},\\
c_{1}(g_{0}^2)&=&-\frac{1}{12}-0.02521g_{0}^{2},\\
c_{2}(g_{0}^2)&=&-0.00441g_{0}^{2},\\
c_{3}(g_{0}^2)&=& 0,
\end{eqnarray}
with leading corrections of order $O(g_{0}^{4}a^{2})$.

\section{Renormalization group improved actions}
In recent years an alternative improvement strategy based on renormalization
group (RG) considerations was developed.
In general a quantum field theory is defined over a large range of
scales from low (relative to the cut-off) physical scale up to the cut-off which goes to
infinity in the continuum limit. 
The goal of a renormalization group transformation (RGT) is to
reduce the number of degrees of freedom by integrating out the very high momenta and taking into account their effect on the remaining variables \footnote{This is the Wilson approach to Renormalization Group \cite{Wilson:1974jj}}.\\
This can be achieved for example introducing a blocked lattice with lattice spacing $a'=2a$ and blocked fields as average of the original fields; integrating out the latter keeping the block averages fixed, one obtains a new action $S'$ which describes the interaction between the block variables. 
The important observation is that the partition function in unchanged, so that the long-distance behaviour of the Green-functions is expected to remain unchanged as well.
On the other hand, the lattice spacing is increased by a factor 2 and the 
dimensionless correlation length, which represents the inverse mass of the lightest 
state in the theory, is reduced by a factor 2
$$
\xi\rightarrow\frac{\xi}{2}.
$$
In momentum space, performing this transformation is equivalent to integrate out all momenta between a given scale $q/2$ and $q$.\\
In general, the new action $S'$ will contain every possible interaction term, even if
the original action $S$ had a simple form. It is then useful to start with a general
action of the form
\begin{equation}
S(\phi)=\sum_{\alpha}K_{\alpha}\theta_{\alpha}(\phi),
\end{equation}
where $\phi$ are the original fields \footnote{In this example for simplicity
  we can use scalar fields defined on the lattice sites.}, $\theta_{\alpha}$ are all possible operators (interactions) and $K_{\alpha}$ the
corresponding couplings. The transformed action $S'(\chi)$, where $\chi$ are the blocked fields, is also expanded in terms of
these operators
\begin{equation}
S'(\chi)=\sum_{\alpha}K'_{\alpha}\theta_{\alpha}(\chi).
\end{equation}
A RGT can be seen as a motion in the coupling constant space: $\{K_{\alpha}\}\rightarrow
\{K'_{\alpha}\}$. Under repeated RGTs a coupling constant flow (RG flow) is generated
\begin{equation}
\{K_{\alpha}\}\rightarrow\{K'_{\alpha}\}\rightarrow\{K"_{\alpha}\}\rightarrow ...,
\end{equation}
while the dimensionless correlation length is reduced at every step 
\begin{equation}\label{correlation_flow}
\xi\rightarrow\frac{\xi}{2}\rightarrow\frac{\xi}{4}... .
\end{equation}
The physical theory is preserved under each RGT, and thus along the whole flow. 
There will be special points called \emph{fixed points} (FP) defined by
\begin{equation}
\{K^{*}_{\alpha}\}\rightarrow\{K^{*}_{\alpha}\},
\end{equation}
where the theory reproduces itself under a RGT. Due to \eq{correlation_flow}, at a fixed point
the correlation function must be 0 or $\infty$.\\
Now consider a point $\{K_{\alpha}\}$ close to the FP $\{K^{*}_{\alpha}\}$, with 
$\Delta K_{\alpha}=K_{\alpha}-K^{*}_{\alpha}$ small.
Applying a RGT and expanding around $\{K^{*}\}$ we have 
\begin{equation}\label{rg_linear}
\Delta K_{\alpha}\rightarrow\Delta K'_{\alpha}=K'_{\alpha}-K^{*}_{\alpha}=\sum_{\beta}
\mathcal{T}_{\alpha\beta}\Delta K_{\beta}+O((\Delta K)^2),
\end{equation}
with
\begin{equation}
\mathcal{T}_{\alpha\beta}=\frac{\partial K'_{\alpha}}{\partial K_{\beta}}|_{K=K^*}.
\end{equation}
Let us denote the eigenvectors and eigenvalues of the matrix $\mathcal{T}$ by $h_{a}$
and $\lambda_{a}$ respectively:
\begin{equation}
\sum_{\beta}\mathcal{T}_{\alpha\beta}h_{\beta}^{a}=
\lambda^{a}h_{\alpha}^{a},\quad a=1,2...
\end{equation}
The eigenvectors $h^{a}$ define the eigenoperators
\begin{equation}
h^{a}(\phi)=\sum_{\alpha}h_{\alpha}^{a}\theta_{\alpha}(\phi),
\end{equation}
which define a new basis in terms of which the action can be expanded
\begin{equation}
S(\phi)=S^{FP}(\phi)+\sum_{a}c^{a}h^{a}(\phi),
\end{equation}
where $c^{a}$ are the corresponding coefficients and are called scaling fields.
By analogy, $h^{a}$ are called scaling operators.\\
Iterating the RGT, one obtains that the coupling $c^{a}$ goes over $(\lambda^{a})^{n}c^{a}$
after $n$ RGT steps. For $|\lambda^{a}|>1(|\lambda^{a}|<1)$ the coupling $|c^{a}|$ is 
increasing (decreasing) under repeated RG steps. The corresponding interaction $h^{a}$ is
called relevant (irrelevant). For $|\lambda^{a}|=1$ the behavior of the operator (called
marginal) is decided by higher order corrections in \eq{rg_linear}.\\

These general ideas can now be applied to Yang-Mills theory; due to asymptotic
freedom, the continuum renormalized theory is obtained as $g_{0}\rightarrow 0 (\beta
\rightarrow\infty)$ in which limit the correlation length goes to infinity, the lattice 
spacing $a\rightarrow 0$, the resolution becomes infinitely good and the cutoff artifacts
disappear from the predictions.
Introducing an infinite dimensional coupling constant space,
\begin{equation}\label{coupling_space}
S=S(\beta,K_{1},K_{2},...),
\end{equation}
under a RGT we will have $\{\beta,K_{1},K_{2},...\}\rightarrow\{\beta',K'_{1},K'_{2},...\}$.
The $\beta=\infty$ hyperplane is the so called
critical surface (where $\xi=\infty$). Starting from a critical point, a RGT produces another
critical point. On the critical surface there is a FP with coordinates $K^{*}_{1},K^{*}_{2},
...$. 
The FP defines a basin of attraction, i.e. a set of critical points that converge to
it under RGT. The basin of attraction determines the universality class and all theories belonging to them show the same long-distance behavior.
 The flow along a relevant scaling field, whose end-point is the FP, is called
renormalized trajectory (RT). In Yang-Mills theory, all the eigenoperators lying in the
critical surface are irrelevant, and there is one marginal direction, which becomes relevant
by higher order corrections, pointing out of this surface. The RT is the attractor of all
flows terminating in the critical points that lie in the basin of attraction of the FP.
Along the RT there are no scaling violations as, by construction, all irrelevant fields are
zero. Thus, simulations done using an action along the exact RT will reproduce the
continuum physics without discretization errors.
Simulations on finite lattices of size $L$ are done in a region of coupling space where 
$\xi\ll L$, that is they lie on flows that terminate at critical points that may be quite
far from the FP. Corrections to scaling along these flows may therefore
be large. A method to reduce these corrections is to adjust the action
to lie on another
flow that starts closer to the FP; since the flows
are attracted by the RT, its location may be estimated by starting with the simple plaquette
action and performing some blocking transformation.\\
A usual assumption is that the FP action is local, that is the
strength of the couplings 
fall off exponentially with the size of the loops corresponding to the operators;
nevertheless, it still involves
an infinite number of couplings, but in practice one is forced to truncate the action to a finite number of interactions.
%%%%%%%%%%%%%%%%%%%%%%%%%%%%%%%%%%%%%%%%%%%%%%%%%%%%%%%%%%%%%%%%%%%%%%%%%%%%%%%%%%%%%%%%%%%%%%%%

%%%%%%%%%%%%%%%%%%%%%%%%%%%%%%%%%%%%%%%%%%%%%%%%%%%%%%%%%%%%%%%%%%%%%%%%%
\subsection{RG gauge improved actions in 2-parameter space}
Investigations on the Renormalization Group in $\sigma$-models and Yang-Mills
theories showed that a satisfactory treatment of RGT's can be achieved by restricting the action to a two-parameter space. The most popular examples are the Iwasaki and DBW2 actions where, in addition to the usual plaquette term, planar rectangular $(1\times 2)$ loops are included
\begin{equation}\label{impr_action}
S=\beta\sum_{x}\left(c_{0}\sum_{\mu<\nu}\left\{1-\frac{1}{3}\Re W_{\mu\nu}^{1\times 1}(x)\right\}+c_{1}\sum_{\mu,\nu}\left\{1-\frac{1}{3}\Re W_{\mu\nu}^{1\times 2}(x)\right\}\right),
\end{equation} 
with the normalization condition $c_{0}=1-8c_{1}$.
Note that \eq{impr_action} has the same form as the tree-level Symanzik action 
\eq{lw_action}, \eq{tree_level}. The coefficients $c_{0},c_{1}$ are now determined by RG considerations.

Iwasaki \cite{Iwasaki:1983ck} introduced a block-spin transformation with
the purpose to obtain a gauge action that after a few blockings comes
close to the RT. The strategy is the following:
(i) one assumes to know the lattice action on the RT;
(ii) the action can be expanded in perturbation theory ;
(iii) the correlation functions on the RT are calculated in perturbation theory.
In principle, if one would know the (iii) then
one can follow the opposite direction (iii) $\rightarrow$ (i) and determine
the action on the RT.
In practice, one truncates \eq{coupling_space} to a two-parameter space
and chooses as improved action a  lattice action which is located near
the RT. Iwasaki obtained for the coefficients in \eq{impr_action} the numerical value $c_{1}=-0.331$.

The QCD-TARO collaboration has investigated the flow of the
Wilson action under blocking \cite{deForcrand:1997bx,Takaishi:1996xj}. 
In particular, they used a blocking
transformation with scale factor $b=2$ and determined the couplings on
blocked lattices using the Schwinger-Dyson equation.
The so called DBW2 action is obtained after a Double Blocking of the
Wilson action in 2-dimensional coupling space; 
starting from the coupling $(\beta=6.3)$ on a $32^3\times 64$ lattice,
one obtains $\beta_{1\times 1}=\beta c_{0}\sim 7.986,\beta_{1\times 2}=\beta c_{1}\sim -0.9169$; the corresponding coefficient in \eq{impr_action} is then
$c_{1}=-1.4088$.

To summarize, in a two-parameter space, the coefficient $c_{1}$ in \eq{impr_action} takes different values for various choices of alternative actions
\begin{equation}\label{c1_impr}
c_{1} =\left\{\begin{array}{ll}
-1/12  & \textrm{Symanzik, tree level impr.}\\
-0.331 & \textrm{Iwasaki, RG}\\
-1.4088 & \textrm{DBW2, RG}
\end{array}\right.
\end{equation}
Notice that the strength of the rectangular loops for RG improved actions is larger than what is expected in order to cancel the ${\rm O}(a^2)$ at tree level, i.e. this method suggests an over-correction; this feature will be discussed later in chapter \ref{improved}. 
%%%%%%%%%%%%%%%%%%%%%%%%%%%%%%%%%%%%%%%%%%%%%%%%%%%%%%%%%%%%%%%%%%%%%%%%%%%%%
\subsection{FP action}
Hasenfratz and Niedermayer \cite{Hasenfratz:1994sp} used the framework of the renormalization group to propose a \emph{classically perfect action}.
The basic idea is to consider the FP described before, to evaluate the action at the FP, and then allow
$\beta$ to move away from $\infty$. This does not represent
a RG flow but defines an action for every value of
$\beta$. This defines a ``classical perfect action'', in the sense that its
classical predictions agree with those in the continuum, independently of whether the lattice is fine or coarse. On can say that the FP action is an on-shell tree-level Symanzik improved action to all orders in $a$.
The detailed form of the FP action and the RT will depend on the form and parametrization of the block transformation; the theoretical properties of the FP action and RT are, however, independent of these details.\\
 The next step is the observation that in asymptotically free theories like 4-dimensional $\SUn$, the FP lies at $\beta=\infty$ and in this limit the path integral describing the kernel of the blocking transformation can be calculated in the saddle point approximation. This leads to an equation in classical field theory which determines the FP action. However, this step is too demanding to be performed in a Monte Carlo simulation and one needs a sufficiently fast but at the same time accurate method to approximate the FP action. 
The first works on $\SUthree$ gauge theory proposed a parametrization which used plaquettes, rectangular loops and their powers. Afterwards, \cite{Niedermayer:2000yx} a new parametrization using plaquettes made by original and smeared (``fat'') links was investigated.

%%% Local Variables: 
%%% mode: latex
%%% TeX-master: t
%%% End: 

%% file: potential.tex
\chapter{The static quark potential}\label{chapt_potential}
The potential between a static quark and anti-quark is an object of
considerable theoretical interest that has been studied for almost 30 years.
It defines the energy of gauge fields in the presence of two static color
sources separated by a distance $r$. From the phenomenological point of view, in the limit where the mass of a quark $q$
becomes large compared to typical QCD scale, bound states $q\bar{q}$ are
expected to be described by an effective nonrelativistic Schr\"odinger
equation; the nonrelativistic potential is given by the static quark potential
between charges at distance $r$. In the real world, this description is
expected to be approximately valid for the $b\bar{b}$ and maybe for the
$c\bar{c}$ spectra.
\footnote{ 
The fact that the interaction energy within a heavy quark bound state whose effective Hamiltonian contains a kinetic term will approach the static 
potential defined through the Wilson
  loop in the limit of infinite quark masses is still not clarified.
 Hence, in principle one should distinguish between static and heavy
  quark potential.}
Due to asymptotic freedom of Yang-Mills theory, one expects at short distances a Coulomb-like behavior and the reliability of perturbation theory.
At large distances quark confinement shows up and perturbative methods are
no longer able to describe the behavior of physical observables; in this
context, lattice gauge theories play an important r\^{o}le providing a full non
perturbative approach. Lattice results can be used to investigate
non-perturbative features of the static quark potential as well as to test the
range of validity of perturbation theory, which is still an important
phenomenological issue which has to be investigated more precisely.\\

In this chapter the static potential will be defined and the parameter-free
predictions from perturbation theory and the bosonic string picture will be discussed.
Then the lattice formulation,
the simulation details and the continuum extrapolation will be explained and
the results will be discussed and compared with perturbative and string predictions.

\section{Definition} 
\begin{figure}
\begin{center}
\includegraphics[width=5cm]{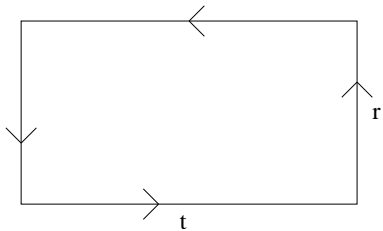}
\end{center}
\caption[A Wilson loop $\mathcal{C}_{r,t}$.]{\footnotesize{A Wilson loop $\mathcal{C}_{r,t}$.}\label{f_wilson_loop}.}
\end{figure}

In the framework of the Euclidean functional integral, observables are gauge
invariant functions of the field variables; as already mentioned, particular choices are the traces
of parallel transporters around closed loops, which are given on the lattice by
\begin{equation}
\Tr_{j}U(\mathcal{C}),
\end{equation}
where $\Tr_{j}$ denotes the trace in some representation $j$ of the gauge
group and $U(\mathcal{C})$ indicates the product of link variables along the
closed curve $\mathcal{C}$.  The expectation values of these loop variables 
\begin{equation}\label{wilson_loop}
W(\mathcal C)=\langle \Tr U(\mathcal{C})\rangle
\end{equation}
are called Wilson loops;
their special r\^{o}le is revealed by the fact that every observable which depends continuously on the link variables can be approximated arbitrarily well by expressions of the form
\begin{equation}
\sum_{n\geq 0}\sum_{\mathcal{C}_{1},...,\mathcal{C}_{n}}c(\mathcal{C}_{1},...,\mathcal{C}_{n})\Tr U(\mathcal{C}_{1})...\Tr U(\mathcal{C}_{n}).
\end{equation}
As a special case consider a rectangular loop $\mathcal{C}_{r,t}$ of side lengths $r$ and
$t$ (\fig{f_wilson_loop}); the static quark potential is related to the large $t$ behavior of the
corresponding Wilson loop in the fundamental representation
\footnote{The same static quark potential could also be defined through the correlation of Polyakov loops, which are products of time-like links at a certain ${\vec x}$ that wind around the lattice in the periodic time direction.} 
\begin{equation}\label{potqq}
V(r)\equiv- \lim_{t\rightarrow\infty}\frac{1}{t} \ln W(\mathcal C_{r,t}),
\end{equation}
so that 
\begin{equation}\label{potqq_asympt}
W(\mathcal C_{r,t})\sim_{t\rightarrow\infty}Ce^{-tV(r)}.
\end{equation}

The meaning of \eq{potqq} can be understood in the Hamiltonian
formalism (see \app{app_trans}), where states with several static external charges in
different representations can be characterized by means of the corresponding
behavior under gauge transformation.
Let $\Psi[U]$ be a wave function depending on the gauge links at time zero, and
fix the temporal gauge
\begin{equation}\label{t_gauge}
U(x,0)=1. 
\end{equation}
The transfer matrix $\mathbb{T}=e^{-\mathbb{H} a}$ is a bounded hermitian operator which acts on these functions.
The only gauge
transformations left which respect this gauge are the time-independent ones;
in this framework the gauge-invariant wave functions describe the physical
states in the vacuum sector, while states with two color indices which
transform under a gauge transformation \eq{gauge_transf} according to
\begin{equation}\label{wavef_qq}
\Psi_{\alpha\beta}[U']=V_{\alpha\gamma}({\vec x})V^{-1}_{\delta\beta}({\vec y})\Psi_{\gamma\delta}[U].
\end{equation}
describe a state  with a static quark at ${\vec x}$ and a static anti-quark at
${\vec y}$. In the following we consider $\vec{y}=\vec{x}+\vec{r}=\vec{x}+r\hat{\mu}$, with $\mu=1$.\\
Due to the gauge invariance of the Hamiltonian, sectors with different distributions of external static charges decouple completely.
The states \eq{wavef_qq} form an Hilbert space $\mathcal{H}_{{\vec x}\vec y}$, 
and the static potential coincides with the ground state energy.
The Wilson loop can be then interpreted as\\
\begin{equation}\label{pot_spectrum}
W(\mathcal C_{r,t})=\langle\Psi|e^{-t\mathbb{H}}|\Psi\rangle=\sum_{n}|\langle\Psi^{(n)}|\Psi\rangle|^{2}e^{-tE_{n}}\sim_{t\rightarrow\infty}|\langle\Psi^{(0)}|\Psi\rangle|^{2}e^{-tV(r)},
\end{equation}
where $\Psi^{(n)}$ is a complete set of eigenstates of the Hamiltonian and $\Psi$ is a state of type \eq{wavef_qq} $\in\mathcal{H}_{{\vec x}\vec y}$.\\

In the continuum, the static quark potential can be described also through
\eq{potqq}; the Wilson loop is defined in the continuum gauge theory as
\begin{equation}\label{pot_continuum}
W(\mathcal C_{r,t})=\Bigg{\langle} \Tr\mathcal{P}\exp\left(\oint_{\mathcal{C}_{r,t}} dx^{\mu}A_{\mu}\right)\Bigg{\rangle},
\end{equation}
where $\mathcal{P}$ denotes the path ordering along the loop $\mathcal C_{r,t}$, and the average is taken with respect to the Euclidean Yang-Mills action \eq{ym_cont}.

%%%%%%%%%%%%%%%%%%%%%%%%%%%%%%%%%%%%%%%%%%%%%%%%%%%%%%%%%%%%%%%%%%%%%%%%%%%%%%%%%%%%%%%%%%%%%%%%%%%%%%%%%%%%%%%%%%%%%%%%%%%%%%%%%%%%%%%%%%%%%%%%%%%%%%%%%%%%%%
\section{The static quark potential in perturbation theory}\label{s:pot_pert}
At high energies, corresponding to small distances between the static quark-antiquark pair, asymptotic freedom justifies the application of perturbative methods. In this section we will consider the continuum Euclidean formulation of Yang-Mills theory.\\

%By analyzing the potential \eq{pot_continuum} in a perturbative way,
%one has first to fix a gauge. Since the Wilson loop is
%gauge-invariant, the final result will not depend on the particular
%choice: a convenient one is in this case the Coulombe gauge,

In a perturbative analysis one can show
\cite{Fischler:1977yf,Appelquist:1977tw,Appelquist:1978es} that at least up to
two loop, after a convenient gauge fixing,
all contributions in \eq{pot_continuum} containing connections to the spatial components of the gauge fields $A_{i}({\vec r},\pm t/2)$ vanish in the limit $t\rightarrow\infty$.\\
The definition of the potential can then be reduced to
\begin{equation}
V_{{\rm pert}}(r)=-\lim_{t\rightarrow\infty}\frac{1}{t}\log
\Bigg{\langle}\Tr\mathcal{T}\exp\left(-\int d^{4}xJ_{\mu}^{a}A_{\mu}^{a}\right) \Bigg{\rangle},
\end{equation}
where $r=|{\vec r}|$, $\mathcal{T}$ is the time-ordering operator and
\begin{equation}
J_{\mu}^{a}=ig_{0}\delta_{\mu 0}T_{a}[\delta({\vec x})-\delta({\vec x}+{\vec r})]\theta(t^2/4-x_{0}^2)
\end{equation}
defines a static quark in ${\vec x}$ and a static antiquark in $({\vec x}+{\vec r})$
that couple to the gauge field $A_{\mu}$.

From this definition, once a regularization scheme is fixed, for
example dimensional regularization in $d=4-\epsilon$ dimensions,
one can perform the renormalization, subtracting the divergences by
substituting the bare coupling with the renormalized one, $\gbar$. For
example, at two loop one obtains \cite{Fischler:1977yf,Appelquist:1977tw,Appelquist:1978es}
\begin{equation}
V_{{\rm pert}}(r)={\int\frac{d^{3}q}{(2\pi)^3}}e^{i {\vec q}{\vec
    r}}\tilde{V}_{{\rm pert}}(q) +V_{self},
\end{equation}
with $q\equiv q_{1}\equiv|\vec{q}|$,
\begin{equation}\label{v_perturba1}
\tilde{V}(q)_{{\rm pert}}  =   -\frac{C_{F}4\pi}{q^2}\alphaMSbar(\mu)
\big\{1+f_{1}(q,\mu)\alphaMSbar(\mu)+f_{2}(q,\mu)\alphaMSbar^{2}(\mu)+...\big\}
\end{equation}
with $C_{F}=\frac{4}{3}$ for $\SUthree$, $\alpha=\frac{\gbar^{2}}{4\pi}$;
the coefficients $f_{1},f_{2}$ are known and will be discussed later.
The constant part $V_{self}$ corresponds to the self energy
of the static sources; it has to be fixed by imposing an additional condition
on the potential, for example that it vanishes at a given distance $r$.\\
From \eq{v_perturba1} one can see that, once the self energy has been
subtracted, the perturbative potential can be rewritten in a Coulomb-like form 
(the subscript is now omitted)
\begin{equation}
\tilde{V}(q)=-4\pi C_{F}\frac{\alphav(\mu)}{q^2},\quad \mu^2=q^2,
\end{equation}
where $\alphav(\mu)$ is a renormalized running coupling ("effective
charge"), which defines the so called V-scheme, and is connected to
$\alphaMSbar(\mu)$ by the relation \footnote{By matching perturbatively two
  schemes one always encounters a relative scale ambiguity: in this case for
  examples one has to specify the relation between $q$ and $\mu$.}
\begin{equation}\label{relation_partic}
\alphav(q)=\alphaMSbar(\mu)\left(1+f_{1}(q,\mu)\alphaMSbar(\mu)+f_{2}(q,\mu)\alphaMSbar(\mu)^{2} +...  \right).
\end{equation}
Alternatively, one can define a coupling from the potential in the coordinate space ($\overline{V}$-scheme)
\begin{equation}\label{e_alphavbar}
V(r)=-C_{F}\frac{\alphavbar(\mu)}{r},\quad \mu=\frac{1}{r},
\end{equation}
or use the force ($\qqbar$-scheme)
\begin{equation}\label{qqbar_scheme}
F(r)=\frac{d V}{d r}=C_{F}\frac{\alphaqqbar(\mu)}{r^2},\quad \mu=\frac{1}{r}.
\end{equation}
An important observation at this point is that only the coupling
$\alphaqqbar(\mu)$ defined through the force is also non-perturbatively
defined, because the self energy term of the potential drops out. Non
perturbatively it is not clear how to subtract the constant part, and
hence $\alphav$ and $\alphavbar$ are restricted to perturbation
theory. This is a first reason to consider the force as
natural observable for comparison between perturbation theory and non-perturbative QCD.
Furthermore, although the couplings $\alphav,\alphavbar,\alphaqqbar$
differ only by kinematics (differentiation, the Fourier transformation), it makes a big difference 
for the applicability of perturbation theory which one is chosen to represent
the potential. This problem will be discussed in details in sect.\ref{comparison_pert}.\\

In general, if $\gbar(\mu)$ is a renormalized coupling, its
running is described by the $\beta$-function
\begin{equation}\label{beta_function}
\beta(\gbar)=\mu\frac{\partial}{\partial\mu}\gbar,
\end{equation}
which has a perturbative expansion 
\begin{eqnarray}\label{beta_pert}
\beta(\gbar) &\sim_{\gbar \rightarrow 0}& - \gbar^3 
                     \{ b_0 + b_1 \gbar^2 + b_2 \gbar^4 + \ldots \}.
\end{eqnarray}
The first two coefficients are universal in the sense that they do not
depend on the renormalization scheme in which $\gbar$ is
defined.
For $\SUthree$ gauge theory and $\nf$ massless quarks one obtains \cite{Jones:1974mm}, \cite{Caswell:1974gg}
\begin{eqnarray}
              && b_0=\frac{1}{(4\pi)^2}\left(11 - \frac{2}{3}\nf\right),
              \, b_1=\frac{1}{(4\pi)^4}\left( 102 - \frac{38}{3} \nf\right).
\end{eqnarray}
The three-loop coefficient of the $\beta$-function can be expressed as
\be\label{e_b2gen}
b_{2}^{S}=b_{2}^{S}|_{\nf=0}+\frac{1}{(4\pi)^6}\left(e_{1}\nf+e_{2}\nf^2+e_{3}\nf^3\right)
\ee
with $e_i$ listed in \tab{t_ei}; some of them could be taken directly from the
literature
\cite{Peter:1997me,Schroder:1998vy,Fischler:1977yf,Billoire:1980ih,Larin:1993tp,pert:2loop},
others such $b_{2}^{\qqbar}$ had to be computed by straightforward
algebra. For comparison we 
included also the Schr\"odinger functional scheme \footnote{In this discussion on different running couplings, we will include also
a particular scheme used by the lattice community to solve the problem of 
matching low energy hadronic scales with high energy perturbative regimes.
In the last ten years the ALPHA collaboration applied a finite size technique, the so called Schr\"odinger functional (SF) scheme, with the goal to compute the running coupling at short distances in units of the low energy scales of the theory \cite{alpha:sigma,alpha:SU2,alpha:SU3}. The strategy is the following: one begins by formulating the theory in a finite spatial volume with linear extension $L$ and periodic boundary conditions in all spatial directions. Next one introduces some renormalized coupling $\gbar(L)^{2}$ which depends only on the scale $L$ and which can hence be considered a running coupling. This coupling is then computed over a range of $L$ through numerical simulations of the lattice theory, using a recursive procedure which allows one to go from large values of $L$ (where one can make contact with the non-perturbative scales) to the perturbative domain where $L$ is small. At this scale $\alphaSF(L)$ can be related analytically to other more commonly used 
schemes, like $\MSbar$, which are defined in infinite volume. 
The method has been applied successfully also to the case with two massless flavors \cite{DellaMorte:2002vm}.}.

%%%%%%%%%%%%%%%%%%%%%%%%%%%%%%%%%%%%%%%%%%%%%%%%%%%%%%%%%%%%%%%%%%%%%%%%%%%%%%%%%%%
\begin{table}[ht]
\begin{center}
\begin{tabular}{c| c| c| c }
\hline &&&\\[-1ex]
$S$ & $e_{1}$ & $e_{2}$ & $e_{3}$  \\[1ex]
\hline&&&\\[-1ex]
$\MSbar$ & $-\frac{5033}{18}$ & $325/54$  & 0\\[2ex]
$\rm{V}$      &  $\frac{3}{2}\pi^4-24\pi^2-\frac{2239}{6}-\frac{704}{3}\zeta(3)$ & $\frac{377}{54}+\frac{104}{9}\zeta(3)$ & 0\\[2ex]
$\overline{\rm V}$ & $-\frac{2239}{6}-\frac{704}{3}\zeta(3)+\frac{3}{2}\pi^4-\frac{314}{3}\pi^2$ & $\frac{377}{54}+\frac{104}{9}\zeta(3)+\frac{44}{9}\pi^2$ & $-\frac{8}{81}\pi^2$\\[2ex]
$\rm{q\bar{q}}$  &  $\frac{3}{2}\pi^4-\frac{314}{3}\pi^2+\frac{3569}{6}-\frac{704}{3}\zeta(3)$ & $\frac{2791}{54}+\frac{44}{9}\pi^2+\frac{104}{9}\zeta(3)$ & $\frac{32}{27}-\frac{8}{81}\pi^2$\\[2ex] 
$\rm{SF}$  & $-0.275(5)\times(4\pi)^{3}$ & $ 0.0361(4)\times(4\pi)^{3}$  
& $-0.00175(1)\times(4\pi)^{3}$ \\[1ex]
\hline
\end{tabular}
\end{center}
  \caption[Coefficients $e_i$ in $b_{2}$ for several schemes]{\footnotesize \label{t_ei} Coefficients $e_i$ of \protect\eq{e_b2gen}.}
\end{table}
%%%%%%%%%%%%%%%%%%%%%%%%%%%%%%%%%%%%%%%%%%%%%%%%%%%%%%%%%%%%%%%%%%%%%%%%%%%%%%%%%%%

From \eq{beta_function} follows that the behavior of $\gbar$ at high
energies is given by
\begin{equation}\label{gbar_high_en}
\gbar^2\sim_{\mu\rightarrow\infty}\frac{1}{b_{0}t}-\frac{b_{1}\log
  t}{b_{0}^3 t^2}+{\rm O}\left(t^{-3}(\log t)^2\right),
\end{equation}
where 
\begin{equation}
t=\log\left(\frac{\mu^2}{\Lambda_{S}^{2}}\right),
\end{equation}
and $\Lambda_{S}$ in an integration constant, the so-called
$\Lambda$-parameter associated with the given scheme, with the
dimension of a mass. Although classical non-Abelian gauge field theory
does not contain any mass scale, the
renormalization introduces this mass scale into the quantized
theory: this fact is called \emph{dimensional transmutation}.\\
The solution of \eq{beta_function},
\be\label{e_lambda}
\lambdas=\mu(b_{0}\gbar^{2})^{-b_{1}/2b_{0}^2}
\rme^{-1/b_{0}\gbar^{2}}\exp
\bigg\{-\int_{0}^{\gbar}dx
\left[\frac{1}{\beta(x)}+\frac{1}{b_{0}x^{3}}-
\frac{b_{1}}{b_{0}^{2}x}\right]\bigg\},
\ee
relates the coupling $\gbar=\gbar(\mu)$ to the $\Lambda$-parameter and will be used to obtain a parameter free perturbative prediction for $\gbar$.\\

\Eq{relation_partic} can be generalized: given two running
couplings $\alphas$ and $\alphasp$ defined in two different 
renormalization schemes, they can be matched in perturbation theory by
\be \label{e_match}
  \alphasp(s\mu) = \alphas(\mu) + f_1^{S'S}(s) \alphas(\mu)^2 + 
                      f_2^{S'S}(s) \alphas(\mu)^3 + \ldots,
\ee
where $s$ is a relative scale which is a free parameter (for example
$s=q/\mu$ in \eq{relation_partic}).

The one-loop coefficient in \eq{e_match} assumes the general form
\be \label{e_f1}
f_{1}^{S'S}(s)=\frac{1}{4\pi}(a_{1}+a_{2}\nf)-8\pi b_{0}\log(s),
\ee
where $a_{1}$ and $a_{2}$ are listed in \tab{t_ai}, with $\MSbar$ 
as reference scheme. The other coefficients can be evaluated by
$f_{1}^{SS'}(s)=f_{1}^{SS''}(s)+f_{1}^{S''S'}(s)$.
%%%%%%%%%%%%%%%%%%%%%%%%%%%%%%%%%%%%%%%%%%%%%%%%%%%%%%%%%%%%%%%%%%%%%%%%%%%%%%%%%%%%%%%%
\begin{table}[ht]
\begin{center}
\begin{tabular}{ c  c  c}
\hline\\[-1ex]
$S'$         & $a_{1}$ & $a_{2}$      \\[1ex]
\hline\\[-1ex]
$\rm V$              & $31/3$  &$-10/9$              \\[1ex]
$\overline{\rm V}$  & $31/3+22\gamma_{E}$ & $-10/9-4\gamma_{E}/3$  \\[1ex]
$\rm{q\bar q}$              & $-35/3+22\gamma_{E}$ & $2/9-4\gamma_{E}/3$ \\[1ex]
$\rm{SF}$   & $-1.25563(4)/(4\pi)$   &  $-0.03983(2)/(4\pi)$   \\[1ex]
\hline 
  \end{tabular} 
\end{center}
  \caption[Matching coefficients for different schemes.]{\footnotesize \label{t_ai} Coefficients $a_i$ of \protect\eq{e_f1} for $S=\MSbar$}
  \end{table}
%%%%%%%%%%%%%%%%%%%%%%%%%%%%%%%%%%%%%%%%%%%%%%%%%%%%%%%%%%%%%%%%%%%%%%%%%%%%%%%%%
\\
Using \eq{gbar_high_en} and the universality of $b_{0}$ and $b_{1}$,
one finds that the $\Lambda$- parameters in two schemes $S$ and $S'$ are related by
\begin{equation}
\Lambda_{\rm S'}=\Lambda_{\rm S}\exp\left\{\frac{f_{1}^{S'S}(1)}{8\pi b_{0}}\right\}
\end{equation}
exactly. The fact that this matching involves only one-loop coefficients is
due to the fact that it is related to the asymptotic behavior at $\mu\rightarrow\infty$.\\
For the 3-loop coefficient $b_{2}^{S}$ of the $\beta$-function, we
have
\begin{equation}\label{beta2_rel}
b_{2}^{S'} = b_{2}^{S}-b_{1}\frac{f_{1}^{S'S}(s)}{4\pi}+\frac{b_{0}}{(4\pi)^2}\left(f_{2}^{S'S}(s)- f_{1}^{S'S}(s)^{2} \right).
\end{equation}
This relation, together with \tab{t_ei} and \tab{t_ai}, can be used to
obtain the two-loop coefficient in \eq{e_match}
\begin{equation}
f_{2}^{S'S}(s)=\frac{(4\pi)^2}{b_{0}}\left\{b_{2}^{S'}-b_{2}^{S}+b_{1}\frac{f_{1}^{S'S}(s)}{4\pi}-b_{0}\left[\frac{f_{1}^{S'S}(s)}{4\pi}^{2}\right]\right\}.
\end{equation}
As a particular example, one can consider the evaluation of the coefficient $b_{2}^{\qqbar}$; the starting point is the relation
\begin{equation}\label{alphaqq-v}
\alphaqqbar(\mu)=\alpha_{\overline{V}}(\mu)\left(1+2\frac{\beta^{\overline{V}}(\bar{g}(\mu))}{\sqrt{4\pi\alpha_{\overline{V}}(\mu)}}\right),
\end{equation}
which follows directly from the definition of the force.\\
\Eq{alphaqq-v} defines the connection between the $\qqbar$ and $\Vbar$ schemes; in particular, one can read off the matching coefficients
\begin{equation}
f_{1}=-2 b_{0}4\pi,\quad f_{2}=-2b_{1}(4\pi)^2,
\end{equation}
which are related to the case $s=1$ in \eq{e_match}  \footnote{Note that the values of $b_{2}$
  computed via \protect\eq{beta2_rel} must be independent on the factor $s$,
  so that one is allowed to consider a particular case.}, with $S=\Vbar$ and $S'=\qqbar$.
From \eq{beta2_rel} one can easily obtain 
\begin{equation}
b_{2}^{\qqbar}=b_{2}^{\Vbar}-b_{0}^3,
\end{equation}
which is the simplest way to obtain $b_{2}^{\qqbar}$ from $b_{2}^{\Vbar}$, which is known from the literature.\\

In the perturbative matching of different schemes the choice of the relative scale factor $s$
plays an important r\^ole for the  
quality of the perturbative prediction \cite{Peter:1997me,alpha:SU2impr}.
The only viable criterion for fixing $s$ appears to be to demand that
the coefficients $f_{i}(s)$ are small ("fastest apparent
convergence"). For instance, the choice 
\begin{equation}\label{eq_s0}
s=s_{0}=\frac{\Lambda_{S'}}{\Lambda_{S}}
\end{equation}
yields
\begin{eqnarray}
f_{1}(s_{0}) & = & 0,\\
f_{2}(s_{0}) & = & \frac{(4\pi)^2}{b_{0}}\left[b_{2}^{S'}- b_{2}^{S}\right],
\end{eqnarray}
and $f_{2}(s_{0})$ can be shown to be close to the minimum of $|f_{2}(s)|$.\\
This relative scale ambiguity is related to the fact that we are matching two
couplings at finite $\mu$; alternatively, if one knows the
$\Lambda$ parameter in a given renormalization
scheme, then truncating in \eq{beta_pert} after the term $b_{n-1}$ and
solving \eq{e_lambda} (numerically) for $\gbar$ at given $\mu$ (in
units of $\Lambda$) defines the $n$-loop RG solution for the
coupling. 
In single scale problems, such as the static potential
depending only on the separation $r$, this 
is expected to be the best perturbative prediction. 
In contrast to the expansion \eq{e_match}, the matching is now
realized at $\mu=\infty$  and one does not need to fix a relative
scale parameter $s$.

\begin{figure}[ht]
\begin{center}
\includegraphics[width=9cm]{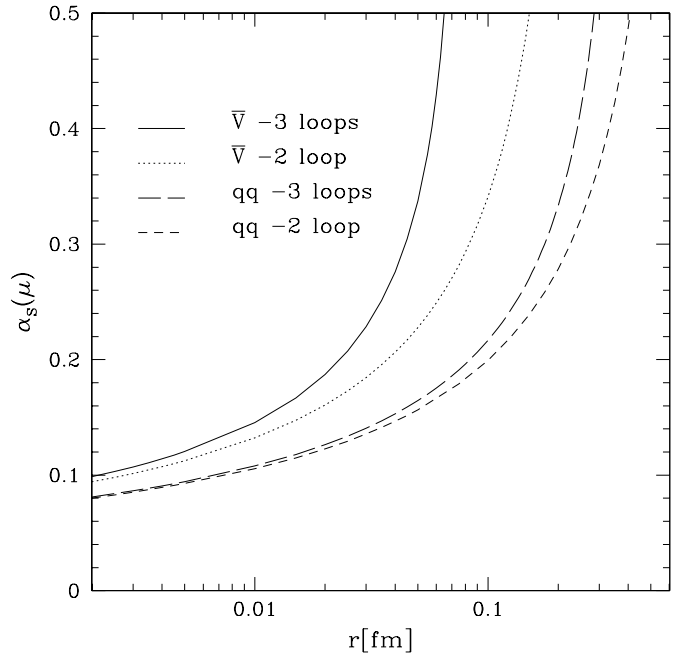}
\end{center}\vspace{-0.8cm}
\caption[ Running coupling obtained by integration of the 
RG with truncation of the $\beta$-functions at 2- and 3-loop.]{\footnotesize \label{f_alpha_pert} Running coupling obtained by integration of the 
RG with truncation of the $\beta$-functions at 2- and 3-loop. 
We use $\Lambda_{\MSbar}=238\MeV$ \protect\cite{mbar:pap1}}
\end{figure}

The ALPHA collaboration evaluated $\Lambda_{\MSbar}$ for $\nf=0$ from lattice calculations employing the Schr\"odinger functional scheme \cite{mbar:pap1}, extracting it at a sufficiently high scale $\mu$ where the perturbative error is negligible and converting it to the $\MSbar$ scheme. In terms of the length scale $\rnod\approx 0.5\fm$, which will be defined in section \ref{setting_the_scale}, the result is
\begin{equation}\label{e_lambda_rnod}
\Lambda_{\MSbar}\rnod=0.602(48),
\end{equation}
corresponding to 
\begin{equation}
\Lambda_{\MSbar}\approx 238 \,\MeV.
\end{equation}
This result, together with the ratios $\Lambda_{S}/\Lambda_{\MSbar}$ and the 3-loop coefficient of the $\beta$-function $b_{2}^{S}$, allows to use \eq{e_lambda} for a parameter-free perturbative prediction for $\alpha_{S}$ from 2- and 3-loop RG. For completeness the numerical values of $\Lambda_{S}/\Lambda_{\MSbar}$ and $b_{2}^{S}$ for $\nf=0$ are collected in \tab{t_lambda_b2},

%%%%%%%%%%%%%%%%%%%%%%%%%%%%%%%%%%%%%%%%%%%%%%%%%%%%%%%%%%%%%%%%%%%%%%%%%%%%%%%%%%%%%%%%%%%
\begin{table}[ht]
\begin{center}
\begin{tabular}{ c |  c  c  c  c }
\hline\\[-1ex]
scheme $S$:       &  $\qqbar$  & $V$  & $\overline{V}$  &  $\rm SF$  \\[1ex]
\hline\\[-1ex]
$\Lambda_{S}/\Lambda_{\MSbar}$ &  $\exp(\gamma_{E}-35/66)$ & $\exp(31/66)$ & 
   $\exp(31/66+\gamma_{E})$ &$0.48811(1)$\\  
$b_{2}^{S}\times(4\pi)^3$   & $1.6524$ & $2.1287$ & $4.3353$ & $0.483(9)$\\[1ex]
\hline 
  \end{tabular} 
\end{center}
 \caption[ Ratio of $\Lambda$-parameters and 2-loop coefficient of the $\beta$-function for various schemes for $\nf=0$.]{\footnotesize \label{t_lambda_b2} Ratio of $\Lambda$-parameters and 2-loop coefficient of the $\beta$-function for various schemes for $\nf=0$. These results follow from \protect\cite{Peter:1997me,Schroder:1998vy,Fischler:1977yf,Billoire:1980ih,Larin:1993tp,Bode:1998hd,pert:2loop,Tarasov:1980au}.}
  \end{table} 
%%%%%%%%%%%%%%%%%%%%%%%%%%%%%%%%%%%%%%%%%%%%%%%%%%%%%%%%%%%%%%%%%%%%%%%%%%%%%%%%%%%%%%%%%%%%%
The couplings $\alphaqqbar$ and $\alphavbar$ 
from 2- and 3-loop RG are illustrated in 
\fig{f_alpha_pert}, using the central value 
$\Lambda_{\MSbar} \rnod=0.602$ (the $8\%$ overall uncertainty of this number 
corresponds to a common small horizontal shift of all curves in the figure). 
By analyzing the relative difference between the 2- and 3-loop behavior 
of $\alphaqqbar$, one observes that the  perturbative expansion appears 
quite well behaved up to distances $r\sim 0.25 \fm$. At $r\sim 0.2 \fm$
one would expect the 3-loop curve to have an accuracy of about 10\%.
Since the force is completely equivalent to $\alphaqqbar$
it is given with the same relative accuracy; the potential
may be obtained by integration of the force. 
On the other hand, we note that the 3-loop coefficients $b_2$ 
for $V$ and $\overline{V}$-schemes are larger 
in comparison to $b_{2}^{\qqbar}$ (see \tab{t_lambda_b2}), in particular
in the $\overline{V}$ scheme. As a consequence, the 
difference between the 2-loop and the
3-loop running coupling in this scheme 
is only small at very short distances and this
perturbative expansion appears to be applicable only up to 
$\alpha \sim 0.15$. This is also illustrated in \fig{f_alpha_pert}.

%%%%%%%%%%%%%%%%%%%%%%%%%%%%%%%%%%%%%%%%%%%%%%%%%%%%%%%%%%%%%%%%%%%%%%%%%%%%%
%%%%%%%%%%%%%%%%%%%%%%%%%%%%%%%%%%%%%%%%%%%%%%%%%%%%%%%%%%%%%%%%%%%%%%%%%%%%%
Furthermore, if one observes the numerical values of the matching coefficients $f_2(s_0)$ \eq{e_match} for $\nf=0$ in \tab{t_coeff}, one can notice that the SF-scheme is very close to
the $\MSbar$-scheme, the $\qqbar$ scheme is not very far, but the other 
schemes have quite large values of $f_2(s_0)$ in relation to the 
$\MSbar$-scheme. In particular, the large coefficient between the $\MSbar$ 
scheme and
the $\Vbar$ scheme means that the direct expansion
of the coordinate space potential in terms of $\alphaMSbar$ (or $\alphaSF$)
 is badly behaved,
as it has been pointed out in \cite{Peter:1997me,Melles:2000dq}.

\begin{table}[ht]
\begin{center}
\begin{tabular}{ c|  c  c  c  c  }
\hline\\[-1ex]
$f_{2}^{\rm {S'S}}(s_{0})$ & $\MSbar$         & $\qqbar$  & $V$     & $\Vbar$    \\[1ex]
\hline\\[-1ex]
$\qqbar$      & $1.0653$ &              &              &                   \\
$V$             & $1.6095$ & $0.5441$ &               &                      \\
$\Vbar$         & $4.1303$ & $3.0650$ &  $2.5208$   &                      \\
SF            & $-0.271(10)$  & $-1.336(10)$  & $-1.880(10)$      & $-4.401(10)$             \\[1ex]
\hline
  \end{tabular} 
\end{center}
  \caption[ 2-loop coefficients for $s=s_0=\Lambda_{S'}/\Lambda_{S}$ and $\nf=0$.]{\footnotesize\label{t_coeff} 2-loop coefficients for $s=s_0=\Lambda_{S'}/\Lambda_{S}$ and $\nf=0$.}
  \end{table}

For the Schr\"odinger functional coupling, $\alphaSF(\mu)$, it was observed in \cite{mbar:pap1} that it is in remarkable qualitative agreement with perturbation theory for $\alpha<0.3$ and for $\alpha<0.2$ the 3-loop expression describes $\alphaSF(\mu)$ within better than $2\%$. It is an interesting question, whether the coupling in this scheme is a special case or whether this is a more ``general property of the theory''. For this purpose the perturbative prediction for the potential and the force will be compared with our non-perturbative computation in 
section \ref{comparison_pert}.

%%%%%%%%%%%%%%%%%%%%%%%%%%%%%%%%%%%%%%%%%%%%%%%%%%%%%%%%%%%%%%%%%%%%%%%%%%%%%%%
%%%%%%%%%%%%%%%%%%%%%%%%%%%%%%%%%%%%%%%%%%%%%%%%%%%%%%%%%%%%%%%%%%%%%%%%%%%%%%%
\section{Quark confinement and the bosonic string picture}\label{sec_string}
Perturbation theory can be applied to gauge theories in the continuum as well
as on the lattice. Several interesting aspects, however, are inaccessible to
a perturbative treatment and require non-perturbative methods.
One of these is the strong coupling expansion, which amounts to expanding in
powers of the inverse coupling and can be used in the lattice formulation to investigate the region where
$g_{0}$ is large.
If one applies this method to the Wilson loops \eq{potqq_asympt} \cite{munster_strong} one finds that the leading term is
\begin{equation}\label{area_law}
W(\mathcal C_{r,t})=2e^{-\sigma rt}+..., 
\end{equation}
where $\sigma=-\log\frac{1}{24g_{0}^2}$ is the so-called string tension. 
By analyzing higher orders in the expansion it was shown \cite{Osterwalder:1978pc} that the strong coupling expansion of $W(\mathcal C_{r,t})$
has a finite range of convergence and within this range the Wilson loop obeys
\eq{area_law}, which is called the \emph{area law}
and establishes that the potential $V(r)$ asymptotically rises linearly with $r$
\begin{equation}
V(r)\sim_{r\rightarrow \infty}\sigma r .
\end{equation}
This behavior is called linear quark confinement \cite{wilson}.\\

The strong coupling expansion has given an important contribution to an intuitive picture of the confinement mechanism, where a tube of non-Abelian chromoelectic flux holds quarks and antiquarks together. In the limit of very large $r$, one can expect that these flux tubes behave like strings with fixed ends, and the expectation values of large Wilson loops could be predicted by an effective bosonic string theory \cite{nambu}.
In this theory, the Wilson loops are expressed in terms of functional integrals over all two-dimensional surfaces bounded by the loops.
For $r,t$ large, the configuration with the minimal area will dominate; the deviation of the fluctuating surface from the minimal one is described by a bosonic vector field, whose action defines the effective string theory.
Under this assumption, the static quark potential $V(r)$ has the asymptotic
expansion for large $r$ \cite{Luscher:1980fr,Luscher:1981ac} \footnote{We
  assume as leading order of the effective string theory $V(r)=\sigma
  r+\rm{const}$. \Eq{pot_string} hence represents the next-to-leading order expression.}
\begin{equation}\label{pot_string}
V(r)=\sigma r +\mu+\frac{\gamma}{r}+{\rm O}\left(\frac{1}{r^2}\right),
\end{equation}
where $\mu$ is a mass constant depending on the regularization and
\begin{equation}
\gamma=-\frac{\pi}{24}(d-2)
\end{equation}
is a quantum effect that is characteristic of the relativistic bosonic string.
$\gamma$ is universal in the sense that it depends only on the dimension $d$ of the space-time and is common to a large class of effective string actions.
The lattice evaluation of the static quark potential $V(r)$ should allow in
principle a test of the validity of \eq{pot_string}. However, it turns out
that the statistical and systematic errors increase rapidly at large $r$, and
thus it is very difficult to obtain reliable values for $V(r)$ at large distances.
L\"uscher and Weisz proposed a new type of simulation algorithm \cite{Luscher:2001up} for the $\SUn$ gauge theory that leads to an exponential reduction of the statistical errors in numerical calculations of Polyakov loop correlation functions.
They evaluated the force $V'(r)$ and the second derivative $V^{''}(r)$
up to $r\approx 1\fm$ \cite{Luscher:2002qv} with sufficient precision to test the string predictions. For $d=3$ their numerical results at large $r$ are in full agreement with \eq{pot_string} without need of further corrections. For $d=4$, if one allows for a higher order term
\begin{equation}\label{string_corrected}
V(r)=\sigma r +\mu-\frac{\pi}{24r}(d-2)\left(1+\frac{b}{r}\right)
\end{equation}
one finds agreement in the range $r>0.5\fm$ by setting $b=0.04 \fm$.

As pointed out by the authors, this good agreement seems surprising, because
 at such distances one can strongly doubt the physical picture of a thin
 fluctuating flux tube.\\
 A further important test is related to the energy spectrum of the
excited string levels, which at next to leading order of the effective
string theory are given by \cite{Dietz:1983uc,Ambjorn:1984yu,string:caselle}
\begin{equation}
E_{n}=\sigma r +\mu+\frac{\pi}{r}\left[-\frac{1}{24}(d-2)+n \right].
\end{equation} 
In particular, to this order the energy gap $\Delta E= E_{1}-V(r)$
is equal to $\frac{\pi}{r}$. It was observed that the spectrum $E_{n}$ does not seem to agree with the one obtained in the effective string picture \cite{Juge:1998nc,Morningstar:1998da,Juge:1999ar,Juge:2001mj,Juge:2001rb},
although this argument alone can not exclude the validity of the effective
theory, since higher order corrections could be responsible for the large
discrepancy. Moreover, one has to take into account the technical difficulties
in the evaluation of the spectrum.
% recently, the L\"uscher-Weisz algorithm was applied to the 3-dimensional $\SUtwo$ gauge theory to investigate both large
%Wilson and Polyakov loop \cite{Majumdar:2002mr}. In this work is pointed out that the finite
%$t$ corrections are crucial for comparing the data with the expected string spectrum. \\
More investigations are then needed to understand if the string picture is really able to describe the large distance properties including static quark confinement and the excited states.

\section{Evaluation of the potential on the lattice}
Since the original work by Creutz in the $\SUtwo$ gauge theory, many
computations of the potential from lattice gauge theories have been performed
in $\SUthree$, but the short distance regime has not
yet been investigated convincingly. The reason is that by "the potential"
one means the potential in infinite volume; in practice one considers systems with finite extension $L$ and one
knows that for the
force at distances up to $r\sim 0.5\fm$, the deviation from infinite volume
are small on an $L^4$ torus with $L=1.5\fm$. With such a physical value of $L$,
an investigation of distances of $r\sim0.05\fm$ with control over
${\rm O}(a/r)$ effects requires very large lattices. 
Setting the number of flavors $\nf=0$ (for
our purposes this means the pure Yang-Mills theory), large lattices may
nowadays be simulated. We studied the static potential in the range 
$0.05~\fm\leq r \leq 0.8~\fm$, employing a sequence of lattices up to $64^4$.
This study of the potential from short to intermediate distances was performed 
using the standard Wilson action and extrapolating the results to the continuum.
In chapter \ref{improved} we will consider the potential for improved actions
with the purpose to investigating the lattice artifacts, and we will restrict
ourselves to coarser
lattices leaving out the goal to perform a continuum extrapolation.

\subsection{Setting the scale in Yang-Mills lattice theories}\label{setting_the_scale}
The first step in  the evaluation of renormalized quantities from lattice QCD
is to set the scale.
In a pure gauge theory the coupling $g_{0}$ is the only parameter in the
Lagrangian, then one has to specify one physical quantity in order to
renormalize the theory. It turns out that it is convenient to choose  a
dimensionful long distance observable, which introduces \emph{the} scale into
the theory: predictions of all other dimensionful quantities are then
expressed in units of this scale. They are well defined renormalized
quantities and can be extrapolated to the continuum limit from results at 
finite resolution.\\
 It is clear that the scale should be chosen with care since
it influences the precision of many predictions. 
From the confinement picture, one expects a linearly growing potential at large distances, with a certain slope which defines the string tension. 
One possible choice to set the scale is then
\begin{equation} 
\sigma=\lim_{r\rightarrow\infty}F(r),
\end{equation}
where $F(r)=\frac{dV}{dr}$.
This limiting procedure is not easy to perform because statistical and
systematic errors in the force increase with the distance. 
Moreover, if one believes the string picture described in sec. \ref{sec_string}, one can expect that higher order terms proportional to $b$ in \eq{string_corrected} can be neglected only at very large distances and this should be taken into account in the extraction of the string tension from the potential.\\ 
In most cases, the string tension is extracted from lattice data by fitting the potential with a three-parameter formula
\begin{equation}
V(r)=a_{1}+a_{2}r+\frac{a_{3}}{r},
\end{equation}
but to identify $a_{2}$ with the string tension can be a source of large and uncontrolled systematic errors.\\
For this reason it is 
desirable to use the force at intermediate distances to determine the scale.
One can observe that both $b\bar b$ and $c\bar c$ spectra can be accurately described by
one effective potential in the range of $r\approx 0.2\,\fm$ to $r\approx 1\,\fm$, and
this suggests that the range where one has the best information on
the force $F(r)$ between static quarks is at distances of around $0.5\,\fm$. One then calculates $r(c)$ satisfying the equation \cite{pot:r0}
\begin{equation}\label{e_rnod}
r^{2}F(r)\big|_{r=r(c)}=c\,,\quad \rnod=r(1.65)\approx 0.5\,\fm.
\end{equation}

When one is interested in short distance properties of the theory, it is
convenient to choose a smaller reference length scale, because large distances
on fine lattices are very expensive from the computational point of view.
We decided to separately consider two regions, $0.05\,\fm\leq r\leq 0.3\,\fm$ and
 $0.2\,\fm\leq r\leq 0.8\,\fm$, where the first requires very large lattices and
 small lattice spacings and the second one had been simulated before on
 coarser lattices \cite{pot:r0_SU3}. The region of overlap serves for calibration.
 We therefore introduced an additional scale
\be  \label{e_rc}
  \rc=r(0.65)\,
\ee
and evaluated its relation to $\rnod$.

\subsection{Definition of the force}\label{def_of_the_force}
In the continuum, the definition \eq{e_rnod} is unique, but at finite lattice
spacing one has to specify which discretization of the force is to be used.
Given the potential $V(r)$ in a certain direction, in general one has
\be\label{finite_force}
F(r')=\left[V(r)-V(r-a) \right]/a.
\ee
The naive choice for $r^{\prime}$ would be 
\be\label{r_naiv}
r'=r_{n}\equiv r-\frac{a}{2}.
\ee
Instead of $r_{n}$, we decided to use a distance $\rI$ chosen such that
the force on the lattice has no deviations from the force in the 
continuum when evaluated at tree level.
On the lattice, the force at tree level is given by
\begin{equation}\label{force_treelevel}
F_{tree}(r)=-\frac{4}{3}\frac{g_{0}^{2}}{a}\int_{-\pi}^{\pi}\frac{d^{3}k}{(2\pi)^3}\frac{\cos(rk_{1}/a)-\cos((r-a)k_{1}/a)}  {4\sum_{j=1}^{3}\sin^{2}(k_{j}/2)} 
\end{equation}
and one would introduce  $\rI$ such that
\be
F(\rI)=\frac{4}{3}\frac{g_{0}^{2}}{4\pi\rI^2}+\rm{O}(g_{0}^4).
\ee
The lattice artifacts remain only in the ${O}(g_{0}^4)$ terms; $F(\rI)$ is
called a \emph{tree-level improved observable}.

In the $\SUtwo$ theory it has been observed that by adopting this definition the remaining
lattice artifacts in the force are surprisingly small
\cite{pot:r0} and here we shall again find no evidence for them 
as long as $r>2a$.

Explicitly we have from \eq{force_treelevel}
\be
\label{e_rI}
(4\pi \rI^2)^{-1}=-\left[G(r,0,0)-G(r-a,0,0)\right]/a,
\ee
where 
\begin{equation}
G({\vec r})=\frac{1}{a}\int_{-\pi}^{\pi}\frac{d^{3}k}{(2\pi)^3}\frac{\prod_{j=1}^{3}\cos(x_{j}k_{j}/a)}{4\sum_{j=1}^{3}\sin^{2}(k_{j}/2)} 
\end{equation}
is the (scalar) lattice propagator in 3 dimensions, for general ${\vec r}= (x_1,x_2,x_3)$.
In \cite{coord_space} a method was proposed to construct it directly 
in coordinate space using a recursive relation. We extended the method to 
3-d finding the necessary invariant of the recursion relation  \eq{recursion}.
Formulae are listed  in \app{app1}. 

\subsection{Computation of the potential and force}
The evaluation of the potential from \eq{pot_spectrum} is technically difficult
because of the presence of higher order excitations, so that the ground state
becomes dominant only at large values of $t$,
where the statistical uncertainties can be very large.
In order to enhance the overlap with the physical ground state one can use
smearing techniques, which consists of building operators which describe in our case static quarks and antiquarks in an ``optimized'' way.\\
In particular, we adopted a smearing operator
$\mathcal{S}$ which acts on the spatial components of the gauge fields via \cite{smear:ape}
\be\label{smearing}
\mathcal{S}U(x,k)=\mathcal{P}_{\SUthree}\Bigg\{U(x,k)+\alpha\sum_{j\neq k}
\big[U(x,j)U(x+a\hat{j},k)U^{\dagger}(x+a\hat{k},j)
\ee
$$
+U^{\dagger}(x-a\hat{j},j)U(x-a\hat{j},k)U(x+a\hat{k}-a\hat{j},j)\big]\Bigg\},
$$
where $\mathcal{P}_{\SUthree}$ denotes the projection into the group $\SUthree$ and
$\alpha$ is an adjustable parameter.
\footnote{For any $3\times 3$ matrix $V$, $\mathcal{P}_{\SUthree}V$ denotes the
matrix $W\in\SUthree$ which maximizes $\Re\Tr(WV^{\dagger})$}.

For different smearing levels $l=0,...,M-1$ one then constructs smeared spatial
links according to 
\be
U_{l}(x,k)=\mathcal{S}^{n_{l}}U(x,k).
\ee
In particular we chose
\be
\alpha=\frac{1}{2},\quad M=3.
\ee
In \cite{pot:r0_SU3} $n_{l}$ was set to
\be\label{smear_level}
n_l\sim\frac{l}{2}\left(\frac{\rnod}{a}\right).
\ee 
According to this choice, $n_{2}$ corresponds to what was estimated to be the optimal smearing in \cite{pot:r0_SCRI}. 
Since we want to compute the potential for shorter distances, where one can reach high precision, we adopted in general smaller values than these.
 
Another difficulty is the fact that the Wilson loops
\eq{pot_spectrum} decay exponentially for large $t$. In order to measure
these values in a Monte Carlo simulation with statistical significance, also
the variance of the matrix should decrease exponentially with $t$. To achieve
this, one can apply a so-called multi-hit procedure \cite{PPR} to the time-like links, which are substituted by their expectation values in the fixed configuration of the other field variables
\be\label{multi_hit}
\overline{U}(x,0)=\frac{\int dU(x,0)U(x,0)e^{-S[U(x,0)]}}{\int dU(x,0)e^{-S[U(x,0)] }},
\ee
where $S[U(x,0)]$ denotes the part of the action depending on $U(x,0)$.
For a general observable, the substitution of $U(x,0)$ with
$\overline{U}(x,0)$ is allowed under the following restrictions: (i) the
observable must depend linearly on the considered links and (ii) no pair of substituted links may be coupled through the action.\\
For $\SUtwo$, an exact expression for \eq{multi_hit} can be obtained in terms of modified Bessel functions. For $\SUthree$, there is no analytic expression and \eq{multi_hit} is evaluated numerically by means of Monte Carlo integration. \footnote{We evaluated the average using 15 Cabibbo-Marinari iterations.}
At fixed $r$, one then forms a $M\times M$ correlation matrix of Wilson loops
\be\label{corrma}
C_{lm}(t)=\Bigg\langle\Tr\left\{V_{l}(0,r\hat{1})\overline{V}(r\hat{1},r\hat{1}+t\hat{0})V_{m}^{\dagger}(t\hat{0},r\hat{1}+t\hat{0})\overline{V}^{\dagger}(0,t\hat{0}) \right\}\Bigg\rangle
\ee
$$
=C_{ml}(t),
$$
where
\begin{eqnarray}
V_{l}(x,x+r\hat{1}) &=& U_l(x,1)U_l(x+a\hat{1},1)...U_l(x+(r-a)\hat{1},1),\\
\overline{V}(x,x+t\hat{0}) &=& \overline{U}(x,0)\overline{U}(x+a\hat{0},0)...\overline{U}(x+(t-a)\hat{0},0).
\end{eqnarray}
The correlation matrices were computed on field configurations generated by a hybrid algorithm with a mixture of $N_{\rm{or}}$ over-relaxation steps per heat bath step \cite{HOR1,HOR2}. 
The ratio $N_{\rm{or}}$ increases with the coupling $\beta$, and in practice we adopted 
\begin{equation}\label{n_or}
N_{\rm{or}}\approx 1.5 (\rnod/a).
\end{equation}
In a Monte Carlo simulation the configurations are correlated, hence the measurements are not statistically independent.
We decided to perform a measurement of Wilson loops every $N_{it/meas}$ sweeps, where $N_{it/meas}$ is chosen such that the time needed for the updating is about half of the time used for the measurement.\\
By collecting the data in bins and applying the so-called \emph{jackknife} procedure, it is possible to check that with this criterion 
the measurements are practically uncorrelated.\\
The simulation parameters for $5.7\leq \beta\leq 6.4$ are reported in \cite{pot:r0_SU3}, while for the new simulations in the range $6.57\leq \beta\leq 6.92$
they are listed in \tab{sim_par}.
We used $L\times L\times L\times T$ lattices, with $T=L$. The parameters $\beta$ and $L/a$ have been chosen such that 
\begin{equation}\label{e_size}
L/\rnod\sim 3.3.
\end{equation}
This condition is necessary because the physical results are obtained in the limit $L\rightarrow\infty$, and by using a finite volume one has to be sure that  finite-size effects are negligible. In \cite{pot:r0_SU3} it was shown that 
\eq{e_size} is safe and no finite volume effect is observed.\\
We evaluated the Wilson loops in \eq{corrma} for each value of the loop size $r,t$, with $a\leq r\leq r_{\rm{max}}$ and $a\leq t\leq t_{\rm{max}}$.
\begin{table}
\caption[Simulation parameters]{\footnotesize\label{sim_par}Simulation parameters.}
\begin{center}
\begin{tabular}{cccccccc}
\hline
$L/a$ & $\beta$ & $n_0$ & $n_1$ & $n_2$ & $N_{\rm{or}}$ &  $r_{\rm{max}}/a$ & $t_{\rm{max}}/a$ \\ 
\hline
40 &  6.57 & 0  & 77 & 153 & 18 & 7 & 11 \\
48 &  6.69 & 0  & 53 & 106 & 22 & 10 & 12  \\
56 &  6.81 & 0  & 72 & 144 & 25 & 12 & 14 \\
64 &  6.92 & 0  & 94 & 188 & 29 & 12 & 16 \\
\hline
\end{tabular}
\end{center}
\end{table}
Due to the high precision that one can reach at short distances, only 40 to 75 measurements were sufficient to extract the observables, and with relatively modest computational effort we could extend the calculations to $a\sim 0.025\,\fm$ where a $64^4$ lattice had to be simulated.

\subsection{Analysis details}
The correlation matrices \eq{corrma} were analyzed using a variational method, which was proposed in \cite{varia:michael,phaseshifts:LW}.
For any $r$ and $t_{0}$ fixed ($t_{0}=0$ in practice), one solves a generalized eigenvalue problem
\begin{equation}\label{gen_eigv}
C_{lm}(t)v_{\alpha,m}(t,t_0)=\lambda_{\alpha}(t,t_0)C_{lm}(t_0)v_{\alpha,m}(t,t_0),\quad\lambda_{\alpha}>\lambda_{\alpha+1},
\end{equation}
which corresponds to finding the eigenvalues of the symmetric matrix \\
$\overline{C}=C(t_0)^{-1/2}C(t)C(t_0)^{-1/2}$.
For the Wilson action, the positivity of the transfer matrix ensures that $C(t)$ is positive definite for all $t$ (see appendix \ref{app_trans}).\footnote{For gauge actions of kind \eq{impr_action} containing not only the plaquette, but also other operators, this property is not satisfied, and the implications for the applicability of the variational method are discussed in chapter. \ref{improved}.}
One can prove that the energies $E_{\alpha}$ are related to the generalized eigenvalues by the relation
\begin{equation}\label{eigv_energy}
aE_{\alpha}=\ln(\lambda_{\alpha}(t-a,t_0)/\lambda_{\alpha}(t,t_0))+{\rm O}(e^{-t\Delta E_{\alpha}}),
\end{equation}
where $\Delta E_{\alpha}={\rm min}_{\beta\neq\alpha}|E_{\alpha}-E_{\beta}|$.
The static quark potential corresponds to the ground state energy, $V(r)\equiv E_{0}$.\\
By using this method one expects that the coefficients of the higher order terms in \eq{eigv_energy} are suppressed and one can read off the energy values already at moderately large $t$. Solving the generalized eigenvalue problem \eq{gen_eigv} is indeed equivalent to looking for a linear combination of operators appearing in the correlation function such that the state with the lowest energy has the maximal weight.\\
We decided to extract the potential $V(r)$ projecting the Wilson loop correlation function
\begin{equation}
W(t)=v_{0}^{T}C(t)v_{0},
\end{equation}
where $v_{0}$ is the ground state eigenvector computed for $t=t_{0}+a$ in
\eq{gen_eigv} corresponding to the ground state, and then constructing
the effective potential
\begin{equation}\label{pot_extract}
aV(r)=-\ln\left(\frac{W(t)}{W(t-a)}\right).
\end{equation}

\begin{figure}
\begin{center}
\includegraphics[width=9cm]{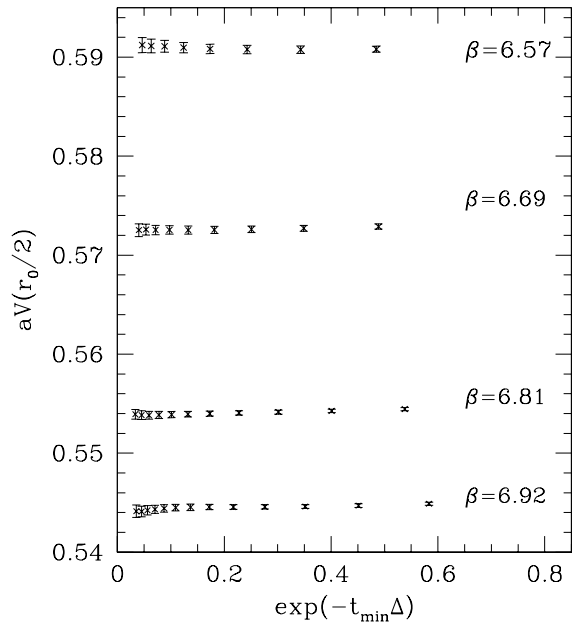} 
\end{center}
\caption[The potential evaluated at $r\sim\rnod/2$ for different
    couplings as a function of $\exp(-t_{min}\Delta)$.]{\footnotesize{The potential evaluated at $r\sim\rnod/2$ for different
    couplings as a function of $\exp(-t_{min}\Delta)$; $t_{min}$ will be determined such that $\exp(-t_{min}\Delta)<0.3$. Similar results hold for other values of $r$. }\label{f_error}}
\end{figure}

Beside the potential, one is also interested in the energy gap between the ground state and the first excited state 
\begin{equation}
\Delta E_{0}=E_{1}-E_{0}, 
\end{equation}
which dominates the finite time corrections in the extraction of $V(r)$.
We estimated them by
\begin{equation}
a\Delta E_{0}=\ln(\lambda_{1}(t-a,t_0)/\lambda_{1}(t,t_0))-
\ln(\lambda_{0}(t-a,t_0)/\lambda_{0}(t,t_0)).
\end{equation}
To demonstrate that these corrections are small, we computed $aV(r)$
from \eq{pot_extract} for $r\approx \rnod/2$ and  $r\approx \rnod/4$
and we plotted it as a function of $\exp(-t\Delta E_{0})$; as in \cite{pot:r0_SU3} we found that for
$t>t_{\rm{min}}$ excited states contaminations are very small when
$\exp(-t_{min}\Delta E_{0})<0.3$ (\fig{f_error}), and we used this criterion to
establish at which $t$ the systematic errors in \eq{pot_extract} are
negligible with respect to the statistical uncertainties.
In \cite{pot:r0_SU3} a continuum extrapolation of the quantity
$\rnod\Delta E_{0}$ was performed, yielding
\begin{equation}
\rnod\Delta E_{0}|_{r=\rnod}=3.3(1).
\end{equation}
Since the gap grows towards small $r$ from $\approx 3/\rnod$ at
$r=\rnod$ to $\approx 7/\rnod$ at $r=0.1\rnod$, the short distance
region is easiest in this respect.\\
 All errors were computed by
jackknife binning, and the results for the potential and force are listed
in appendix \ref{numerical_results}, in particular in tables \ref{t_force2a}, \ref{t_force1a}.

\subsection{Interpolation}\label{interpo}
Solving \eq{e_rnod} and evaluating the force and potential 
at distances $r=x\rc$ for given $x$, requires the interpolation of those quantities. While
this can in principle be done in many ways, it is advantageous
to use physics motivated interpolation formulae.

For the force we followed \cite{pot:r0} choosing the interpolation function
\be\label{force_interp}
F(r)=f_{1}+f_{2}r^{-2},
\ee
between two neighboring points. 
The systematic error arising from the 
interpolation was estimated by adding a term $f_{3}r^{-4}$ and taking a third point. 
The difference between the two interpolations was added (linearly) to
the statistical uncertainty. We observed that at least for $r\gtrsim 0.4\rnod$ the 
interpolation error is smaller than the statistical uncertainty. For small distances the systematic error increases and can be of the order of the statistical one
or even bigger.

The potential is interpolated by the corresponding 
ansatz
\be\label{pot_interp}
\VI(r)=v_{1}+v_{2}r+v_{3}r^{-1}.
\ee
Here two points are chosen such that the desired value of $r$ is in between.
For the choice of a neighboring third point one has two possibilities, leading to 
two results.
Their difference was taken as the interpolation error. Also in this case 
the systematic 
errors are larger than the statistical ones at short distances  while
at large distances the situation is reversed.\\
It is important to stress that the interpolation formulas
\eq{force_interp}, \eq{pot_interp} provide excellent approximations of the $r$
dependence only locally, and to use instead a global fit in a large
range could be a source of large systematic errors.

\subsection{The ratio $\rc/\rnod$ and parametrization of $\rnod/a$}\label{rc_para}
\begin{figure}
\begin{center}
\includegraphics[width=9cm]{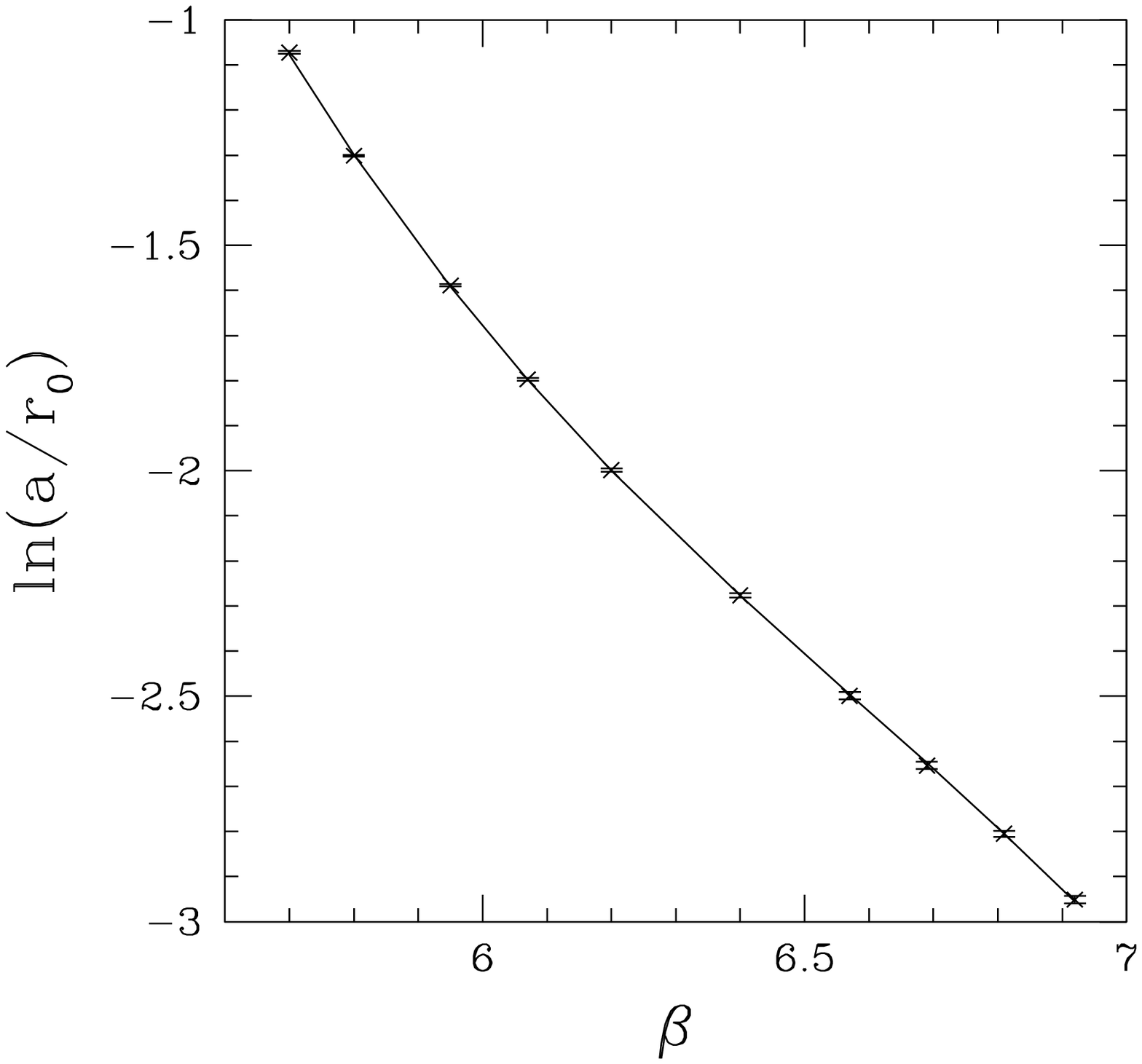}
\caption[Interpolation of $\rnod/a$.]{\footnotesize{Interpolation of $\rnod/a$.}\label{f_fitrnod}}
\end{center}
\end{figure}

Following the procedure described in the previous sections, we
evaluated $\rc/a$ and $\rnod/a$ in the region 
$5.95\leq\beta\leq 6.57$ where both quantities are accessible.
%%%%%%%%%%%%%%%%%%%%%%%%%%%%%%%%%%%%%%%%%%%%%%%%%%%%%%%%%%%%%%%%%%%%%%%%%%
\begin{figure}
\begin{center}
\includegraphics[width=9cm]{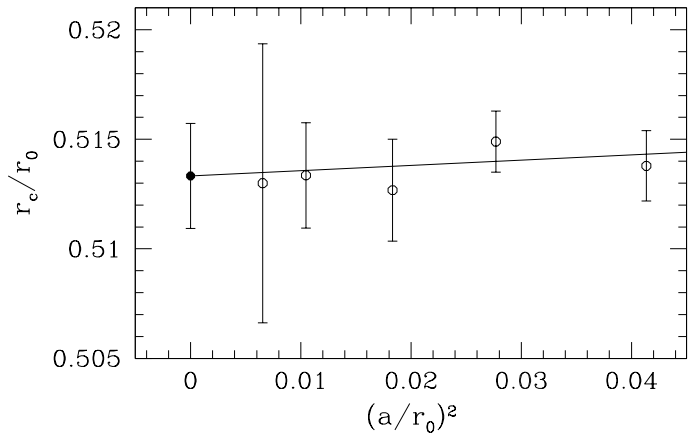}
\caption[The ratio $\rc/\rnod$ for $5.95\leq\beta\leq 6.57$ including the continuum extrapolation.]{\footnotesize{The ratio $\rc/\rnod$ for $5.95\leq\beta\leq 6.57$ (circles) including the continuum extrapolation (solid point). The data for $5.95\leq\beta < 6.57$ are taken from \protect\cite{pot:r0_SU3}.}\label{f_rcrnod}}
\end{center}
\end{figure}
%%%%%%%%%%%%%%%%%%%%%%%%%%%%%%%%%%%%%%%%%%%%%%%%%%%%%%%%%%%%%%%%%%%%%%%%%%%%%
\Fig{f_rcrnod} shows the ratio $\rc/\rnod$ for several lattice
spacings as a function of $(a/\rnod)^{2}$. No dependence on the resolution
is seen within the errors of below 1\%. A continuum extrapolation gives
\be\label{e_rcrnodcont}
{\rc/\rnod}=0.5133(24)
\ee
and we note that it is safe to use \eq{e_rcrnodcont} also at finite lattice 
spacings starting around $\beta=6.4$.\\
The direct determination of $\rnod/a$ for $5.7\leq\beta \leq 6.4$ \cite{pot:r0_SU3} \
\footnote{In \cite{pot:r0_SU3} the simulation for $\beta=6.57$
was performed with a single smearing level, yielding less control
over excited state contaminations. As a result
the error in $\rnod/a$ seems to be somewhat underestimated. Although the
statistical significance of this
small effect is not clear, we decided to use our new data for $\beta=6.57$
instead.}
and the new computations of $\rc/a$ in the range $6.57\leq\beta\leq 6.92$ may be combined 
with $\rc / \rnod$=0.5133(24) to obtain
$\rnod/a$ in the whole range $5.7\leq\beta\leq 6.92$. 
The results are reported in \tab{tab_rnod}.\\
Once $\rnod/a$ is known for several $\beta$ values, it is useful to attempt a phenomenological representation in terms of an interpolating formula giving $\rnod/a$ in the whole range of couplings. 
%%%%%%%%%%%%%%%%%%%%%%%%%%%%%%%%%%%%%%%%%%%%%%%%%%%%%%%%%%%%%%%%%%%%%%%%%%%%
\begin{table}
\caption[Results for $\rnod/a$ and $\rc/a$.]{\label{tab_rnod}\footnotesize{Results for $\rnod/a$
    \protect\cite{pot:r0_SU3} and $\rc/a$. For comparison also the values from
    the interpolation formula \protect\eq{e_rnodfit} are shown.}}
\begin{center}
\begin{tabular}{c c c c c c c }
\hline
$\beta$ & $\rnod/a$ & \eq{e_rnodfit} &   & $\beta$ & $\rc/a$  & \eq{e_rnodfit},\eq{e_rcrnodcont} \\[1ex]
\hline
5.7 &  2.922(9)  & 2.938     &~~~~~~& 6.57 & 6.25(4) & 6.22  \\
5.8 &  3.673(5)  & 3.665     &~~~~~~& 6.69 & 7.29(5) & 7.25  \\
5.95 & 4.898(12) & 4.912    &~~~~~~& 6.81 & 8.49(5)  & 8.48  \\
6.07 & 6.033(17) & 6.038    &~~~~~~& 6.92 & 9.82(6)  & 9.86  \\
6.2 & 7.380(26)  & 7.383   \\
6.4 & 9.74(5)    & 9.741   \\
\hline
\end{tabular}
\end{center}
\end{table}
%%%%%%%%%%%%%%%%%%%%%%%%%%%%%%%%%%%%%%%%%%%%%%%%%%%%%%%%%%%%%%%%%%%%%%%%%%%%
One can start from the renormalization group equation for the bare coupling;
on the lattice the $\beta$ function associated to $g_{0}$ is defined as,
\begin{equation}\label{beta_lattice}
\beta(g_{0})=-a\frac{\partial g_{0}}{\partial a}= -b_{0}g_{0}^{3}-b_{1}g_{0}^{5}+...
\end{equation}
Solving \eq{beta_lattice} yields the $\Lambda$-parameter associated to the
lattice 
\begin{equation}
\Lambda_{LAT}=\lim_{g_{0}\rightarrow 0}\frac{1}{a}e^{-1/(2b_{0}g_{0}^2)}(b_{0}g_{0}^2)^{-b_{1}/(2b_{0}^2)}.
\end{equation}
In pure gauge theory every physical quantity with dimensions of a mass must
 be proportional to $\Lambda_{LAT}$ in the continuum limit; therefore the mass
 (or $a/\rnod$ in our case) in lattice units, which is the quantity that one
 evaluates in numerical simulations, varies as a function of 
$g_{0}$ according to
% in pure gauge theory every physical quantity $m$ with dimensions of a mass satisfies
%\begin{equation}\label{ren_eq_g0}
%\left\{a\frac{\partial}{\partial a}-\beta\frac{\partial}{\partial g_{0}}\right\}m={\rm O}(a^{2}m^{2}),
%\end{equation}
%where $\beta(g_{0})=-a\frac{\partial g_{0}}{\partial a}$ is the
%lattice $\beta$ function associated to $g_{0}$. \\
%The solution of \eq{ren_eq_g0} for the quantity $\rnod^{-1}$ yields the expression
\begin{equation}
\frac{a}{\rnod}=A e^{-1/(2b_{0}g_{0}^{2})}(b_{0}g_{0}^{2})^{-b_{1}/(2b_{0}^{2})}e^{-c{_1}g_{0}^2-c_{2}g_{0}^4-...},
\end{equation}
where $A$ is a constant and the terms with $c_{1},c_{2},...$ are due
to higher order contributions to the $\beta$ function. This behavior is also called \emph{asymptotic scaling}.\\
The leading behavior is then given by
\begin{equation}
\frac{a}{\rnod}\propto e^{-\beta/(12b_{0})},\quad \beta=6/g_{0}^2,
\end{equation}
then one is justified to attempt a phenomenological representation of
$(\ln(a/\rnod))$ as a polynomial in $\beta$. In order to avoid 
cancellations in the fit parameters, it is convenient to shift the
value of $\beta$ to $(\beta-x)$, such that $(\beta-x)$ is small in the
interval of couplings that we considered. We then chose the following
ansatz:
\begin{equation}
\ln(a/\rnod)=\sum_{k=0}^{p}a_{k}(\beta-6)^{k},
\end{equation}
and found that
\begin{eqnarray} \label{e_rnodfit}
\ln(a/\rnod)&=&-1.6804-1.7331(\beta-6)+0.7849(\beta-6)^{2}-0.4428(\beta-6)^{3}\,, 
     \nonumber \\ 
  && {\rm for} \;\; 5.7\leq\beta\leq 6.92, 
\end{eqnarray}
is an excellent approximation to the MC results, as shown in \fig{f_fitrnod}. 
The accuracy of $\rnod/a$ in 
\eq{e_rnodfit} is about $0.5\%$ at low $\beta$ decreasing to $1\%$ 
at $\beta=6.92$.

%%%%%%%%%%%%%%%%%%%%%%%%%%%%%%%%%%%%%%%%%%%%%%%%%%%%%%%%%%%%%%%%%%%%%%%%%%

%%%%%%%%%%%%%%%%%%%%%%%%%%%%%%%%%%%%%%%%%%%%%%%%%%%%%%%%%%%%%%%%%%%%%%%%%%%%%

\section{Continuum force and potential}
%%%%%%%%%%%%%%%%%%%%%%%%%%%%%%%%%%%%%%%%%%%%%%%%%%%%%%%%%%%%%%%%%%%%%%%%%%%%%%%
One can now determine the continuum force and potential by
extrapolation of the MC-results at finite values of
the lattice spacing to the continuum, which corresponds to $a=0$ ($\beta=\infty$).\\
As mentioned in sect. \ref{interpo}, the lattice results are first
interpolated in $r$, then renormalized dimensionless quantities are evaluated.
If the lattice spacing is small enough, one expects renormalized dimensionless quantities to be independent of $\beta$ (and thus of $a$); this feature is called \emph{scaling}. Any deviation from this scaling behavior is due to the presence of lattice artifacts. The size of the scaling violating corrections will in general depend on which quantities are being considered.\\
In order to approach the continuum limit, one needs a theoretical
understanding of the lattice artifacts; the standard framework for this
discussion is Symanzik's effective theory, which is expected to give the
asymptotic expansion of suitable lattice observables in integer powers of the
lattice spacings (up to at least logarithmic further suppressions as
$a\rightarrow 0$) in an asymptotically free theory \cite{impr:Sym1,impr:Sym2}.
In the 2-dimensional O(3) $\sigma$ model it predicts the expansion to start at
order $a^2$ but unexpectedly numerical results are described better by a
dominant linear term in $a$ for a range of lattice spacing
\cite{sigma:unexpected}. While there is no evident contradiction with the
result of an analysis \`a la Symanzik which is supposed to describe the {\em asymptotic} behavior, the standard picture
should be tested in 4-d gauge theories and QCD as much as possible.\\
The leading artefacts in potential
differences are expected to be $\rmO(a^2)$, but this is not obvious
since the potential is usually defined in terms of a
Wilson loop, which does not fall into the category of
correlation functions of local fields discussed by Symanzik.\\
In \cite{Necco:2001gh} a clean argument supporting this expectation is given:
in short, the motivation is that the heavy quark 
effective theory \cite{stat:eichhill1}
formulated on the lattice is $\Oa$-improved without
adding any additional operators to the Lagrangian \cite{zastat:pap1}.
The potential is an energy of the effective theory and thus 
has no linear $a$-effects. Of course, for this statement to be meaningful,
the self energy has to be eliminated by 
considering potential differences or the force.
%We want to also mention that our argument is
%based on the assumption that the effective theory is renormalized as 
%usual by the addition of local operators. Explicit perturbative
%computations support this assumption 
%but it has not been proven so far.

\begin{figure}
\begin{center}
\includegraphics[width=10.5cm]{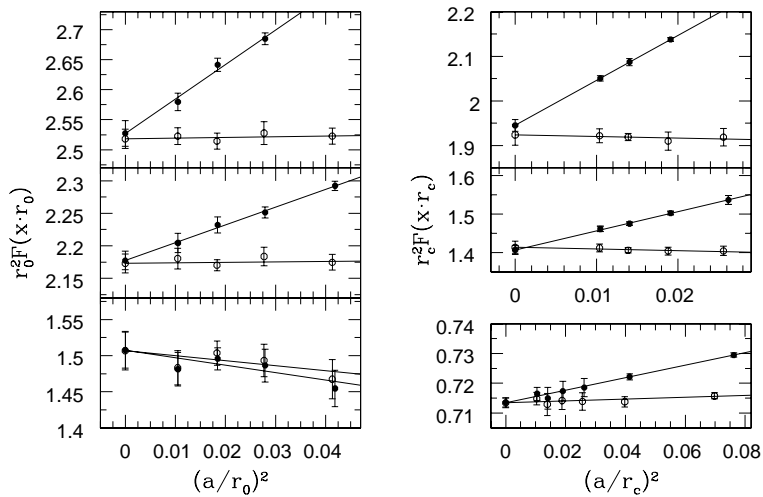}
\vspace{-0.5cm}
\caption[Continuum
extrapolation of $\rc^2 F(x\rc)$.]{\footnotesize
Continuum
extrapolation of $\rc^2 F(x\rc)$, for $x=0.4,~0.5,~0.9$ from top to bottom
and of $\rnod^2 F(x\rnod)$, for $x=0.5,~0.6,~1.5$ from top to bottom. 
The data are from our new computations and from \protect\cite{pot:r0_SU3}. 
Filled circles correspond to the naive value $r'=r-\frac{a}{2}$ instead
of \protect\eq{e_rI}.\label{f_force_extr}}
\end{center}
\end{figure}

\Fig{f_force_extr}  shows the dependence of the force in
units of $\rc$ on the resolution
for some selected values of the separation $r$. Fitting~\footnote{In general,
   the discretization errors parametrized by $s$ depend also on
   the orientation ${\vec r}/r$, which in our calculation was fixed to
   $ {\vec r}/r = (1,0,0)$, however.}
\begin{eqnarray}  \label{e_force_extr}
  \rc^2 F(r) = \left.\rc^2 F(r)
   \right|_{a = 0}\left[1 + s(r/\rc) \times (a/\rc)^2\right]\,
\end{eqnarray}
for fixed $r/\rc$ and for $r/a>2$, the slope $s$ is statistically not significant throughout our range of $r$ and $\beta$.
 Our statistical precision
allows to quote
\begin{eqnarray} \label{e_slope}
 |s(r/\rc)| < 1\;\; {\rm for} \; 0.4\leq r/\rc  \leq 1\,.
\end{eqnarray}
Of course this 1-$\sigma$ bound is valid only when all
details are as discussed above, in particular eqs.~(\ref{finite_force},\ref{e_rI})
are used to define the force at finite $a$. If the naive form
$r' =r-\frac{a}{2}$ is employed instead, the
corresponding slopes $s$ become rather large, as can be seen in the figure.

A similar statement ($r/a>2$),
\begin{eqnarray}
   \rnod^2 F(r) &=& \left.\rnod^2 F(r) \right|_{a = 0}
                    \left[1 + s(r/\rnod) \times (a/\rnod)^2\right]\,,  \\
  |s(r/\rnod)| &<& 1.2\;\; {\rm for}\; 0.5\leq r/\rnod  \leq 1.5\,,
\end{eqnarray}
can be made for larger $r$.\\
One can conclude that using a tree-level improved definition of the force,
lattice spacing effects are below the level of our small statistical errors
when one restricts oneself to $r> 2a$ and to the lattice spacings considered
here. By contrast, with a naive midpoint rule for the force, the $a$-effects
are quite sizeable. The difference is by construction a pure power series in
$a$, starting with $a^2$. For $r> 2a$ this series is well approximated by
the leading term. Thus both data sets in \fig{f_force_extr} contain the same
essential information: full compatibility with Symanzik's theory of
discretization errors.

\begin{figure}
\begin{center}
\includegraphics[width=9cm]{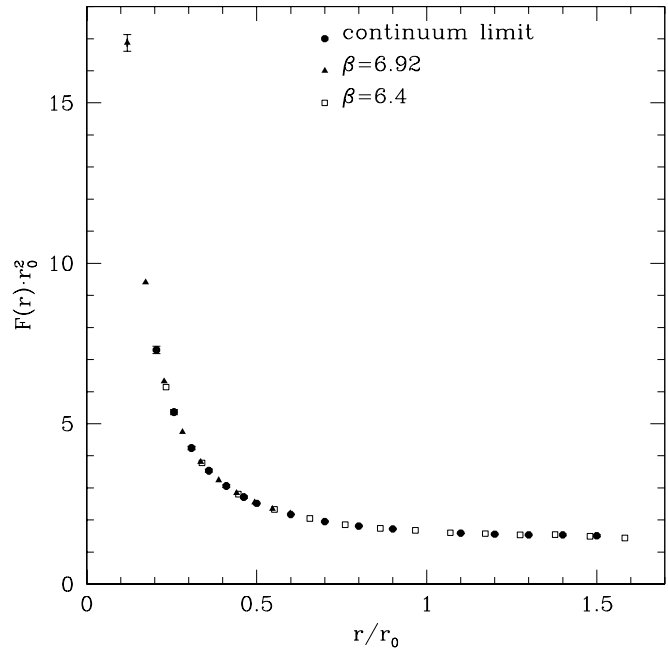}
\caption[The force in the continuum
limit and for finite resolution, where the discretization
errors are estimated to be
smaller than the statistical errors.]{\footnotesize{The force in the continuum
limit and for finite resolution, where the discretization
errors are estimated to be
smaller than the statistical errors.}\label{f_force}}
\end{center}
\end{figure}

The continuum
force is plotted in \fig{f_force} using \eq{e_rcrnodcont} to 
combine the two regimes of $r$.
Some data at finite $\beta$ (in particular $\beta=6.4,6.92$) are included in the figure. In these
cases we used our ``bounds'' on $s$ to estimate that the discretization 
errors are smaller than the statistical ones. One can observe that for large distances the force reaches a constant value, as sign of the confinement;
a detailed discussion on the comparison with the predictions of the
perturbative theory and the bosonic string model will be presented in
section \ref{sec_stringcom}.\\

%%%%%%%%%%%%%%%%%%%%%%%%%%%%%%%%%%%%%%%%%%%%%%%%%%%%%%%%%%%%%%%%%%%%%%%%%%%%
%%%%%%%%%%%%%%%%%%%%%%%%%%%%%%%%%%%%%%%%%%%%%%%%%%%%%%%%%%%%%%%%%%%%%%%%%%%%
\begin{figure}
\begin{center}
\includegraphics[width=11cm]{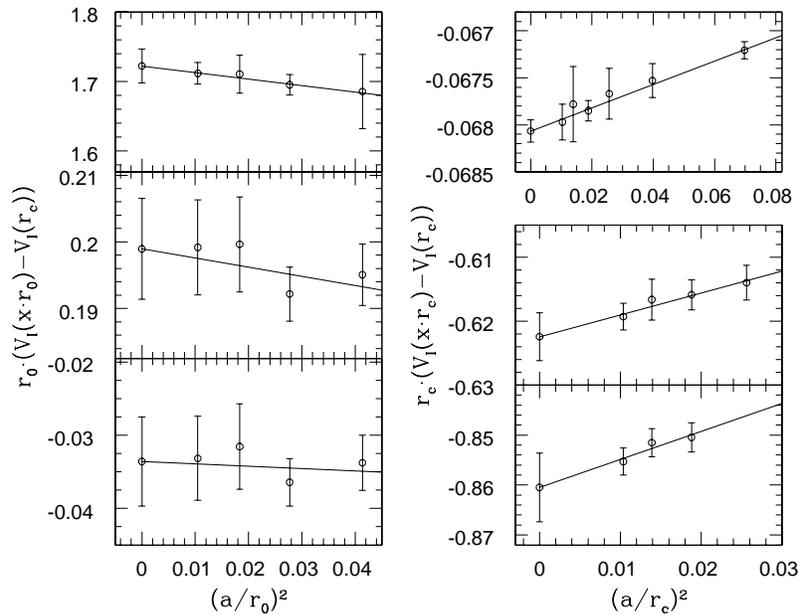}
\caption[Continuum extrapolation 
of $\VI(r)$.]{\footnotesize 
Continuum extrapolation 
of $\VI(r)$, for $r/\rnod=1.5,\,0.6,\,0.5$ on the left hand side and
for $r/\rc=0.9,\,0.4,\,0.3$ on the right hand side.
\label{f_pot_extr}}
\end{center}
\end{figure}
%%%%%%%%%%%%%%%%%%%%%%%%%%%%%%%%%%%%%%%%%%%%%%%%%%%%%%%%%%%%%%%%%%%%%%%%%%%%

The potential contains the same physical 
information as the force but statistical and systematic errors 
in the lattice determination
are different.
From the non-perturbative point of view, the potential is defined only up to a constant, which represents the self energy of the static quark-antiquark pair.
To eliminate this self energy contribution
we consider $\VI(r)-\VI(\rc)$, where we apply the same improvement 
as for the force,
\be\label{pot_impr}
\VI(d_{I})=V(d)\,\, ,\,\, (4\pi d_{I})^{-1}= G(d,0,0).
\ee 
In analogy to $\rI$, $d_{I}$ is defined such that the potential has no lattice artifacts when evaluated at three level.
The continuum extrapolation is performed exactly as for the 
force. In this case the slopes, $s(r/\rc)$, defined as above are statistically
significant for both our smallest distances ($r\approx 0.3\rc$) and the
largest one ($r\approx 1.5\rnod$). As a consequence, at small $r$ the 
combination $[\VI(r)-\VI(\rc)]\rc$
has about $1\%$ discretization errors at $\beta=6.92$ while
in the large distance region these errors go up to $1\%$ in
$[\VI(r)-\VI(\rc)] \rnod$ at $\beta=6.4$. 
These are the $\beta$-values corresponding to
the smallest lattice spacings available in the two different regions.\\
One concludes that here, small $a$-effects could also be observed for the tree
level improved definition. As for the force -but not shown in the figure- the
standard definition without tree-level improvement shows quite large
$a$-effects compatible with an $a^2$-behavior, and this confirms again the validity of the Symanzik picture.\\

\begin{figure}
\begin{center}
\includegraphics[width=9cm]{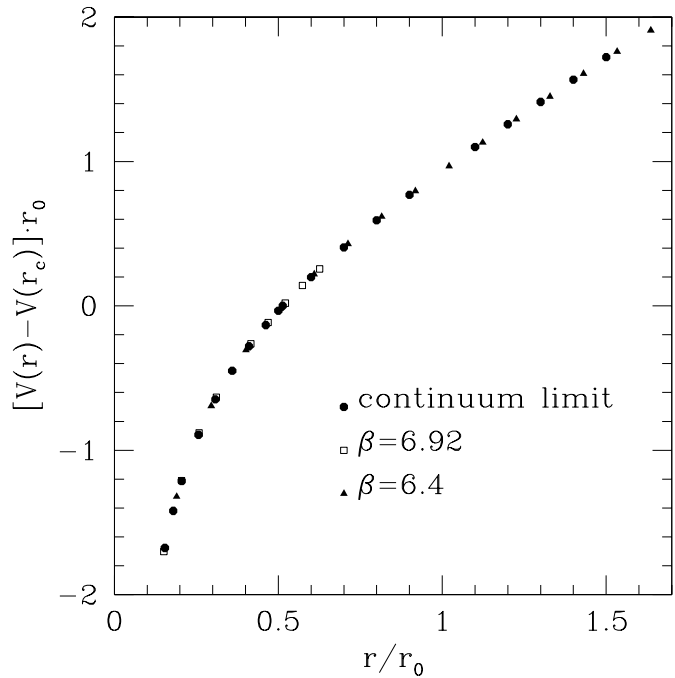}
\caption[The static potential in the continuum limit]{\footnotesize{The static potential in the continuum limit. }\label{v_pot}}
\end{center}
\end{figure}

The continuum potential is plotted in \fig{v_pot}, with some data at finite 
$\beta$. In this figure the statistical as well as the discretization errors
are below the size of the symbols. 
The results for continuum force and potential are listed in
\tab{t_cont}.
\input tab_cont.tex

%%%%%%%%%%%%%%%%%%%%%%%%%%%%%%%%%%%%%%%%%%%%%%%%%%%%%%%%%%%%%%%%%%%%%%%%%%
%%%%%%%%%%%%%%%%%%%%%%%%%%%%%%%%%%%%%%%%%%%%%%%%%%%%%%%%%%%%%%%%%%%%%%%%%%

%%%%%%%%%%%%%%%%%%%%%%%%%%%%%%%%%%%%%%%%%%%%%%%%%%%%%%%%%%%%%%%%%%%%%%%%%%%%%
\section{Comparison with perturbation theory} \label{comparison_pert}
The perturbative expansion in the coupling $\alpha_{s}$ is the most
important theoretical tool for analyzing strong interactions effects
in high energy scattering experiments. In practical applications however, in
particular in the determination of the running coupling itself, it is
tempting to apply the perturbative series already where the coupling
is not so small to justify the applicability of perturbation theory.
Popular examples are the determinations for $\alpha_{s}$ from
$\tau$-lepton decays and the perturbative evolution of deep inelastic
scattering structure functions starting at a renormalization point
below 1 $\GeV$. In order to establish the reliability of
perturbation theory, it would be very desirable to study such
processes systematically as functions of the energy, but either they
involve a fixed energy ($\tau$-decays, hadronic $Z$-decays) or the
precision is not sufficient over a larger energy range (deep inelastic
scattering, $e^{+}e^{-}$ total cross section, Adler function). The
main phenomenological test of perturbation theory, therefore, is the
overall consistency of the determination of $\alpha_{s}$ from
different processes (see \cite{Bethke:2000ai}).

Complementary information may be obtained from observables that can be
computed via lattice QCD, like the static quark potential. As already mentioned
in section \ref{s:pot_pert}, in perturbation theory the potential has been
computed at two loops, but at the same time the usefulness of perturbation
theory even at distances as short as $0.1\,\fm$ has been questioned \cite{Peter:1997me,Schroder:1999sg}.
Comparison between the perturbative prediction and the non-perturbative results will show that this is a
question of a suitable renormalization scheme. When the most natural scheme,
the $\qqbar$ defined in terms of the force \eq{qqbar_scheme} is adopted,
perturbation theory is well behaved at such distances. This is a rather
stringent test of perturbation theory, since we can use the knowledge of the
$\Lambda$-parameter together with the RG perturbative
evaluation of the coupling, which predicts not only the scale dependence of
the coupling, but also its absolute value.

In an exploratory investigation, concentrated on the question, whether there are
''large non-perturbative'' terms in the potential at short distances
\cite{pot:bali99} they had been argued to exist \cite{Akhoury:1998by}.\\
In particular, non-perturbative ultraviolet contributions to the running
coupling proportional to $\Lambda^2/q^2$ were investigated. In the case of the
static quark potential, such a contribution to the running coupling results in
a term proportional to the quark separation, $r$.

\begin{figure}[ht]
\begin{center}
\includegraphics[width=9cm]{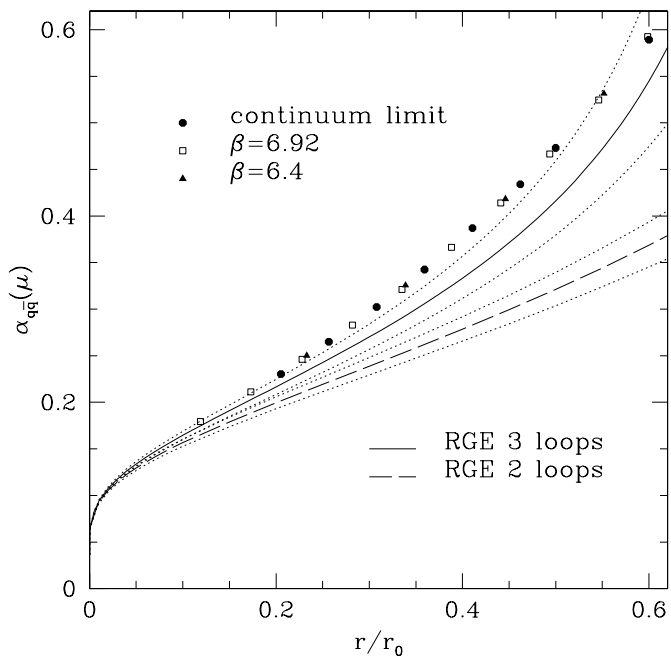}
\end{center}\vspace{-0.9cm}
\caption[Running coupling in the $\qqbar$ scheme.]{\footnotesize \label{f_alphaqq_comp} Running coupling in the $\qqbar$ scheme. Results for the continuum
  limit as well as additional points at finite $\beta$, corresponding
  to finite lattice spacing are shown. The perturbative curves use $\Lambda_{\MSbar} \rnod$
  from \protect\cite{mbar:pap1} with the dotted lines corresponding to
  the 1-$\sigma$ uncertainties of this number.}
\end{figure}

The first important step to start our analysis is to give a definition of ''large non-perturbative
term''. We assume that the following is understood by this statement.
\begin{enumerate}
\item A certain quantity, here the potential $V(r)$, is considered in
  a region where its perturbative expansion looks well behaved, i.e.
  the $n$-loop contribution is a small correction and significantly
  smaller than the $(n-1)$-loop contribution (unless the latter is
  accidentally small itself).
\item The difference between the full non-perturbative observable and
  the truncated perturbative series is much larger that the last term
  in the series.
\end{enumerate}
With such a definition, necessarily somewhat phenomenological in
character, we shall demonstrate that there are definitely no large
non-perturbative terms in the potential. To the contrary, perturbation
theory works remarkably well where criterion 1. is satisfied.\\ In
section \ref{s:pot_pert} we established that for $\alphaqqbar$ this
condition is valid up to distances $r\sim 0.25\,\fm$, corresponding to
$\alphaqqbar\lesssim 0.3$.  In \fig{f_alphaqq_comp} we compare the
perturbation theory predictions for $\alphaqqbar(r)$ with our non
perturbative results, which were obtained from the force extrapolated
to the continuum limit by the simple relation
\begin{equation}
  \alphaqqbar(r)=\frac{3}{4}F(r)r^2.
\end{equation}
We included also additional points at finite $\beta$, where the
discretization errors were estimated to be smaller than the
statistical ones.  The 3-loop RG expression with $\Lambda_{\MSbar}$ at
the upper end of the error bar of \eq{e_lambda_rnod} is in very close
agreement with the non-perturbative coupling. In fact the agreement
extends up to values of $\alphaqqbar$ where perturbation theory is not
to be trusted a priori. For $\alphaqqbar\lesssim 0.3$ criterion 1.
is satisfied and there is no evidence for non-perturbative terms in
this region.

%%%%%%%%%%%%%%%%%%%%%%%%%%%%%%%%%%%%%%%%%%%%%%%%%%%%%%%%%%%%%%%%%%%%%%%%%%
%%%%%%%%%%%%%%%%%%%%%%%%%%%%%%%%%%%%%%%%%%%%%%%%%%%%%%%%%%%%%%%%%%%%%%%%%%%%

\begin{figure}[ht]
\begin{center}
\includegraphics[width=9cm]{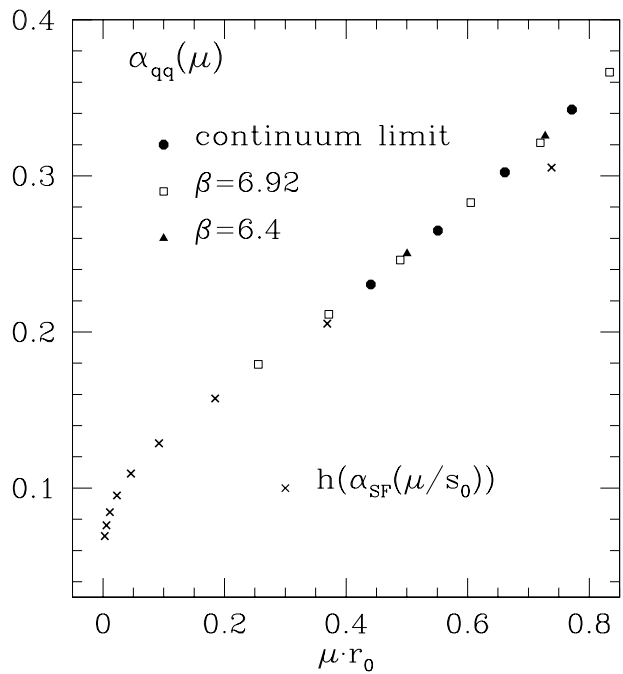}
\end{center}\vspace{-0.9cm}
\caption[Perturbative matching between the $\qqbar$ and the SF-scheme.]{\footnotesize \label{f_alphaqqsf}
  Test of \eq{e_match}. The uncertainty in the combination $\mu\rnod$ has been translated into an uncertainty for $h(\alphaSF(\mu/s_{0}))$ and $\alphaqqbar(\mu)$. The non-perturbative values for $\alphaSF(\mu)$ are constructed from the data of \protect\cite{mbar:pap1}. Errors are smaller than the size of the symbols.   }
\end{figure}

One can perform an additional test by considering the relation between
$\alphaSF(\mu)$ and $\alphaqqbar(\mu)$ at finite $\mu$: using
\eq{e_match} and the coefficient in \tab{t_coeff}, we define the
3-loop expression \footnote{Note that the term proportional to $\alphaSF(\mu)^2$ in not present due to our specific choice of the relative scale factor $s$, \protect\eq{eq_s0}.}
\begin{equation}
  h(\alphaSF(\mu))=\alphaSF(\mu)+1.336\left[\alphaSF(\mu)\right]^{3}
\end{equation}
such that
\begin{equation}\label{matching_sfqq}
  \alphaqqbar(\mu)=h(\alphaSF(\mu/s_{0}))+{\rm
    O}(\left[\alphaSF(\mu/s_{0})\right]^{4}).
\end{equation}
Then the non-perturbative values of $\alphaqqbar(\mu)$ and
$h(\alphaSF(\mu/s_{0}))$ are compared in \fig{f_alphaqqsf}. If the
higher order terms in \eq{matching_sfqq} are negligible, the two
different quantities should agree. At $\alphaqqbar(\mu)\approx 0.3$ a
difference is visible but this is only about $3\times \alpha^4$, not
so far from the expected size of the next order term in the series. At
$\alphaqqbar(\mu)\approx 0.2$, the difference
$\alphaqqbar(\mu)-h(\alphaSF(\mu/s_{0}))$ is not significant at all.
We conclude that also in \eq{matching_sfqq} a large non-perturbative
term at short distances is excluded.\\ 

One has to note here that 
the static quark potential defined through the Wilson loop suffers
from infrared (IR) divergences when computed at finite orders of
perturbation theory \cite{Appelquist:1977tw}; this is related to the
non-Abelian nature of $\SUthree$, which allows massless gluons to
self-interact at arbitrarily small energy scales. The leading IR
singularities in the Wilson loop were argued to exist at order
$\alpha_{s}^4$ in \cite{Appelquist:1978es}. These singularities can be
regulated by a resummation, and the result is that the next order
correction in \eq{matching_sfqq} is formally enhanced by a logarithm
of $\alpha$ and reads
\cite{Appelquist:1977tw,Appelquist:1978es,Schroder:1999sg}
\begin{equation}
  (A\log(\alpha)+B)\alpha^4.
\end{equation}
While $A=\frac{9}{4\pi}$ has been calculated
\cite{Brambilla:1999xf,Brambilla:1999qa}, $B$ is not known. The
$A\alpha^4\log(\alpha)$ term by itself constitutes a small correction
in the \fig{f_alphaqqsf}, which would slightly enlarge the difference
between $h(\alphaSF(\mu/s_{0}))$ and $\alphaqqbar(\mu)$.\\

\begin{figure}[ht]
\begin{center}
\includegraphics[width=9cm]{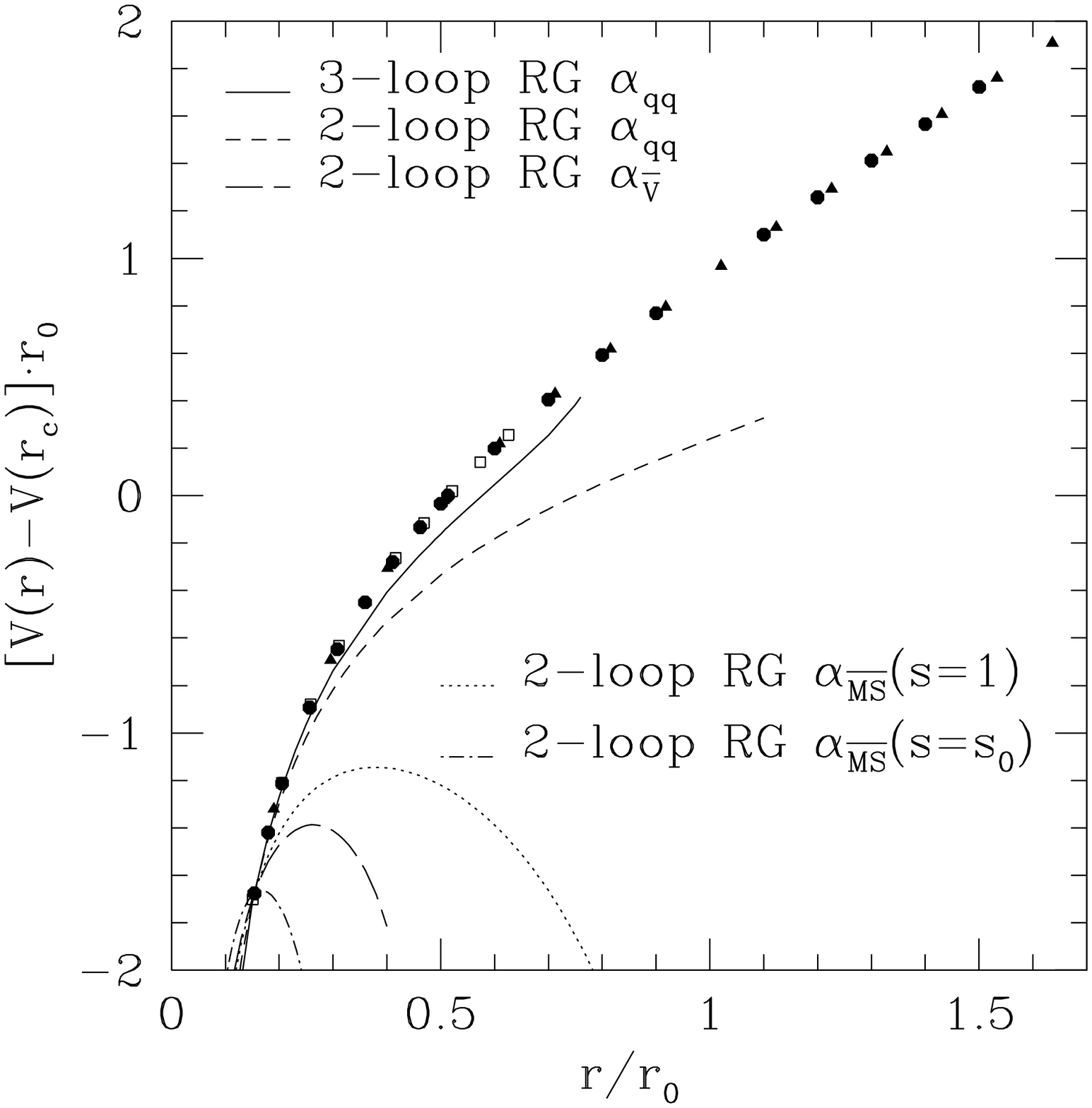}
\end{center}\vspace{-0.9cm}
\caption[The potential compared to different perturbative expressions.]{\footnotesize \label{f_pot_comparison}
  The potential compared to different perturbative expressions. $\rc$
  is related to $\rnod$ by \protect\eq{e_rcrnodcont}.}
\end{figure}

Finally, \fig{f_pot_comparison} shows the static potential itself compared to different perturbative approximations. In this case we can not use the coupling $\alphavbar(\mu)$ for the comparison, because it does not have a non-perturbative definition, as already mentioned in \sect{s:pot_pert}. The potential is defined up to a constant, which has to be eliminated for example by considering the quantity
$$
[V(r)-V(\rc)]\rnod,
$$
where $\rc$ is related to $\rnod$ by \eq{e_rcrnodcont}.\\
Full line and short dashes are given by integrating the force
\begin{equation}\label{e_vint}
  V(r)=V(0.3\rc)+\int_{0.3\rc}^r \rmd y F(y)\,,\;F(r)=C_{F} r^2
  \alphaqqbar(1/r)
\end{equation}
with the 3-loop and 2-loop RG-solution for $\alphaqqbar$. The
free constant
was fixed such that at $r=0.3\rc$ the
perturbative prediction coincides with the non-perturbative
result. This comparison is meaningful only with the assumption that
the perturbative expansion is well behaved at that distance.  
Since we
know that the 3-loop RG-solution for $\alphaqqbar$ 
is accurate, this also hold
for $V(r)$ computed through \eq{e_vint}. Again, the full line moves
very close to the data points ($r<0.5\rnod$), when ${\Lambda_{\msbar}}\rnod$ at
the upper end of the error bar is inserted.\\

Another possibility is to compare the potential directly with \eq{e_alphavbar}
by using the RG-solution for $\alphavbar$.
From the considerations of section \ref{s:pot_pert} we expect that perturbation theory is applicable only up to $\alphavbar\lesssim 0.15$; by solving the RG equation in the $\Vbar$ scheme with the central value \eq{e_lambda_rnod}, one observes that at $r=0.3\rc$ one has $\alphavbar({\rm 2 loop})\simeq 0.28$, $\alphavbar({\rm 3 loop})\simeq 0.9$. This means that one is already at those small distances outside the region where perturbation theory can be trusted. The long-dashed line in \fig{f_pot_comparison} represents \eq{e_alphavbar} with the 2-loop RG solution for $\alphavbar$; we decided not to use the 3-loop prediction due to the fact that the coupling is very large at the distance $r\sim 0.15\rnod$.
As it was to
be expected due to the missing
stability of this perturbative expression, it fails in describing the
non-perturbative potential.\\
A similarly bad perturbative expression 
is the direct expansion of the potential in terms of $\alphaMSbar$;
for this comparison we used the matching equation \eq{e_match}, with
$S'=\Vbar$, $S=\MSbar$. For the same reason explained before, we decided to 
use the 2-loop RG prediction for $\alphaMSbar$.
The dotted line corresponds to the choice $s=1$ in \eq{e_match}, while the
dashed-dotted line corresponds to $s=s_{0}$ (\eq{eq_s0}).\\

%%%%%%%%%%%%%%%%%%%%%%%%%%%%%%%%%%%%%%%%%%%%%%%%%%%%%%%%%%%%%%%%%%%%%%%%%%%%%%
%%%%%%%%%%%%%%%%%%%%%%%%%%%%%%%%%%%%%%%%%%%%%%%%%%%%%%%%%%%%%%%%%%%%%%%%%%%%%%

%%%%%%%%%%%%%%%%%%%%%%%%%%%%%%%%%%%%%%%%%%%%%%%%%%%%%%%%%%%%%%%%%%%%%%%%%%%%%%
%%%%%%%%%%%%%%%%%%%%%%%%%%%%%%%%%%%%%%%%%%%%%%%%%%%%%%%%%%%%%%%%%%%%%%%%%%%%%%
In summary, care has to be taken which perturbative scheme is adopted
to describe the potential. However, perturbation theory does its best
in the following sense. As usual in an asymptotic expansion, one
should first investigate the apparent ''convergence'' by comparing
subsequent orders and checking that they decrease significantly. If
this is not the case, one is obviously outside the domain of
applicability of perturbation theory or has chosen a bad truncation
(scheme). According to this criterion the $\beta$-function in the
$\qqbar$-scheme may be trusted up to $\alphaqqbar\approx 0.3$. Other
truncations of perturbation theory for the potential are applicable
for much smaller values of the coupling only; for example, a scheme
with a large 3-loop coefficient such as $\Vbar$ is of no use in the
region $\alpha>0.15$. Therefore, perturbation theory suggests that the
$\qqbar$ scheme should be used in order to obtain a reliable
perturbative expression.  The comparison with non-perturbative
results, obtained in the continuum limit of lattice simulations
($\nf=0$), does confirm that such a perturbative analysis is a good
guideline -at least in the case at hand.  At rather high values of the
coupling, $\alpha\approx 0.3$, we only expect that the 3-loop
perturbative prediction is good to within about $10\%$ and indeed in
\fig{f_alphaqqsf} one sees explicitly that that the truncated
perturbative series has errors of this order. Similar results have
been found for the coupling obtained with $\nf=2$ in the SF-scheme
\cite{Bode:2001jv,DellaMorte:2002vm}.\\ If one considers the full QCD, the first
observation is that, compared to $\nf=0$, the relevant perturbative
coefficients, $b_{2}$ and $f_{2}(s_0)$ which are listed in \tab{t_ei}
and \tab{t_ai}, are roughly a factor two smaller in magnitude for
$\nf=3$. This suggests that including the quarks the perturbative
prediction for the potential computed through $\alphaqqbar$ is also
applicable up to $\alpha\approx 0.3$ and furthermore in full QCD the
issue of the appropriate scheme is somewhat less important.  A direct
lattice QCD check of these expectations is not possible at present and
one has to boldly generalize from the $\nf=0$ case.  In addition,
these remarks apply only to the massless theory (mass effects were not
investigated).  Current phenomenological research concentrates on
application of a velocity dependent potential beyond the static limit
for applications to top-quark physics \cite{Hoang:2000yr}. On the one
hand, in this application the potential is needed for quite short
distances, where perturbation theory is intrinsically more precise
\cite{Kuhn:1981gw}, on the other hand, with the velocity entering as a
new scale, this represents a more difficult multiscale problem. Indeed
the renormalization group has already been applied to deal with this
complication \cite{Hoang:2001rr}.\\ Nevertheless, the lessons learnt
in our investigation may be useful in this context as well; the type
of renormalization group improvement which we found to increase the
reliability of perturbation theory has not been applied in
\cite{Hoang:2000yr} so far.

%%%%%%%%%%%%%%%%%%%%%%%%%%%%%%%%%%%%%%%%%%%%%%%%%%%%%%%%%%%%%%%%%%%%%%%%%%%%
%%%%%%%%%%%%%%%%%%%%%%%%%%%%%%%%%%%%%%%%%%%%%%%%%%%%%%%%%%%%%%%%%%%%%%%%%%%%
\section{Comparison with the bosonic string model}\label{sec_stringcom}
\begin{figure}
\begin{center}
\includegraphics[width=9cm]{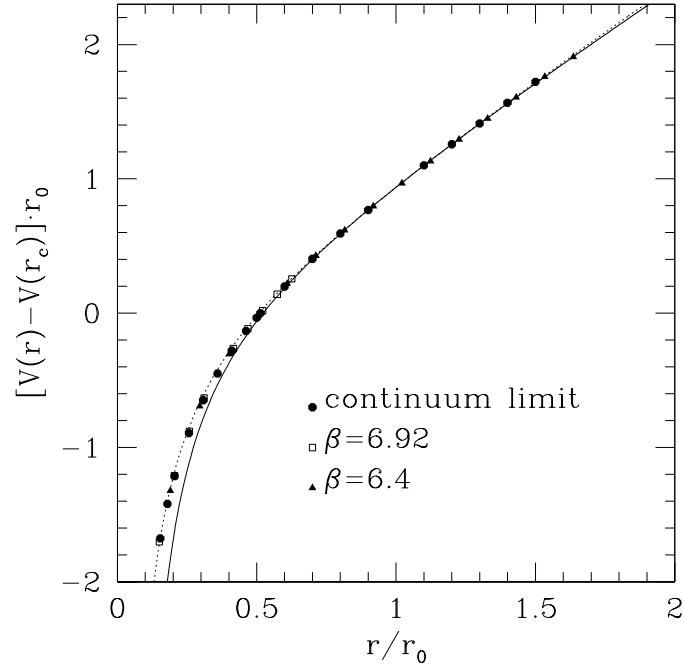}
\caption[The potential compared with
  the parameter-free prediction of the bosonic string model.]{\footnotesize \label{f_pot_string}The potential compared with
  the parameter-free prediction of the bosonic string model, from
  \protect\eq{pot_string} (dashed line) and
  \protect\eq{string_corrected} (solid line). The constant term has been fixed such that the predictions coincide with the non-perturbative results at $r=0.9\rnod$. }
\end{center}
\end{figure}

Although we concentrated on the evaluation of the potential at short and
intermediate distances, where the validity of the string picture described in
section \ref{sec_string} is a priori not expected, it is interesting to compare our
results with this model.\\ 
If one assumes that the bosonic string picture described is valid already at
$r=\rnod$, at next-to-leading order (NLO) in the effective theory one obtains the parameter free predictions
\begin{equation}\label{pot_string_prediction}
V(r)=\sigma r-\frac{\pi}{12r}+const,
\end{equation}
\begin{equation}\label{force_string_prediction}
F(r) =  \sigma+\frac{\pi}{12 r^2},
\end{equation}
where $\sigma$ is fixed by the \eq{e_rnod},
\begin{equation}
\sigma\rnod^2=1.65-\frac{\pi}{12}.
\end{equation}
In \fig{f_pot_string} one can see that \eq{pot_string_prediction} (dotted line) is in excellent agreement with our results for $r\geq 0.8 \rnod$.\\
As already mentioned in sec. \ref{sec_string}, in the same region of $r$, excited potentials do not at all follow the expectations from the effective bosonic string theory. This may suggest that the agreement with \eq{pot_string_prediction}
is rather accidental. In any case one would expect corrections to this formula
to be negligible only for much larger $r$.\\
If one includes the higher order correction (NNLO) \eq{string_corrected} founded in \cite{Luscher:2002qv} (solid line), one 
can notice that at short distances the agreement
is worse, while at our largest distances it seems to be better, although 
our precision here is not good enough to make any definitive statement. This fact seems to confirm that the agreement at short distances is 
coincidental.\\
The fact that at short distances the NLO prediction is able to describe the
data better than the NNLO is not surprising: we are dealing with an asymptotic
expansion in a region where the expansion parameter $\frac{1}{r}$ is large.
Under these circumstances it is well known that the introduction of higher
order terms in the series do not necessarily improve a prediction and can even
make it worse.\\

\begin{figure}
\begin{center}
\includegraphics[width=9cm]{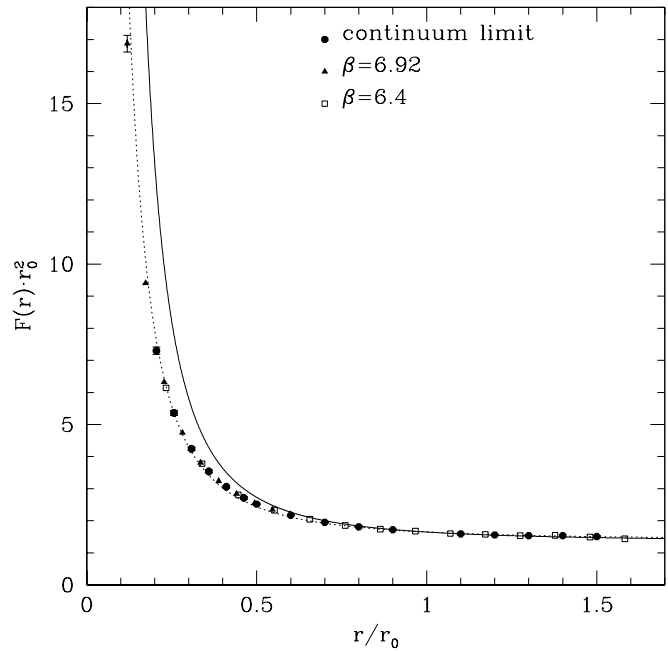}
\caption[The force compared with
  the parameter-free prediction of the bosonic string model.]{\footnotesize \label{f_force_string}The force compared with
  the parameter-free prediction of the bosonic string model. The dashed
  line represents the string prediction at leading order
  \protect\eq{force_string_prediction}, while the solid line include the
  correction \protect\eq{string_corrected} with $b=0.04\fm$.}
\end{center}
\end{figure}

\begin{figure}
\begin{center}
\includegraphics[width=9cm]{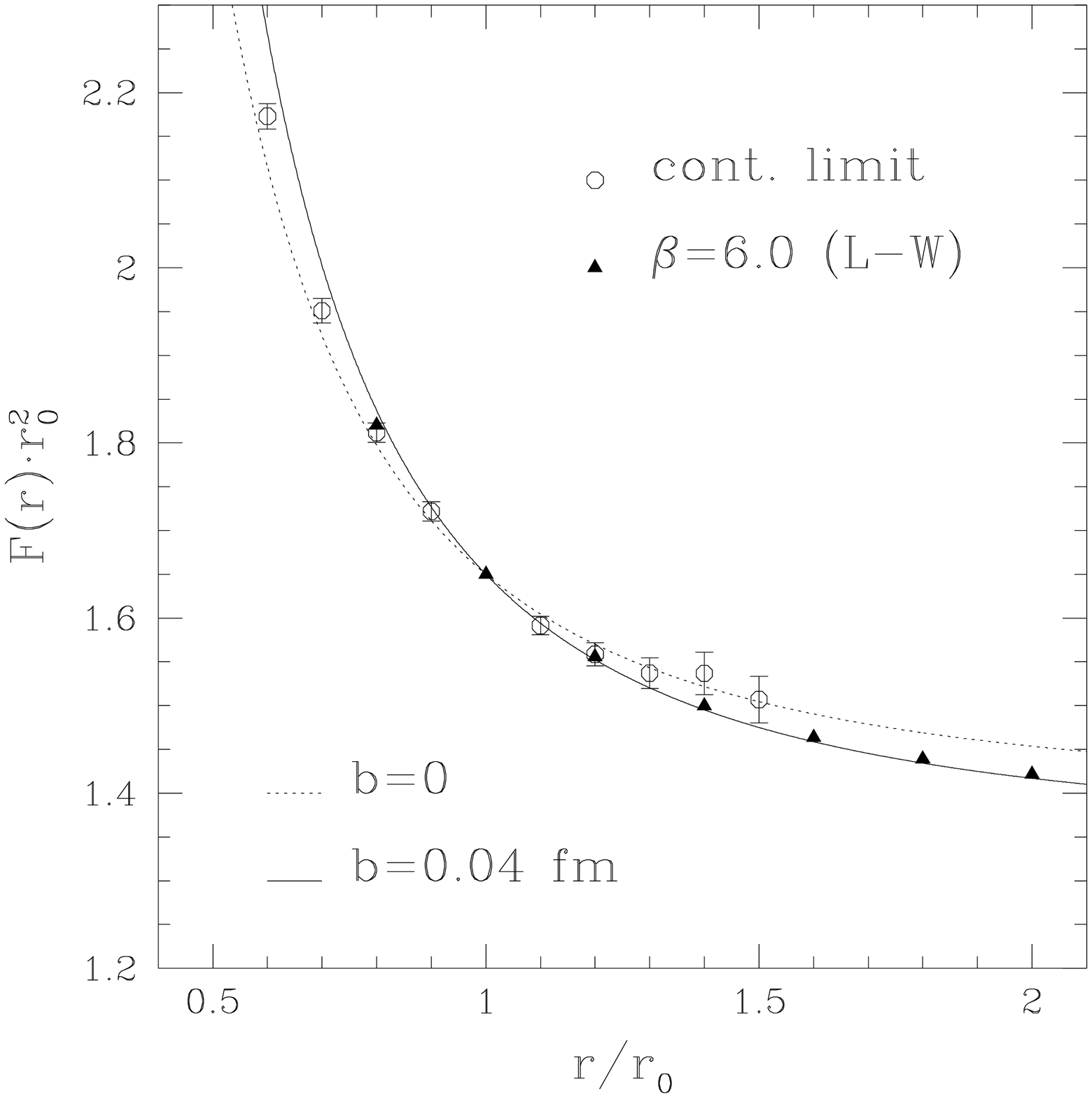}
\caption[The force compared with
  the bosonic string model at large distances.]{\footnotesize \label{f_force_stringcom}The force compared with
  the bosonic string model at large distances.
The points correspond to our computation of the force extrapolated to the continuum, while the triangles are taken from \protect\cite{Luscher:2002qv} for $\beta=6.0$.}
\end{center}
\end{figure}

\Fig{f_force_string} shows the same comparison for the force; finally, in \fig{f_force_stringcom} we compared our results for large $r$ with the ones obtained by L\"uscher and Weisz for their largest $\beta$ \cite{Luscher:2002qv}. One observes that the algorithm for the exponential error reduction \cite{Luscher:2001up} adopted there is highly necessary to obtain a good precision of the data.
On the contrary, with our results for $r>1.2\rnod$ we would not be able to evaluate the correction $b$ in \eq{string_corrected}.

%%%%%%%%%%%%%%%%%%%%%%%%%%%%%%%%%%%%%%%%%%%%%%%%%%%%%%%%%%%%%%%%%%%%%%%%%%%%%%%%%%


%%%%%%%%%%%%%%%%%%%%%%%%%%%%%%%%%%%%%%%%%%%%%%%%%%%%%%%%%%%%%%%%%%%%%%%%%%%%%%%%%%%

From this discussion one can conclude that the understanding of the long 
range properties of the static quark
potential and the testing of bosonic string model is still an important
task and requires the employment of advanced numerical techniques in
order to have a reliable measurement of the observables in that
range.

%%% Local Variables: 
%%% mode: latex
%%% TeX-master: t
%%% End: 

%% file: tab_cont.tex
  \begin{table}[ht] 
\caption[Potential and force after 
continuum extrapolation.]{\label{t_cont}\footnotesize{Potential and force after 
continuum extrapolation. }}
\begin{center}
  \begin{tabular}{ c  c  c  r  c  c}
  \hline 
$r/\rc$ & $\rnod^2 F(r)$ &  $\rnod(V_{\rm I}(r)-V_{\rm I}(\rc))$  &~~~~~~ $r/\rnod$ & $\rnod^2 F(r)$ &  $\rnod(V_{\rm I}(r)-V_{\rm I}(\rc))$\\ 
  \hline 
 $ 0.3  $ & $   $ & $ -1.676 ( 16 )$ & $ 0.5  $ & $ 2.518 ( 16 ) $ & $ -0.0336 ( 61 )$ \\ 
 $ 0.4 $ & $ 7.30 ( 11 ) $ & $ -1.2125 ( 93 )$ & $ 0.6  $ & $ 2.173 ( 15 ) $ & $ 0.1989 ( 76 )$ \\
 $ 0.5 $ & $ 5.363 ( 78 ) $ & $ -0.8926 ( 74 )$ & $ 0.7  $ & $ 1.951 ( 14 ) $ & $ 0.4051 ( 89 )$ \\
 $ 0.6 $ & $ 4.244 ( 53 ) $ & $ -0.6475 ( 54 )$ & $ 0.8  $ & $ 1.812 ( 11 ) $ & $ 0.5930 ( 99 )$ \\
 $ 0.7 $ & $ 3.538 ( 48 ) $ & $ -0.4494 ( 35 )$ & $ 0.9  $ & $ 1.722 ( 11 ) $ & $ 0.769 ( 11 )$ \\
 $ 0.8 $ & $ 3.060 ( 38 ) $ & $ -0.27950 ( 15 )$ & $ 1.1  $ & $ 1.592 ( 10 ) $ & $ 1.101 ( 11 )$ \\
 $ 0.9 $ & $ 2.713 ( 30 ) $ & $ -0.13259 ( 67 )$ & $ 1.2  $ & $ 1.559 ( 13 ) $ & $ 1.258 ( 12 )$ \\
 $     $ & $              $ & $                $ & $ 1.3  $ & $ 1.537 ( 18 ) $ & $ 1.413 ( 13 )$ \\
 $     $ & $              $ & $                $ & $ 1.4  $ & $ 1.537 ( 24 ) $ & $ 1.567 ( 14 )$ \\
 $     $ & $              $ & $                $ & $ 1.5  $ & $ 1.507 ( 27 ) $ & $ 1.722 ( 24 )$ \\
  \hline 
  \end{tabular} 
\end{center}
  \end{table} 

%%% Local Variables: 
%%% mode: latex
%%% TeX-master: t
%%% End: 

%% file: tcr0.tex
\chapter{Scaling properties of RG actions: the critical temperature}\label{improved}
Starting from the Wilson formulation of Yang Mills theory on the lattice, \eq{wilson_action}, it was observed that for many physical observables it leads to sizeable discretization errors. To perform a continuum extrapolation of the physical quantities of interest usually requires a big computational effort which is often not practicable. Therefore it is highly desirable to use a lattice formulation which reduces the discretization errors to have a more reliable estimation of physical quantities. 
% to be reviewed
As we have seen, the Wilson gauge action shows leading lattice artefacts of order ${\rm O}(a^2)$; by formulating fermions on the lattice in general one expects corrections starting already at ${\rm O}(a)$. For this reason in recent years most of the efforts were spent to improve the fermionic action in view of unquenched simulations.\\
For example, by following the Symanzik approach, one can attempt to cancel all
lattice effects of order $a$; this development started some time ago \cite{impr:SW,impr:Wohlert} and the coefficients multiplying the various ${\rm {O}}(a)$ counterterms in the improved theory have been determined
non-perturbatively \cite{impr:pap1,impr:pap2,impr:pap4}.\\
In any case it turns out that also the gauge part plays a very important
r{\^o}le, and the question which is the most convenient gauge action to adopt
has been often addressed.\\
In particular, the Iwasaki action has been used by the CP-PACS collaboration
\cite{AliKhan:2001tx} for the most advanced computation of light hadron
spectrum. Moreover, the RG improved actions have been considered as candidates
to be used in the next simulations on Ginsparg-Wilson/domain wall fermions;
interesting features were observed for these actions, as 
the suppression of small instantons and dislocations and a possible
remedy of the problem of residual chiral symmetry breaking for domain wall fermions \cite{Orginos:2001xa,Aoki:2002vt}.\\
This increasing interest in alternative gauge actions motivates more investigations into their properties, starting from 
the basic ones, like universality and scaling behavior.
There are in principle several quantities that one can use to quantify the lattice artifacts and to test universality by comparing the results with the plaquette action known in the literature. 
The renormalized observables that we have considered are $T_{c}\rnod$, where
$T_{c}$ is the deconfining phase transition temperature, and $m_{G}\rnod$,
where $m_{G}$ represents the glueball mass and will be discussed in the next chapter. In particular, our study required the computation of the scale $\rnod/a$ for RG actions, which was up to now not present in the literature.\\
In this chapter the physical observables will be defined, the details of the lattice calculation will be explained with particular emphasis on the possible difficulties and the results will be discussed.

\section{The critical temperature $T_{c}$}
In chapter \ref{c_lattice} it has been pointed out that Euclidean quantum field theory
in the functional integral formalism, in particular its lattice regularized
version, is formally equivalent to a system of classical statistical mechanics
in four dimensions.\\
Complementary to this, another point of view makes use of Hamiltonian
formalism and the transfer matrix (see appendix \ref{app_trans}); 
from \eq{z_t}, \eq{hamiltonian} one obtains that for a lattice with 
$N_{t}$ points in the time
direction and $N_s$ points in the three space-directions (we assume for
simplicity to be the same for all the space directions) \footnote{In order to avoid notational confusion, in this chapter we will use $T$ for the temperature, and
  $N_{t}$ for the time-extent of the lattice. }, the partition function of the system is given by
\begin{equation}\label{part_f2}
Z=\tr(e^{-aN_{t}\mathbb{H}}).
\end{equation}
In the limit $N_{s},N_{t}\rightarrow\infty$ this corresponds to a 4-dimensional quantum field theory at physical temperature $T=0$.\\
For several applications, for example the study of 
hot hadronic matter or early universe, it would be desirable to consider 
quantum field theory at finite physical temperature $T$ also.\\
In quantum statistical mechanics the partition function of a
finite-temperature system is defined by
\begin{equation}
Z(\frac{1}{T})=\tr (e^{-\frac{1}{T}\mathbb{H}}),
\end{equation}
where the Boltzmann factor has been set equal to 1, 
and the thermal expectation value of any observable $A$ is given by
\begin{equation}
\langle A\rangle =\frac{1}{Z(\frac{1}{T})}\tr(e^{-\frac{1}{T}\mathbb{H}}A).
\end{equation}
Comparing these expressions with \eq{part_f2} and \eq{expect_value},
one recognizes that a ${N_s}^3 N_t$ lattice with periodic boundary conditions
in the time direction can be regarded as 
a system of finite volume $V= (N_s a)^3$ at finite temperature $T$, 
which is related to $N_{t}$ by
\begin{equation}\label{temperature}
\frac{1}{T}=N_{t}a.
\end{equation}
The thermodynamic limit corresponds to $N_{s}\rightarrow\infty$; in order to
approximate this condition in a simulation in a finite spatial
volume, one should respect the relation
\begin{equation}
N_{s}\gg N_{t}.
\end{equation}
%In the limit $N_{t}\rightarrow\infty$ one recovers the zero-temperature field
%theory.\\
It was pointed out in \cite{Polyakov:1978vu,Susskind:1979up} that pure Yang-Mills theory undergoes a first order phase
transition at some temperature $T_{c}$; below the critical temperature one has
the confined phase, while the high temperature region is characterized by a
gluon plasma with freely moving but still interacting gluons.\\
The Polyakov loop \footnote{Note that here ``$\Tr$'' indicates the trace on the gauge group $\SUn$, while in \eq{part_f2} ``$\tr$'' is the trace in the Hilbert space.} 
\begin{equation}
P({\vec x})=\Tr\prod_{x_{0}=0}^{N_{t}-1}U(x,0)
\end{equation}
can be interpreted as an order parameter for the deconfining
phase transition: in the deconfined phase it assumes a non-vanishing
expectation value $\langle P({\vec x})\rangle\neq 0$.\footnote{Strictly
  speaking, only the Polyakov loop correlator 
$\langle P({\vec x})P^{\dagger}({\vec 0})\rangle$ has a physical meaning and can
be used as an order parameter. Assuming cluster decomposition at large
distances one obtains $\langle P({\vec x})P^{\dagger}({\vec
  x})\rangle\sim_{|{\vec 0}|\rightarrow\infty}|\langle P\rangle|^2$, and one
can notice that the phase of $\langle P\rangle$ is not a physically measurable quantity.}\\
In full QCD the low temperature phase corresponds to the usual hadronic QCD
vacuum, and the high temperature phase is a quark-gluon plasma where the
characteristic low energy features, like confinement and spontaneous chiral
symmetry breaking are lost, and the short distance behavior of matter is
governed by the asymptotic freedom of QCD (see \cite{Smilga:1997cm} for a review).\\ 

The first non-perturbative lattice determinations of $T_{c}$ can be
found in \cite{Kuti:1981gh}; the general strategy consists of keeping the relation in 
\eq{temperature} fixed while varying the gauge coupling $\beta$ and therefore
implicitly the lattice spacing $a$. In this way one moves the lattice system
through the phase transition obtaining
\begin{equation}
\frac{1}{T_{c}}=N_{t}a(\beta_{c}).
\end{equation}
There are several methods for determining the critical coupling $\beta_c$; it
is desirable to refer to a definition based on a quantity with a definite
finite size scaling behavior, for instance the Polyakov loop susceptibility,
defined as
\begin{equation}
\chi =V (\langle|P^{2}|\rangle-\langle|P|\rangle^{2}),\quad V=N_{s}^3.
\end{equation}
In the thermodynamic limit, the susceptibility has a delta-function
singularity corresponding to a first order phase transition.
On a finite lattice the singularity is rounded off and one observes a peak for
$\beta_{c}(V)$. The critical coupling for each value of $N_{t}$ can be
extrapolated to the thermodynamic limit using the finite size scaling law for
a first order phase transition,
\begin{equation}
\beta_{c}(N_{t},N_s)=\beta_{c}(N_{t},\infty)-h\left(\frac{N_t}{N_s}\right)^3,
\end{equation}
where $h$ is considered to be a universal quantity independent of $N_t$
\cite{Beinlich:1997ia}.\\

\begin{figure}
\begin{center}
\includegraphics[angle=-90,width=9cm]{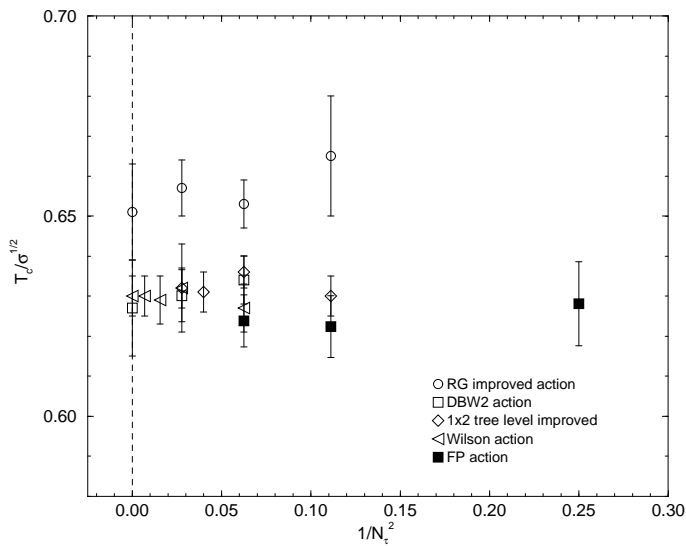}
\end{center}
\caption[$T_{c}/\sqrt{\sigma}$ as function of $1/N_{t}^2$ for different actions.]{\footnotesize{$T_{c}/\sqrt{\sigma}$ as function of $1/N_{t}^2$ for different actions, from \protect\cite{Niedermayer:2000yx}. By ``RG improved'' is meant here the Iwasaki action.}\label{tc_sqrtsigma_fig}}
\end{figure}

In addition to its intrinsic importance as a fundamental non-perturbative
prediction, $T_{c}$ provides also a useful quantity to study the lattice
artifacts for different gauge actions and to test universality. \\
In order to investigate the scaling
properties, one has to measure a dimensionful quantity like $\rnod$ 
at zero temperature for $\beta=\beta_{c}$ and then
build the renormalization group invariant combination $\rnod T_{c}$.
For known values of $\beta_{c}$ at given $N_{t}$ for different actions one can
refer to \cite{Boyd:1996bx,Beinlich:1997ia,Okamoto:1999hi,deForcrand:1999bi,Niedermayer:2000yx};
for Wilson, Iwasaki and DBW2 actions, these values are collected in \tab{tab_betacrit}.\\
Instead of using $\rnod$ to set the scale,
the results that are available in the literature are mostly related to the quantity $T_{c}/\sqrt{\sigma}$;
they are reported in \tab{tc_sqrtsigma}, where the continuum extrapolations are from a reanalysis by Teper \cite{glueb:teper98}. 
\Fig{tc_sqrtsigma_fig} collects the results for Wilson, Iwasaki, DBW2, Symanzik tree level and FP action, and was taken from \cite{Niedermayer:2000yx} which reports the latest evaluation of $T_{c}/\sqrt{\sigma}$ (FP action). 

%%%%%%%%%%%%%%%%%%%%%%%%%%%%%%%%%%%%%%%%%%%%%%%%%%%%%%%%%%%%%%%%%%%%%%%%%%%%%
\begin{table}
\begin{center}
\begin{tabular}{c c c c c}
\hline
        &  $\beta_{c}$ & Wilson \cite{Beinlich:1997ia}  & Iwasaki \cite{Okamoto:1999hi}  & DBW2 \cite{deForcrand:1999bi}  \\
$N_{t}$ &              &                 &                  &               \\
\hline
3       &              &                 & 2.1551(12)       & 0.75696(98) \\  
4       &              & 5.6925(2)       & 2.2879(11)       & 0.82430(95) \\
6       &              & 5.8941(5)       & 2.5206(24)       & 0.9636(25) \\ 
8       &              & 6.0624(12)      & 2.7124(34)       &             \\
12      &              & 6.3380(23)      &                  &               \\
\hline
\end{tabular}
\end{center}
\caption[The critical coupling from the literature for Wilson,
        Iwasaki and DBW2 actions.]{\footnotesize{The critical coupling from the literature for Wilson,
        Iwasaki and DBW2 actions.  \label{tab_betacrit}}}
\end{table}

%%%%%%%%%%%%%%%%%%%%%%%%%%%%%%%%%%%%%%%%%%%%%%%%%%%%%%%%%%%%%%%%%%%%%%%%%%%%%
\begin{table}[h]
\begin{center}
\begin{tabular} {c c c}
\hline
action          & $N_{t}$   & $T_{c}/\sqrt{\sigma}$ \\
\hline
Wilson\cite{Beinlich:1997ia} & $\infty$  &  0.630(5)                     \\
Iwasaki \cite{Okamoto:1999hi} & $\infty$  &  0.651(12)                     \\
DBW2 \cite{deForcrand:1999bi} & $\infty$  &  0.627(12)                     \\
Sym. tree level\cite{Beinlich:1997ia} & $\infty$  &  0.634(8) \\
1 loop tadpole impr. \cite{Bliss:1996wy}  &  $\infty$  &  0.659(8)\\
FP \cite{Niedermayer:2000yx}             & 4         &  0.624(7)                     \\
\hline
\end{tabular}
\end{center}
\caption[Results for the deconfining temperature in units
  of the string tension from different actions.]{\footnotesize{\label{tc_sqrtsigma}Results for the deconfining temperature in units
  of the string tension from different actions \protect\cite{glueb:teper98}. }}
\end{table}
The first observation from \fig{tc_sqrtsigma_fig} is that from this specific quantity it is not possible to arrive at precise 
conclusions about the discretization errors for different actions.\\
Furthermore, one notices that for the Iwasaki and the Wilson action a difference of order $2\sigma$ in the continuum results is observed. 
The most drastic explanation for this discrepancy could be a violation of 
universality, but this scenario seems unrealistic; a
more natural explanation is that the string tension is difficult to determine,
as already pointed out in chapter \ref{chapt_potential}, and it is preferable
to use $\rnod$ to reliably set the scale in pure Yang-Mills theory.\\
For the Wilson action, the values of $\rnod/a$ corresponding to the critical
couplings can be easily obtained by the parametrization formula \cite{pot:r0_SU3}
\footnote{We decided to use the parametrization formula of \protect\cite{pot:r0_SU3}
  instead \eq{e_rnodfit}, which is valid for larger range of $\beta$ but less accurate.}
\begin{equation}\label{fit_wilson}
\ln(a/\rnod)=-1.6805-1.7139(\beta-6)+0.8155(\beta-6)^2-0.6667(\beta-6)^3.
\end{equation}
For the Iwasaki and DBW2 action there was up to now no precise evaluation of
$\rnod/a$ and we decided to perform new numerical simulations with this purpose.

\section{Evaluation of $\rnod/a$ for RG actions}
For the evaluation of $\rnod/a$ we followed the same procedure as for the
usual plaquette action, as described in \ref{setting_the_scale}. We
constructed Wilson loop correlation matrices employing smearing techniques and
multi-hit methods \footnote{Due to the presence of $(1\times 2)$ plaquettes
  in the RG actions, not every temporal link appearing in the Wilson loop can
  be substituted by the one obtained with the multi-hit integration. We
  applied the multi-hit procedure only on alternating links so that two
  multi-hit links can not belong to the same $(1\times 1)$ or $(1\times 2)$ loops.}. 
We will not repeat here the description of the
procedure, but mention only the difficulties with respect to the plaquette
case. The simulation parameters are reported in \tab{simpar_iwasaki} and \tab{simpar_dbw2}.\\
\subsection{Effective potential}
We adopted the same smearing operator as for the Wilson action \eq{smearing},
with $M=4$, $\alpha=\frac{1}{2}$ and the same condition for the smearing levels \eq{smear_level}.\\ 
For the Wilson action it is possible to define a positive transfer
matrix and hence a well-defined Hamiltonian; from \eq{pot_spectrum} one
notices that the contributions to the spectral decomposition of correlation
functions are always positive.
On the contrary, for the Iwasaki and DBW2 action, 
due to the violation of physical positivity
described in appendix \ref{app_trans}, one
expects also negative contributions, so that the plateau in the effective
potential is reached from below. This fact spoils the possibility to find an
unambiguous criterion to define the "optimal" smearing: the ability
of the smearing operator to suppress the higher order excitations can be tested
only for $t\gg t_{min}$, where $t_{min}$ is estimated in \eq{tmin}.\\

\begin{figure}
\begin{center}
\includegraphics[width=9cm]{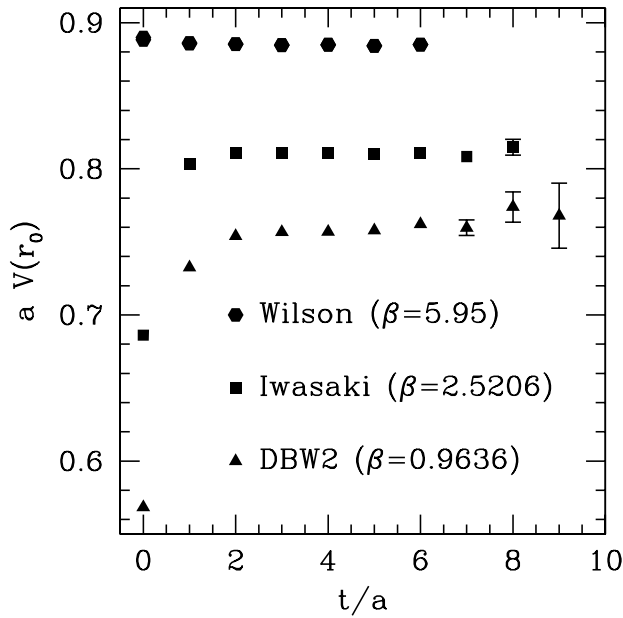}
\caption[The effective potential as function of $t$ for
    $r\approx\rnod$ for the Wilson plaquette action and for
    the RG-improved Iwasaki and DBW2 action.]{\footnotesize{The effective potential as function of $t$ for
    $r\approx\rnod$ ($\rnod\sim 4-5$) for the Wilson plaquette action and for
    the RG-improved Iwasaki and DBW2 action.}\label{f_potr0}}
\end{center}
\end{figure}

We started our analysis by observing the time-dependence of the
effective potential for the different smearing levels.\\
\Fig{f_potr0} shows the effective potential
\begin{equation}\label{v_effic}
a V(r)=-\ln\left(\frac{C_{22}(t)}{C_{22}(t-a)}\right)
\end{equation}
evaluated from the diagonal elements of the correlation matrices \eq{corrma}
for $l=m=2$, which corresponds to what was estimated to be the optimal smearing for the Wilson action.
%%%%%%%%%%%%%%%%%%%%%%%%%%%%%%%%%%%%%%%%%%%%%%%%%%%%%%%%%%%%%%%%%%%%%%%%%%
%%%%%%%%%%%%%%%%%%%%%%%%%%%%%%%%%%%%%%%%%%%%%%%%%%%%%%%%%%%%%%%%%%%%%%%%%%%%%
Notice that the negative contributions for the RG actions are strongly
evident and, as expected, they disappear starting from a certain $t$ which
becomes larger as the coefficient $c_{1}$ in the action increases.\\
One observes however a quite satisfactory plateau in the effective potential,
starting at sufficiently large $t$. We decided to extract the potential from
\eq{v_effic} without applying the variational method, which is mathematically
founded on the positivity of the correlation matrix $C_{lm}$ at a certain
$t=t_{0}$; this condition is verified only for $t_{0}\gg t_{min}$, but on the
other hand $t_{0}$ can not be arbitrarily large because of the statistical
errors, which increase with $t$, that make the inversion of $C_{lm}$ 
impracticable.\\
We extracted the potential from \eq{v_effic} at $t/a=(3-4)$ and we estimated
the systematic error by taking the difference between this value and what one
would obtain extracting the potential at $(t+a)$. The systematic errors were
linearly added to the statistical one; at small $r$ the total uncertainty is dominated by the systematic error, while at large distances the situation is in general reversed.

\subsection{The force at tree level}
\begin{figure}
\begin{center}
\includegraphics[width=9cm]{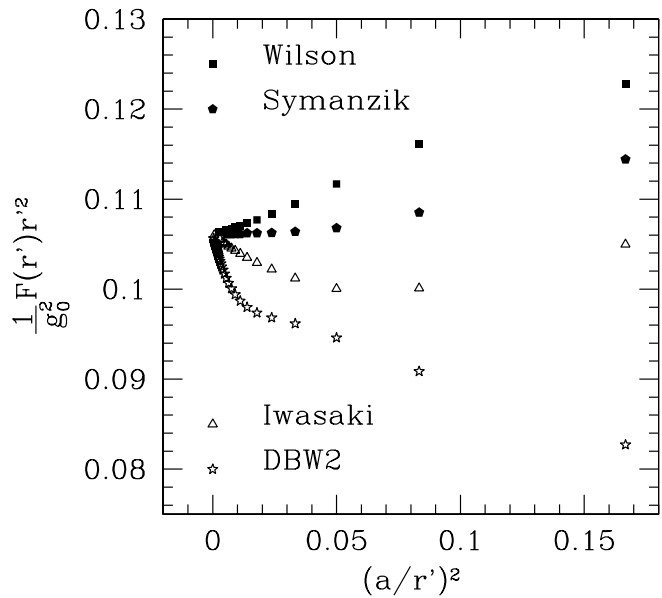}
\end{center}
\caption[The force at tree level for several actions.]{\footnotesize{The force at tree level for several actions.\label{fig_forcetree}}} 
\end{figure}

It is interesting to study the force at tree level for different actions, in
order to investigate how the continuum limit is approached at small coupling $\overline{g}$, where it is legitimate to
expect that the tree-level approximation is quite accurate.\\
Explicitly, the force at tree level is given by \cite{Weisz:1983zw} 
\begin{equation}\label{force_treelevel_improved}
F_{tree}(r^{\prime}) = \frac{V_{tree}(r)-V_{tree}(r-a)}{a} 
\end{equation}
$$
                     = -\frac{4}{3}\frac{g_{0}^{2}}{a}\int_{-\pi}^{\pi}\frac{d^{3}k}{(2\pi)^3}\frac{\cos(rk_{1}/a)-\cos((r-a)k_{1}/a)}  {4\left(\sum_{j=1}^{3}\sin^{2}(k_{j}/2)-4c_{1}\sum_{j=1}^{3}\sin^{4}(k_{j}/2)\right)}. 
$$
Note that for $c_{1}=0$ one obtains the results already mentioned for the
plaquette action \eq{force_treelevel}. For the special choice $c_{1}=-1/12$,
which corresponds to the Symanzik tree-level improved action, one finds \cite{Weisz:1983zw}
%\begin{equation}
%F_{tree}^{sym}(r^{\prime})=\frac{1}{a}\frac{4g_{0}^2}{3}\left(-\frac{1}{4\pi r}+ \frac{1}{4\pi(r-a)}  \right) + {\rm O}(a^4),
%\end{equation}
%so that
\begin{equation}\label{force_tree_sym}
{r^{\prime}}^{2}F_{tree}^{sym}(r^{\prime})=\frac{4}{3}\frac{g_{0}^2}{4\pi}+
{\rm O}\left(\frac{a}{r^{\prime}}\right)^{4},
\end{equation}
if $r^{\prime}=\sqrt{r(r-a)}$; this choice of the action cancels indeed the ${\rm O}(a^2)$ terms at tree level.\\
In \fig{fig_forcetree} the quantity
$\frac{{r^{\prime}}^{2}}{g_{0}^2}F_{tree}(r^{\prime})$ is plotted as function
of $(a/r^{\prime})^2$ for different actions. As expected, in the limit
$r^{\prime}\rightarrow\infty$ all curves converge to the classical value
$\frac{4}{3}\frac{1}{4\pi}$. For the Symanzik tree-level improved action one
finds \eq{force_tree_sym}, that is no ${\rm O} (a^2)$ lattice
artifacts in the force. One can notice that the RG actions, in particular
the DBW2, show at tree level large lattice artefacts at small
$(a/r^{\prime})$. The RG 
actions should reduce the discretization errors at every order in perturbation
theory, but it turns out that at tree level these actions are "overcorrected" and
introduce lattice artefacts of the same order or even larger than what is
expected with the usual plaquette action. This fact should be considered as a
problem if one would for example evaluate the potential at short distances, as
we did for the Wilson action in chapter. \ref{chapt_potential}: in
particular one should consider the continuum extrapolation with great care, because unless $a/r$ is very small, one can not be sure to be in the region where the leading discretization errors are quadratic in $a$.\\
%%%%%%%%%%%%%%%%%%%%%%%%%%%%%%%%%%%%%%%%%%%%%%%%%%%%%%%%%%%%%%%%%%%%%%%%
%%%%%%%%%%%%%%%%%%%%%%%%%%%%%%%%%%%%%%%%%%%%%%%%%%%%%%%%%%%%%%%%%%%%%%%%%%
For our evaluation of $\rnod/a$, the force is defined exactly as in sect. \ref{def_of_the_force}; also in this
case one can introduce a tree-level improved definition of the force such that
\begin{equation} 
F_{tree}(\rI)=\frac{4}{3}\frac{g_{0}^2}{4\pi\rI^2},
\end{equation}
with no lattice artefacts at tree level.\\
From \eq{force_treelevel_improved} one deduces that
\begin{equation}\label{e_rI_new}
 (4\pi\rI^2)^{-1}=-[G(r,0,0)-G(r-a,0,0)]/a,
\end{equation}
where 
\begin{equation}\label{prop_impr}
  G({\vec r})=\frac{1}{a}\int_{-\pi}^{\pi}\frac{d^{3}k}{(2\pi)^3}\frac{\prod_{j=1}^{3}\cos(r_{j}k_{j}/a)}{4(\sum_{j=1}^{3}\sin^{2}(k_{j}/2)-4c_{1}\sum_{j=1}^{3}\sin^{4}(k_{j}/2))}, 
\end{equation}
is the scalar lattice propagator associated to the improved action with
coefficient $c_{1}$ in the $(1\times 2)$ plaquette term.
Due to the fact that $G({\vec r})$ does not satisfy \eq{green} because of 
the presence of higher derivative terms in the action, it is not
possible to use the coordinate space methods described in appendix \ref{app1};
$\rI$ was then computed solving the integral \eq{prop_impr} numerically.
%%%%%%%%%%%%%%%%%%%%%%%%%%%%%%%%%%%%%%%%%%%%%%%%%%%%%%%%%%%%%%%%%%%%%%%%%%
\begin{table}[h]
\begin{center}
\begin{tabular}{c c c c c}
\hline
$N_{s}$ & $\beta$  &   $n_l$    & $N_{or}$ & $N_{meas}$  \\
\hline
8       & 2.1551   &  0,2,4,6     & 3       & 20000\\
12      & 2.2879   &  0,4,9,13    & 4       & 4000\\
24      & 2.5206   &  0,12,25,37  & 8       & 645 \\
32      & 2.7124   &  0,18,36,54  & 9       & 370  \\
\hline
\end{tabular}
\end{center}
\caption[Simulation parameters for the Iwasaki action for the evaluation of $\rnod$.]{\footnotesize{Simulation parameters for the Iwasaki action for the evaluation of $\rnod$}.\label{simpar_iwasaki}}
\end{table}
%%%%%%%%%%%%%%%%%%%%%%%%%%%%%%%%%%%%%%%%%%%%%%%%%%%%%%%%%%%%%%%%%%%%%%%%%%%%
\begin{table}[h]
\begin{center}
\begin{tabular}{c c c c c}
\hline
$N_{s}$ & $\beta$  &   $n_l$    & $N_{or}$ & $N_{meas}$  \\
\hline
10      & 0.75696  &  0,2,4,6     & 3        & 12000  \\ 
12      & 0.8243   &  0,4,9,13    & 4        & 6000\\
16      & 0.9636   &  0,10,20,30  & 8        & 800 \\
24      & 1.04     &  0,18,36,54  & 9        & 220\\
\hline
\end{tabular}
\end{center}
\caption[Simulation parameters for the DBW2 action for the evaluation of $\rnod$.]{\footnotesize{Simulation parameters for the DBW2 action for the evaluation of $\rnod$.}\label{simpar_dbw2}}
\end{table}

%%%%%%%%%%%%%%%%%%%%%%%%%%%%%%%%%%%%%%%%%%%%%%%%%%%%%%%%%%%%%%%%%%%%%%%%%%
\subsection{Results}
Using a local interpolating formula for the force as explained in
sect. \ref{interpo}, we extracted the value of $r_{0}/a$ which are reported in
\tab{r0_iwasaki} and \tab{r0_dbw2}. For the DBW2 action, besides the three
values of $\beta_c$ known in the literature \tab{tab_betacrit}, we decided to
evaluate $\rnod/a$ also for a larger $\beta=1.04$, which should roughly
correspond to $\beta=6$ for the Wilson action and has been used in quenched
simulations \cite{Aoki:2002vt}. Our results for the potential and the force at finite lattice spacing are collected in the tables \ref{tab_iwasaki_results} and \ref{tab_dbw2_results}.

\begin{table}[h]
\begin{center}
\begin{tabular}{c c c}
\hline
$\beta$  &  $\rnod/a$ $(r_{n})$  & $\rnod/a$ $(\rI)$  \\
\hline
2.1551   &   2.311(5)(9)      & 2.320(6)(9)                    \\
2.2879   &   3.026(4)(3)      & 3.026(5)(1)                   \\
2.5206   &   4.535(6)(4)      & 4.511(8)(1)                    \\
2.7124   &   6.020(15)(25)     & 5.999(15)(19)                      \\
\hline
\end{tabular}
\end{center}
 \caption[ Results for $\rnod/a$ evaluated at different
   $\beta=\beta_{c}$ for the Iwasaki action, using the naive definition of the
   force or the tree-level improved.]\footnotesize{ Results for $\rnod/a$ evaluated at different
   $\beta=\beta_{c}$ for the Iwasaki action, using the naive definition of the
   force \protect\eq{r_naiv} or the tree-level improved \protect\eq{e_rI_new}.\label{r0_iwasaki}  }
\end{table}

\begin{table}[h]
\begin{center}
\begin{tabular}{c c c}
\hline 
$\beta$  &  $\rnod/a$ $(r_{n})$  & $\rnod/a$ $(\rI)$  \\
\hline
0.75696   &  2.430(5)(20)             &   2.225(4)(11)                  \\
0.8243    &  3.129(23)(1)             &   3.036(17)(4)                \\
0.9636    &  4.606(13)(17)            &   4.556(17)(20)                    \\
1.04      &  5.500(29)(7)             &   5.452(26)(8)      \\
\hline
\end{tabular}
\end{center}
 \caption[Results for $\rnod/a$ evaluated at different
   $\beta$ for the DBW2 action, using the naive definition of the
   force or the tree-level improved.]{\footnotesize Results for $\rnod/a$ evaluated at different
   $\beta$ for the DBW2 action, using the naive definition of the
   force \protect\eq{r_naiv} or the tree-level improved \protect\eq{e_rI_new}.\label{r0_dbw2}  }
\end{table}

The first error contains the statistical uncertainty summed to the systematic
one due to the interpolation of the force. The second error is the systematic
uncertainty coming from different choices of $t$ in the effective potential
\eq{v_effic}. One can notice that in some cases the last error is the
dominating one.\\
We can notice that for the
DBW2 action the choice of $r_{n}$ (\eq{r_naiv}) or $\rI$ (\eq{e_rI_new}) in the definition of the force
leads to results which can be quite different from each other, above all at small $\rnod/a$; we expected this feature by investigating the force at tree level.
This ambiguity
will make the  discussion of the lattice artefacts difficult, because the
possible conclusions will depend on which definition of the force one has used
and not on intrinsic properties of the action.\\
For the Iwasaki action the two results are not significantly
different.

\subsection{Parametrization of $\rnod/a$}
\begin{figure}
\begin{center}
\includegraphics[width=9cm]{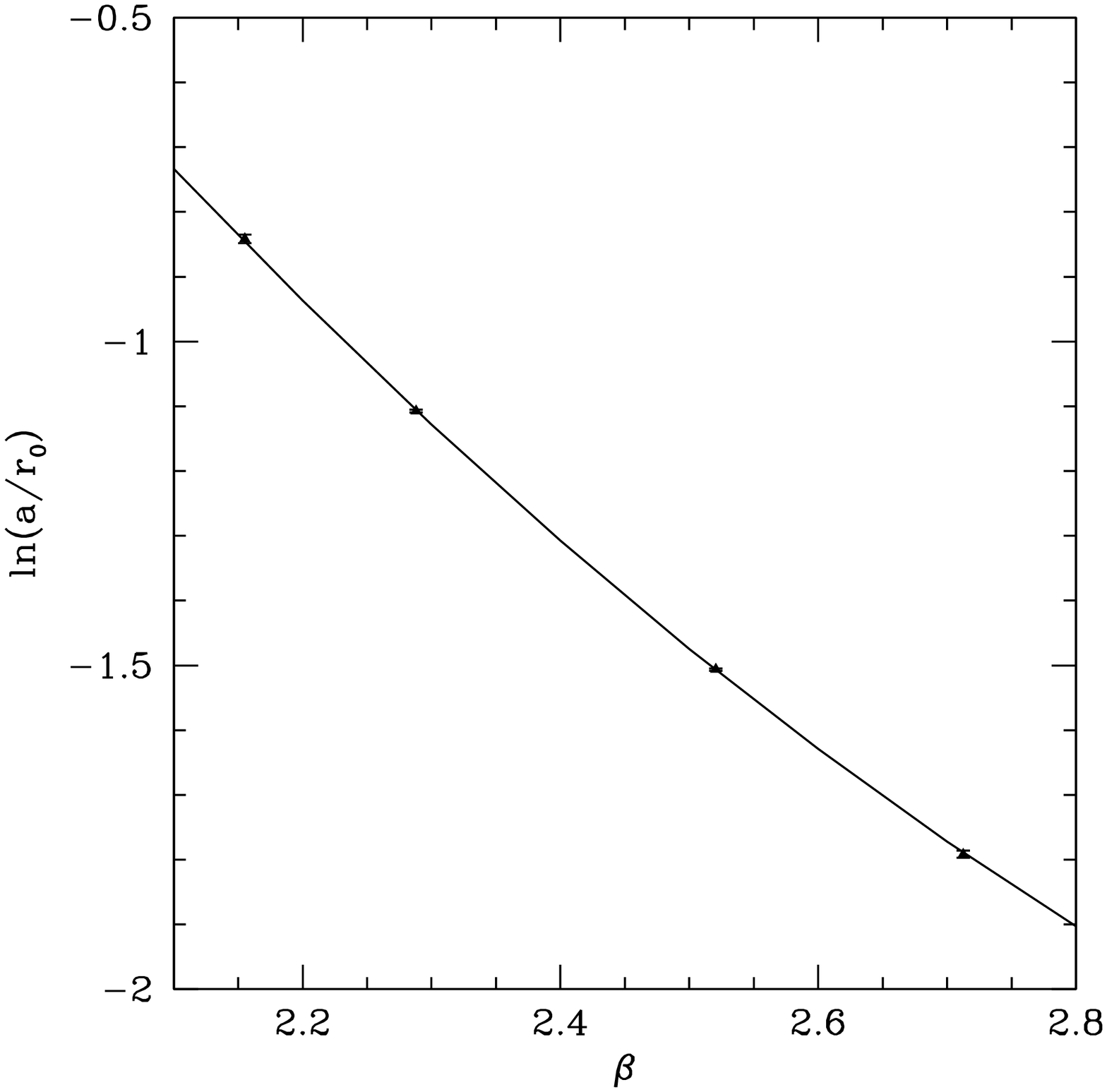}
\end{center}
\caption[Parametrization of $\rnod/a$ for the Iwasaki
    action.]{\footnotesize{Parametrization of $\rnod/a$ for the Iwasaki
    action, \protect\eq{fit_iwasaki}, using the tree-level improved
    definition of the force.}\label{r0_par_iwasaki}}
\end{figure}

Following the strategy of section \ref{rc_para}, one can attempt a
phenomenological parametrization of $\rnod/a$ in the range of couplings that
we considered.\\
For the Iwasaki action the four values of $\rnod$ (obtained by adopting $\rI$)
were fitted in the form
\begin{equation}\label{fit_iwasaki}
\ln(a/\rnod)=c_{1}+c_{2}(\beta-3)+c_3 (\beta-3)^2,
\end{equation}
yielding the numerical results
\begin{equation}
c_{1}= -2.1281   ,\quad c_{2}=-1.0056    ,\quad c_{3} =0.6041        .
\end{equation}
in the range $2.1551\leq\beta \leq 2.7124$.\\
The results and the fit formula are shown in \fig{r0_par_iwasaki}; the
accuracy is about $0.6\%$ at $\beta=2.1551$ and decrease to $0.8\%$ at
$\beta=2.7124$.
%%%%%%%%%%%%%%%%%%%%%%%%%%%%%%%%%%%%%%%%%%%%%%%%%%%%%%%%%%%%%%%%%%%%%%%%%%%%%%%%%
%%%%%%%%%%%%%%%%%%%%%%%%%%%%%%%%%%%%%%%%%%%%%%%%%%%%%%%%%%%%%%%%%%%%%%%%%%%%%%%%%
\begin{figure}
\begin{center}
\includegraphics[width=9cm]{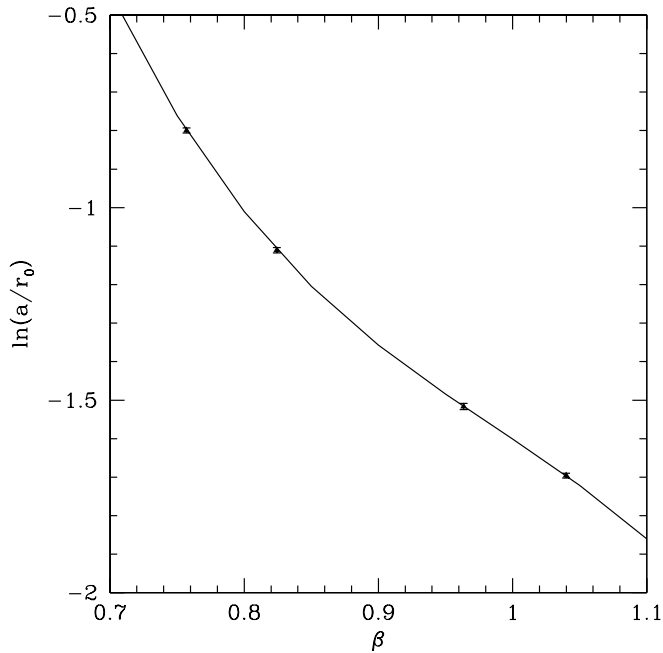}
\end{center}
\caption[Parametrization of $\rnod/a$ for the DBW2
    action.]{\footnotesize{Parametrization of $\rnod/a$ for the DBW2
    action using the tree-level improved definition of the force.}\label{r0_par_dbw2}}
\end{figure}
%%%%%%%%%%%%%%%%%%%%%%%%%%%%%%%%%%%%%%%%%%%%%%%%%%%%%%%%%%%%%%%%%%%%%%%%%%%%%%%%%

For the DBW2 action we have obtained a good accuracy by using a four-parameter
fit 
\begin{equation}\label{fit_dbw2}
\ln(a/\rnod)=d_{1}+d_{2}(\beta-1)+d_{3}(\beta-1)^2 +d_{4}(\beta-1)^3,
\end{equation}
with
\begin{equation}
d_{1}=- 1.6007,\quad d_{2}=-2.3179,\quad d_{3} =-0.8020,\quad d_{4}=-19.8509 ,
\end{equation}
for the range $0.75696\leq\beta\leq 1.04$, where the results always refer to the tree-level improved definition of the
force. The accuracy is between $0.6\%$ at lower $\beta$
and $0.8\%$ at higher $\beta$; it is practically given by the statistical accuracy and  the fit is
plotted in \fig{r0_par_dbw2}.\\
In \cite{Aoki:2002vt} the value $\rnod/a=5.24(3)$ for the DBW2 action
at $\beta=1.04$ is quoted. Our value differs about $3\%$ from that one.
%the reason of the
%discrepancy is not clear and it is maybe due to different methods in the
%evaluation of $\rnod$ and of the force itself.

\subsection{Scaling of $T_{c}\rnod$}
\begin{figure}
\begin{center}
\includegraphics[width=12cm]{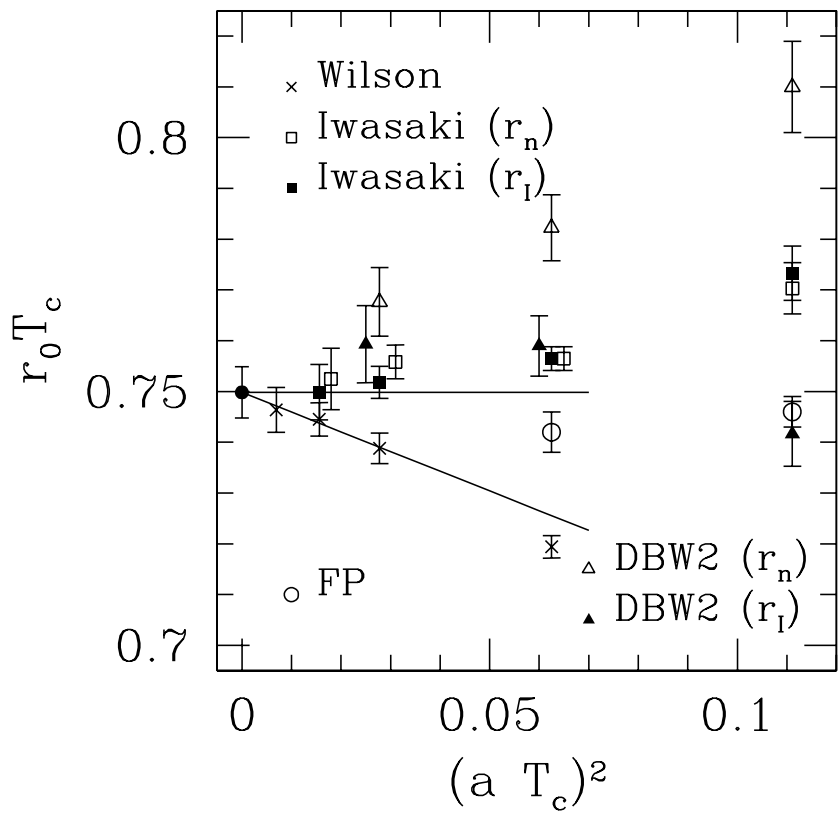}
\end{center}
\caption[$T_{c}r_{0}$ for different actions.]{\footnotesize{$T_{c}r_{0}$ for different actions. The $x$ coordinates were slightly shifted for clarity.}\label{fig_tcr0}}
\end{figure}

Once $\rnod/a$ at the given couplings is known, one can finally consider the
renormalized quantity
\begin{equation}
\rnod T_{c}=\frac{1}{N_{t}}\frac{\rnod}{a}(\beta_c),   
\end{equation}
which has a finite value in the limit $a\rightarrow 0$ (while
$N_t \rightarrow \infty$).\\
The results for Iwasaki and DBW2 actions are given in \tab{tcr0_results},
together with the values obtained using the Wilson action. Also in this case,
we show both the results obtained with the naive and with the tree-level
improved definition of the force. The error in $T_{c}\rnod$ is the quadratic sum of the
error for $\rnod/a$ and the uncertainty in $\beta_c$, which can be translated to an uncertainty in $\rnod$ by using the parametrization formulas
\eq{fit_wilson}, \eq{fit_iwasaki}, \eq{fit_dbw2}, assuming that the coefficients in the
fit are exact.
In our evaluations the error for $\beta_c$ and the uncertainty in $\rnod$ are
roughly of the same order.\\
We expect that the leading lattice artefacts are of order $a^2$, such that the
continuum limit is approached in the following way
\begin{equation}\label{cont_extr_tc}
T_{c}r_{0}=T_{c}r_{0}|_{a=0}+s\cdot (a T_{c})^{2}+ {\rm O} (aT_c)^4.
\end{equation} 
The results for $T_{c}r_{0}$, together with the continuum extrapolation, are shown in \fig{fig_tcr0}.
%%%%%%%%%%%%%%%%%%%%%%%%%%%%%%%%%%%%%%%%%%%%%%%%%%%%%%%%%%%%%%%%%%%%%%%%%%%%%%
%%%%%%%%%%%%%%%%%%%%%%%%%%%%%%%%%%%%%%%%%%%%%%%%%%%%%%%%%%%%%%%%%%%%%%%%%%%%%
\begin{figure}
\begin{center}
\includegraphics[width=9cm]{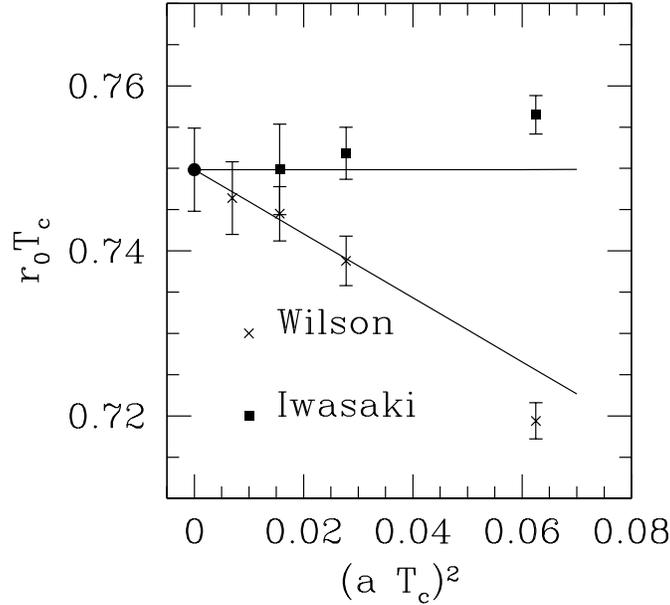}
\end{center}
\caption[Continuum extrapolation of $T_{c}r_{0}$ for the Iwasaki and Wilson action.]{\footnotesize{Continuum extrapolation of $T_{c}r_{0}$ for the Iwasaki and Wilson action, using the constrained fit \protect\eq{cont_extr_tc}.}\label{fig_tcr0_extr}}
\end{figure}

For the Iwasaki action there is
no appreciable difference between the results obtained with
$r_{n}$ and $r_{I}$ and in both cases the data show better scaling properties in
comparison to the Wilson action. Furthermore, the value obtained at
$N_{t} =8$ is in full agreement with the continuum result evaluated
through the Wilson action and hence the universality is confirmed;
this supports the conclusion that the disagreement observed in
$T_{c}/\sqrt{\sigma}$ is indeed due to the difficulty in evaluating the string
tension, particularly at small lattice spacings and it is necessary
to set the scale through a more reliable quantity.\\
Also for the DBW2 action the scaling properties are improved,
although only using $r_{I}$ instead of $r_{naive}$, so that it is more difficult
to make a statement about the lattice artifacts in this case.\\
A constrained fit of the form \eq{cont_extr_tc}
including the points with $N_{t}\geq 6$
for Iwasaki and Wilson actions yields the continuum result (\fig{fig_tcr0_extr})
\begin{equation}
T_{c}r_{0}=0.7498(50).
\end{equation} 
At $N_{t}=6$ the Wilson action shows scaling violations for $r_{0}T_{c}$ of 
about $1.5\%$, while they are $0.3\%$ for the Iwasaki action.\\
For $N_{t}=4$ the discretization errors for the Wilson action increase to
$4\%$, while for the Iwasaki action they remain small ($0.6\%$).\\

In \fig{fig_tcr0} we included also the results obtained with the FP action
\cite{Niedermayer:2000yx}, which also show a good scaling within $1\%$ even on coarse lattices
corresponding to $N_t =3,2$. One has however to mention that for those
lattices the determination of $\rnod/a$ contains large systematic uncertainties.

\begin{table}
\begin{center}
\begin{tabular}{c c c c c c c }
\hline
      &  $T_{c}\rnod$: &  Wilson  & Iwas.($r_{n}$) & Iwas.($\rI$) & DBW2($r_{n}$) & DBW2($\rI$)\\   
$N_t$ &               &          &           &          &               &    \\
\hline
3    &                &         &    0.7703(50)        &  0.7733(53) & 0.8100(90) & 0.7417(90) \\
4    &                & 0.7194(22)  & 0.7565(23)  & 0.7565(23)       & 0.7822(65)  &  0.7590(56) \\ 
6    &                & 0.7388(30)  & 0.7558(33)  & 0.7518(31)       & 0.7676(68) &  0.7593(76)\\
8    &                & 0.7445(33) & 0.7525(60) & 0.7499(55)  &  & \\
12   &                & 0.7464(44) &   &   & & \\ 
\hline
\end{tabular}
\end{center}
\caption[Results for $T_{c}\rnod$.]{\footnotesize{Results for $T_{c}\rnod$.}\label{tcr0_results}}
\end{table}

%%%%%%%%%%%%%%%%%%%%%%%%%%%%%%%%%%%%%%%%%%%%%%%%%%%%%%%%%%%%%%%%%%%%%%%%%%%%%%
\section{Scaling of $\alphaqqbar(\mu)$}
\begin{figure}
\begin{center}
\includegraphics[width=9cm]{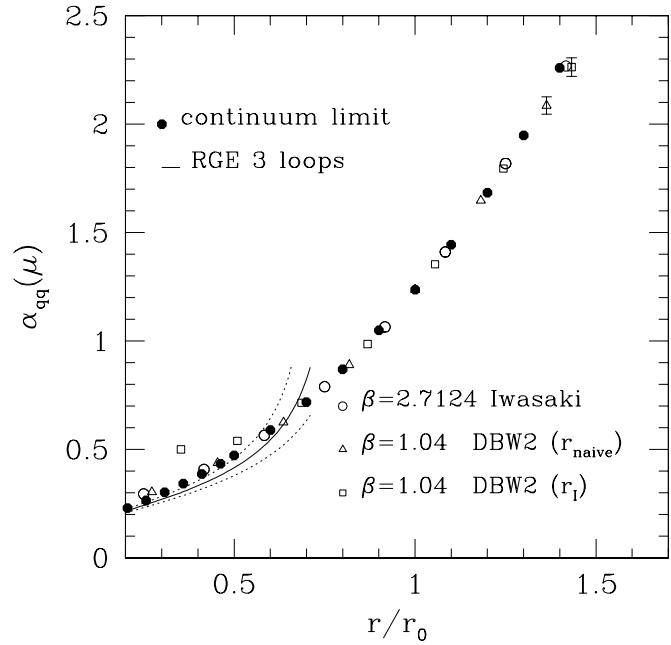}
\end{center}
\caption[$\alpha_{\qqbar}$ at finite lattice spacing for Iwasaki and DBW2 action compared with the continuum result.]{\footnotesize{$\alpha_{\qqbar}$ at finite lattice spacing for Iwasaki and DBW2 action compared with the continuum result. The solid line represents the 3-loop RG perturbative prediction of the running coupling as obtained in sect. \protect\ref{s:pot_pert}; the dashed lines correspond to the uncertainty on the $\Lambda$ parameter in \protect\eq{e_lambda_rnod}.}\label{alphaqq_artifacts}}
\end{figure}

\begin{figure}
\begin{center}
\includegraphics[width=9cm]{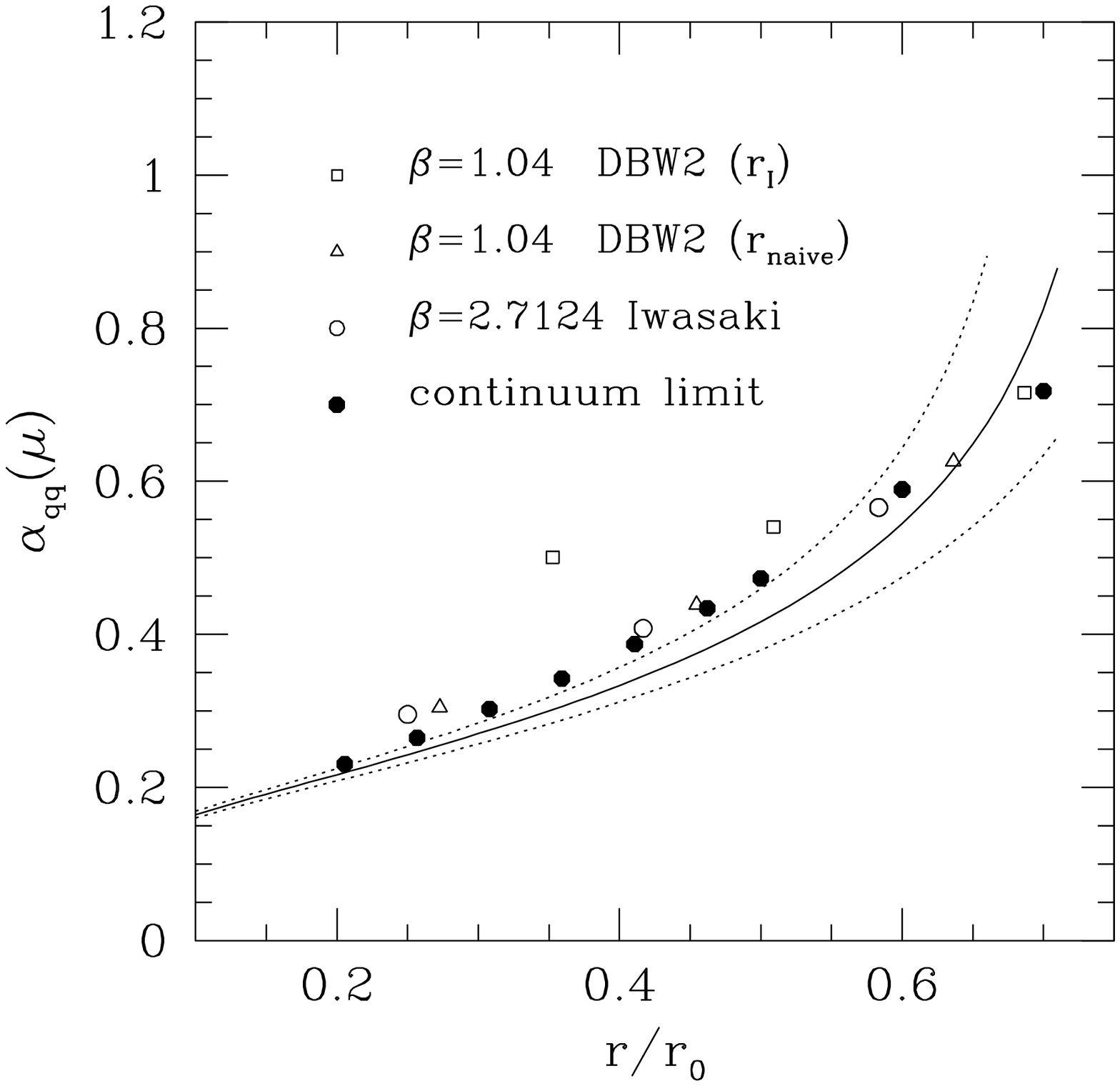}
\end{center}
\caption[$\alpha_{\qqbar}$ at finite lattice spacing for Iwasaki and DBW2 action compared with the continuum result in the short distance region.]{\footnotesize{$\alpha_{\qqbar}$ at finite lattice spacing for Iwasaki and DBW2 action compared with the continuum result in the short distance region.}\label{alphaqq_artifacts_det}}
\end{figure}

Another interesting observable that can be used to test the scaling violations
is the dimensionless coupling $\alphaqqbar$ obtained from the force.
Since in chapter \ref{chapt_potential} we evaluated the force
and the potential very precisely and performed the continuum extrapolation
for those quantities, we can compare these results with our present
determination of the force at finite lattice spacings for the Iwasaki and DBW2
actions at different distances $r$. 
We point out that we only determined the on-axis potential and hence we can not
investigate violations of rotational invariance which would require the
evaluation of off-axis quantities.\\
The coupling $\alphaqqbar$ can be obtained from the force by the simple
relation
\begin{equation}
\alphaqqbar(\mu)=\frac{3}{4}F(r)r^2,\quad \mu=\frac{1}{r}.
\end{equation}
\Fig{alphaqq_artifacts} shows $\alphaqqbar$ in the continuum limit
and the
results obtained with the Iwasaki and DBW2 actions at the largest $\beta$ at our disposal. For the Iwasaki action no appreciable difference in the results obtained with $r_{n}$ and $\rI$ can be seen, while for the DBW2 action the discrepancy becomes large at small distances.\\
At large enough distances one obtains a good scaling in the
coupling, and one does not observe scaling violation within the statistical
errors.
At small $r/\rnod$ one sees deviations from the continuum limit, as one can observe in \fig{alphaqq_artifacts_det}, where only the short distance region is considered. The deviations can be estimated to about $2\%$ for the Iwasaki action at $r\sim 0.4\rnod$; for the DBW2 action they amount to  $4\%$ at $r\sim 0.4\rnod$ if one uses $r_{n}$ to define the force, and reach even $10\%$ by employing $\rI$. This fact shows that the adoption of a tree level improved definition of the force does not guarantee  success in reducing the lattice artifacts, above all for RG actions, which are overcorrected at tree level.

%%%%%%%%%%%%%%%%%%%%%%%%%%%%%%%%%%%%%%%%%%%%%%%%%%%%%%%%%%%%%%%%%%%%%%%%%%%%%
%%%%%%%%%%%%%%%%%%%%%%%%%%%%%%%%%%%%%%%%%%%%%%%%%%%%%%%%%%%%%%%%%%%%%%%%%%%%%

%% file: glueball.tex
\chapter{Scaling properties of RG actions: the glueball 
masses}\label{chapter_glueball}

Osterwalder and Seiler showed in 1978 \cite{Osterwalder:1978pc} that in strongly coupled lattice gauge theory 
the lowest eigenstate of the Hamiltonian above the vacuum has a mass $m$,
which is usually called mass gap. This result can not be obtained in the framework of
perturbation theory: on the classical level the gauge field theory does not
contain any mass term and is scale invariant. Furthermore, the gluon
propagator remains massless to all orders of the perturbative expansion.
This state is hence of purely non-perturbative nature (in the
infinite-volume limit) and states with higher
masses are to be expected too. If the masses scale near the continuum limit
according to
\begin{equation}
am=C_{m}\exp\left(-\frac{1}{2b_{0}g_{0}^2}\right)(b_{0}g_{0}^2)^{-b_{1}/(2b_{0}^2)}\{1+{\rm O}(g_{0}^2)\},
\end{equation}
it means that they are proportional to the $\Lambda$ parameter and hence the
corresponding states survive in the continuum limit and the quantized gauge
theory contains massive physical states.
The existence of these states, called glueballs, has been
predicted in \cite{Fritzsch:2002jv}.\\
Being of non-perturbative nature, Monte Carlo simulations on the lattice
provide a powerful tool to evaluate the glueball masses; they were
indeed among the first quantities that have been computed on the lattice  
\cite{Berg:1980gz}.\\
Since the theoretical discovery of the glueballs, which constitutes one of the most fascinating predictions of the pure gauge sector of QCD, also the experimental search started.
% for example, the $f_{0}(500)$ state has been toutet as a likely
%candidate for a $0^{++}$ glueball \cite{Amsler:1996td}, altough direct evidence is still lacking.
There is indeed an evidence for the existence of exotic glueballs or hybrid
particles consisting of quarks with gluonic excitation. For a review on the light meson spectroscopy see\cite{Godfrey:1998pd}. The difficult task
here is
to distinguish the glueball states from the background of mesons, which have
the same quantum numbers. For this reason the exotic glueballs, e.g $0^{+-}$,
$1^{-+}$, are
particularly interesting because due to their quantum numbers they can
not mix with conventional meson states.\\
Lattice QCD investigations \cite{Bali:1998bj,Bali:1997ec,Bali:2000vr,McNeile:2000xx}, were addressed to study the effects of dynamical quarks and glueball-meson
mixing on the glueball spectrum from lattice QCD.
%but the results are not yet
%$conclusive \cite{}.

Apart from the physical relevance, the mass of the lightest ($0^{++}$) glueball is
particularly interesting since several calculations with the Wilson action
\cite{Bali:1993fb,Michael:1989jr,deForcrand:1985rs,Chen:1994uw} 
showed large lattice artefacts of about $40\%$ at coarse lattice spacings
$a\simeq 0.15 \fm$ and still $20\%$ at $a\simeq 0.10 \fm$. This quantity could
hence in principle provide another stringent test on the scaling behavior of
alternative gauge actions. It is desirable to check the discretization errors
on several physical observables, since they can be quite different depending
on which quantity we are considering.\\ 

As already pointed out since the first works, the calculation of glueball
masses presents a lot of technical difficulties; due to the fact that these
are quite large ($m_{G}\geq 1.6 \GeV$), the signal in the correlation
functions of the gluonic excitations decay fast and disappears in the noise.\\
Several smearing techniques were developed in order to suppress excited
states and to be able to extract the masses at moderately large value of $t$.
It turned out that these techniques are more efficient in the region of small lattice spacings, where one expects for this reason the determination of the glueball masses to be easier.\\

However, as already pointed out, one has to remember that in order to neglect
finite size effects one has to consider a physical lattice extent $L\gtrsim
3\rnod$, so that the simulations at small lattice spacing are very expensive
and it is very difficult to reach a large statistics which is usually required
for those calculations.\\
A possible solution to this problem can be to use anisotropic lattices \cite{Ishikawa:1983xg}, where
the resolution in the time direction is finer that in the space directions
$a_{t}\ll a_{s}$, and hence one can follow the signal over a larger range of
time slices. Morningstar and Peardon \cite{Morningstar:1997ff,Morningstar:1999rf} applied this method to evaluate
the glueball spectrum below $4 \GeV$ in the pure $\SUthree$ gauge theory
and improved the previous determinations of the glueball masses remarkably.
%The other recent calculations of the glueball masses can be founded in \cite{} and
%the results obtained for $m_{0^ {++}}$ and $m_{2^ {++}}$ are is reasonable
%agreement.\\
In this chapter we will present the calculation of $am_{0^{++}}$ and
$am_{2^{++}}$ for RG actions in the range of lattice spacings
$0.11\,\fm\lesssim a\lesssim 0.17\,\fm$, and then make use of the knowledge of
$\rnod/a$ to build the renormalized quantities $\rnod m_{0^{++}}$ and $\rnod m_{2^{++}}$.\\
Our intention here is not to have a more precise determination of the glueball
masses than what can be found in the literature and indeed our total errors on the measurements are quite sizeable; we will focus on the
investigation of discretization effects and on possible further difficulties
that can arise in the extraction of the masses from correlation functions.

\section{Glueball states}
Physical states in the vacuum sector of the Hilbert space of lattice gauge
theory can be obtained by applying gauge invariant operators to the vacuum.
As particular candidates one can take space-like Wilson loops lying in the
$x_{0}=0$ timeslice in the
fundamental representation \eq{wilson_loop}, considered as multiplicative
operators on wave functions. The simplest choice is the plaquette
\eq{plaquette} 
\begin{equation}
W^{1\times 1}_{ij}({\vec x}),
\end{equation}
where ${\vec x}=(x_1 ,x_2, x_3)$, $i,j \in \{1,2,3\}$.\\
Applied to the vacuum the operator yields a state
\begin{equation}\label{wilson_state}
|\tilde{\Psi}_{ij}({\vec x})\rangle=\left\{W^{1\times 1}_{ij}({\vec x})
-\langle 0 |{W^{1\times 1}_{ij}}({\vec x})|0\rangle\right\}|0\rangle,
\end{equation}
where the projection onto the vacuum itself has been subtracted.\\
Eigenstates of the momentum operator are formed by Fourier transformation 
\begin{equation}
|\Psi_{ij}({\vec k})\rangle =L^{-3/2}\sum_{{\vec x}}e^{i {\vec x}\cdot{\vec k}}|\tilde{\Psi}_{ij}({\vec x})\rangle. 
\end{equation}
If one is interested in the determination of masses it is sufficient to
consider zero momentum states, denoted by
\begin{equation}\label{wilson_state2}
|\Psi_{ij}\rangle\equiv|\Psi_{ij}({\vec 0})\rangle.
\end{equation}
We assume that the continuum limit of $\SUthree$ lattice gauge theory exists
and their rotational invariance is restored: the spin of a state is then
characterized by the unitary irreducible representations of $SO(3)$, for
bosonic states.\\
On a cubic lattice the rotation symmetry is broken down to the cubic group $O$; the eigenstates of the Hamiltonian have hence
to be classified according to the unitary irreducible representations of
$O$.\\
For the first application of the representation theory of the cubic group
to lattice gauge theory one can refer to the paper by Berg and Billoire \cite{billoire}.\\
The cubic group contains 24 elements, corresponding to the permutations of the
four space diagonals of a cube. There are five irreducible representations,
denoted by $A_1,A_2,E,T_1$ and $T_2$, which have dimensions 1,1,2,3 and 3.
Since the cubic group is a subgroup of $SO(3)$, any representation $D_{J}$
of $SO(3)$ for a state of spin $J$ will induce a representation on the group
$O$, the so-called subduced representation $D_{J}^{O}$. It will in general no
longer be irreducible and can be decomposed into the irreducible
representations of $O$. Up to $J=4$ one finds
\begin{eqnarray}
D_{0}^{O} & =& A_{1}\\
D_{1}^{O} & =& T_{1}\\
D_{2}^{O} & =& E\oplus T_{2}\\
D_{3}^{O} & =& A_{2}\oplus T_{1}\oplus T_{2}\\
D_{4}^{O} & =& A_{1}\oplus E\oplus T_{1}\oplus T_{2}.
\end{eqnarray}
For example, a spin 2 particle is described in the continuum by a quintuplet
of degenerate states; on the lattice the quintuplet is split into a doublet
$E$ and a triplet $T_{2}$. At finite lattice spacing one expects a mass
splitting between the two representations, so that $m_{E}\neq m_{T_2}$. By
approaching the continuum limit the ratio $m_{E}/m_{T_2}$ will converge to one
in order to restore the full Euclidean rotational symmetry. For the Wilson
action we expect that the leading discretization corrections in mass ratios
are of order $a^2$; for other lattice formulations this is no more obvious and
in fact with our data we will not be able to give conclusions on the form of
the lattice artefacts in the specific range of lattice spacings that we have considered.\\
Given a lattice operator which transforms according to some irreducible
representation $R$ of the cubic group, by applying it to the vacuum one
obtains a state 
\begin{equation}\label{state_rep}
|\Psi_{R}\rangle =\sum_{\alpha}c_{\alpha}^{R}|\Psi_{\alpha}\rangle,
\end{equation}
where $|\Psi_{\alpha}\rangle$ are eigenstates of the Hamiltonian. In the
continuum limit any $|\Psi_{\alpha}\rangle$ belongs to some spin $J$ multiplet,
so that $|\Psi_{R}\rangle$ contains various spins $J$. A spin $J$ can occur in
this superposition only if $R$ is contained in $D_{J}^{O}$.\\
At this point the important assumption is that the lowest spin contained in
$R$ will belong to the lowest mass and will therefore dominate the correlation
functions. The representations of the cubic group, their dimensions and
the lowest spin content are summarized in \tab{rep_tab}.\\
In addition to the symmetry associated with the cubic group, there are two
discrete symmetries: the total space reflection and the charge conjugation.
The eigenvalues of the space reflections correspond to the parity $P=\pm 1$; the cubic group combined with space reflection forms the group
$O_{h}=O\times\mathbb{Z}_{2}$, which contains 48 elements.\\
The charge conjugation corresponds for the Wilson loops to the complex conjugation,
and its eigenvalues are $C=\pm 1$. The states belonging to an irreducible
representation of the lattice symmetry are then labeled by 
\begin{equation}
|\Psi\rangle=|J^{PC}\rangle.
\end{equation}

%%%%%%%%%%%%%%%%%%%%%%%%%%%%%%%%%%%%%%%%%%%%%%%%%%%%%%%%%%%%%%%%%%%%%%%%%%%%%
%%%%%%%%%%%%%%%%%%%%%%%%%%%%%%%%%%%%%%%%%%%%%%%%%%%%%%%%%%%%%%%%%%%%%%%%%%%%%
%%%%%%%%%%%%%%%%%%%%%%%%%%%%%%%%%%%%%%%%%%%%%%%%%%%%%%%%%%%%%%%%%%%%%%%%%%%%%

\section{Irreducible representations of the cubic group on Wilson loops}
\begin{figure}
\begin{center}
\includegraphics[width=10cm]{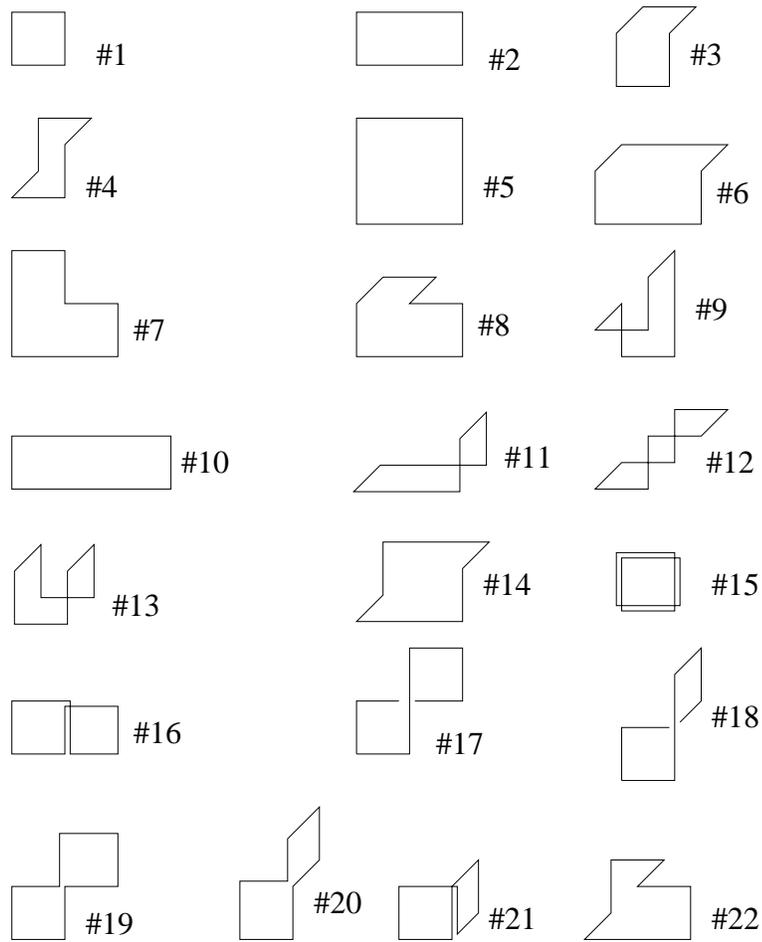}
\end{center}
\caption[Wilson loops un to length 8.]{\footnotesize{Wilson loops un to length 8.}\label{f_shapes}}
\end{figure}

Spatial Wilson loops, being gauge invariant operators acting on the vacuum, create
physical states like \eq{wilson_state2}. 
Here we will consider the space-like Wilson loops up to length 8, for which
all irreducible representations of $O_{h}$ have been constructed
 \cite{billoire}. The loops are shown in \fig{f_shapes}.\\
The first simple rule that one can easily obtain is that real parts of the
Wilson loops have $C$-parity $C=+1$, imaginary parts have $C=-1$. Every 
Wilson loop corresponding to a path
of length $L$ can be represented by an $L$-tuple
\begin{equation}
(\hat{f}_{1},...,\hat{f}_{L}),\quad\textrm{with}\quad \sum_{i=1}^{L}\hat{f}_{i}=0,
\end{equation}
where $\hat{f}_{i}$ are unit vectors corresponding to the space-like
coordinates in positive and negative directions ($\pm\hat{1},\pm\hat{2},\pm\hat{3}$). The equivalence class corresponding to
$(\hat{f}_{1},...,\hat{f}_{L})$ will be denoted by
$[(\hat{f}_{1},...,\hat{f}_{L})]$ and will contain $L$-tuples which are
equivalent up to cyclic permutations.\\
Under $C$-parity one has the transformation
\begin{equation}
C[(\hat{f}_{1},...,\hat{f}_{L})]=[(-\hat{f}_{L},...,-\hat{f}_{1})],
\end{equation}
so that the combinations 
\begin{equation}
[(\hat{f}_{1},...,\hat{f}_{L})]_{\pm}=[(\hat{f}_{1},...,\hat{f}_{L})]\pm[(-\hat{f}_{L},...,-\hat{f}_{1})]
\end{equation}
are even and odd under $C$-parity respectively.\\
On operators of fixed shape we generate a representation $\mathcal{M}$ of
$O_{h}$ by means of
\begin{equation}
\mathcal{M}_{g}[(\hat{f}_{1},...,\hat{f}_{L})]_{\pm}\doteq[(\mathcal{M}_{g}\hat{f}_{1},...,\mathcal{M}_{g}\hat{f}_{L})]_{\pm},\quad\forall
g\in O_{h},
\end{equation}
where $\mathcal{M}_{g}$ is the matrix corresponding to the group element $g$ in the fundamental
representation. Excluding the $C$-parity, the dimension $d$ of the generated
representation will be less or equal than 48, which is the order of the group
$O_{h}$ and which will correspond to the number of different spatial
orientations of the given shape.
In the second column of \tab{tab_irred} the dimensions $d$ of the generated representations on every loop shape are listed.\\
The irreducible contents of the representation $\mathcal{M}$ can be determined
by means of the so-called character relation. The characters are a set of
quantities which are the same for all equivalent representations. For finite
groups they uniquely determine the representations up to equivalence, providing
in particular a complete specification of the irreducible representations
contained in a given representation. In general we will have
\begin{equation}
\mathcal{M}=n_{1}\Gamma^1 \oplus +n_{2}\Gamma^2 \oplus...,
\end{equation}
where $\Gamma^p$ are the irreducible representations listed in \tab{rep_tab} and
$n_{p}$ is the multiplicity of $\Gamma^p$ in $\mathcal{M}$.\\
For fixed $C$-parity one finds
\begin{equation}
n_{p}=\frac{1}{48}\sum_{\mathcal{C}}n_{\mathcal{C}}\chi(\mathcal{C})\chi^{p}(\mathcal{C}),
\end{equation}
where the sum goes over all classes of conjugate elements, $n_{\mathcal{C}}$
is the number of elements in the class $\mathcal{C}$ and $\chi(\mathcal{C})$,
$\chi^{p}(\mathcal{C})$ are the characters of $\mathcal{M}$, $\Gamma^p$
respectively; for matrix representations the character is given by the trace
of the corresponding matrix and is the same for all elements in a given
conjugacy class. The results for the $C$-parity plus representations are given in \tab{tab_irred}.
%%%%%%%%%%%%%%%%%%%%%%%%%%%%%%%%%%%%%%%%%%%%%%%%%%%%%%%%%%%%%%%%%%%%%%%%%%%%%
\begin{table}
\begin{center}
\begin{tabular}{c c c}
\hline
R     & Dimension     & Lowest spin $J$ \\
\hline
$A_{1}$ & 1             & 0 \\ 
$A_{2}$ & 1             & 3  \\
$E$     & 2             & 2  \\
$T_{1}$ & 3             & 1  \\
$T_{2}$ & 3             & 2  \\
\hline
\end{tabular}
\end{center}
\caption[Irreducible representations of the cubic group $O$.]{\footnotesize{Irreducible representations of the cubic group $O$.}\label{rep_tab}}
\end{table}
%%%%%%%%%%%%%%%%%%%%%%%%%%%%%%%%%%%%%%%%%%%%%%%%%%%%%%%%%%%%%%%%%%%%%%%%%%%%%

%%%%%%%%%%%%%%%%%%%%%%%%%%%%%%%%%%%%%%%%%%%%%%%%%%%%%%%%%%%%%%%%%%%%%%%%%%%%%
\section{Wave functions of glueball operators}
In order to perform a Monte Carlo calculation of the glueball masses, for each irreducible representation one has to calculate an orthonormal basis explicitly. This procedure involves the character projection operator, which is defined in general by
\begin{equation}\label{char_projector}
\mathcal{P}^{p}=\frac{d_{p}}{g}\sum_{T\in\mathcal{G}}\chi^{p}(T)^{\star}P(T),
\end{equation}
where $p$ labels the irreducible representations $\Gamma^{p}$ of dimension $d_{p}$ of a finite group of coordinate transformations $\mathcal{G}$ of order $g$, $\chi^{p}(T)$ being the character of $T\in\mathcal{G}$ in $\Gamma^{p}$ and $P(T)$ the unitary operator in the Hilbert space of the coordinate transformations $T\in\mathcal{G}$.\\
$\mathcal{P}^{p}$ projects out of a function $\phi$ of the Hilbert space the sum of all the parts transforming under $\Gamma^{p}$. Having chosen a function $\phi$ such that $\mathcal{P}^{p}\phi$ does not vanish identically, one constructs $P(T)(\mathcal{P}^{p}\phi)$ for each $T\in\mathcal{G}$. Each of these are linear combinations of the $d_{p}$ basis functions of $\Gamma^{p}$. From these functions one extracts $d_{p}$ linearly independent functions and applies the Gram-Schmidt orthogonalization to obtain a set of orthonormal functions of the irreducible representation.\\
What we practically did is to construct from a loop prototype for each shape by respecting the following rule: 
one builds the path first going in 1-, then on 2- and finally, if necessary, in 3-direction. From this reference orientation all others can be generated by applying the group elements of $O_{h}$ in a given order, taking care not to generate orientations that are equivalent up to translations.
Once all the independent orientations were created, we formed the orthonormal
basis systems for each irreducible representation by taking the related
coefficients from the literature \cite{billoire,wenger}.\\
As example we can consider
the single plaquette operator (loop shape $\#1$), for which the orthogonal
wave functions are listed in \tab{shape1_ort}, with the following convention.\\
The numbers in each row denote the contribution of the specific orientation to the wave function in question; the three positive orientations of the single plaquette can be labeled as $O_{12}$, $O_{13}$ and $O_{23}$. The wave function associated to the irreducible representation $A_{1}^{++}$ will be constructed by forming 
\begin{equation}
O^{A_{1}^{++}}=O_{12}+O_{13}+O_{23},
\end{equation}
while the $E^{++}$ functions are
\begin{eqnarray}
O_{1}^{E^{++}} & = & 2O_{12}-O_{13}-O_{23}\\
O_{2}^{E^{++}} & = & O_{13}-O_{23}.
\end{eqnarray}
\newpage

%%%%%%%%%%%%%%%%%%%%%%%%%%%%%%%%%%%%%%%%%%%%%%%%%%%%%%%%%%%%%%%%%%%%%%%%%%%%%%%%%%
\begin{table}
\begin{center}
\begin{tabular}{c c c c c c c c c c c c}
\hline
shape & $d$  &  $A_{1}^{++}$ & $A_{2}^{++}$ & $E^{++}$ &  $T_{1}^{++}$ & $T_{2}^{++}$ & $A_{1}^{+-}$  & $A_{2}^{+-}$ & $E^{+-}$ & $T_{1}^{+-}$ & $T_{2}^{+-}$\\ [1ex]
\hline
$\# 1$ & 3 & 1 &        0  &   1 & 0 &   0 & 0 & 0 & 0 & 0 & 0  \\
$\# 2$ & 6 & 1 &        1  &   2 & 0 &   0 & 0 & 0 & 0 & 0 & 0  \\
$\# 3$ & 12 & 1  &      0  &   1 & 0 &   1 & 0 & 0 & 0 & 1 & 1  \\
$\# 4$ & 4 & 1 &        0  &   0 & 0 &   1 & 0 & 0 & 0 & 0 & 0  \\
$\# 5$ & 3 & 1 &        0  &   1 & 0 &   0 & 0 & 0 & 0 & 0 & 0  \\
$\# 6$ & 12 & 1 &       0  &   1 & 0 &   1 & 0 & 0 & 0 & 1 & 1  \\
$\# 7$ & 12 & 1 &       0  &   1 & 0 &   1 & 0 & 0 & 0 & 1 & 1  \\
$\# 8$ & 48 & 1 &       1  &   2 & 3 &   3 & 1 & 1 & 2 & 3 & 3  \\
$\# 9$ & 24 & 1 &       0  &   1 & 1 &   2 & 1 & 0 & 1 & 1 & 2  \\
$\# 10$ &  6 & 1 &      1  &   2 & 0 &   0 & 0 & 0 & 0 & 0 & 0  \\
$\# 11$ &  24 & 1 &     1  &   2 & 1 &   1 & 0 & 0 & 0 & 2 & 2  \\
$\# 12$ & 12 & 1 &      1  &   2 & 1 &   1 & 0 & 0 & 0 & 0 & 0  \\
$\# 13$ & 6  & 1 &      0  &   1 & 0 &   0 & 0 & 0 & 0 & 0 & 1  \\
$\# 14$ & 12  & 1 &     0  &   1 & 1 &   2 & 0 & 0 & 0 & 0 & 0  \\
$\# 15$ & 3 & 1 &       0  &   1 & 0 &   0 & 0 & 0 & 0 & 0 & 0  \\
$\# 16$ & 6 & 1 &       1  &   2 & 0 &   0 & 0 & 0 & 0 & 0 & 0  \\
$\# 17$ & 6 & 1 &       0  &   1 & 0 &   1 & 0 & 0 & 0 & 0 & 0  \\
$\# 18$ & 24 & 1 &      0  &   1 & 1 &   2 & 1 & 0 & 1 & 1 & 2  \\
$\# 19$ & 6 & 1 &       0  &   1 & 0 &   1 & 0 & 0 & 0 & 0 & 0  \\
$\# 20$ & 24 & 1 &      0  &   1 & 1 &   2 & 1 & 0 & 1 & 1 & 2  \\
$\# 21$ &  12 & 1 &     0  &   1 & 0 &   1 & 0 & 0 & 0 & 1 & 1  \\
$\# 22$ &  48 & 1 &     1  &   2 & 3 &   3 & 1 & 1 & 2 & 3 & 3  \\[1ex]
\hline
\end{tabular}
\end{center}
\caption[Irreducible contents of the representations of the symmetry group of the cube on Wilson loops up to length 8, for $C=+1$.]{\footnotesize{Irreducible contents of the representations of the symmetry group of the cube on Wilson loops up to length 8, for $C=+1$. $d$ is the dimension of the representation of $O_{h}$ on the different loop shapes.}\label{tab_irred}}
\end{table}
%%%%%%%%%%%%%%%%%%%%%%%%%%%%%%%%%%%%%%%%%%%%%%%%%%%%%%%%%%%%%%%%%%%%%%%%%%%%%%%%%
%%%%%%%%%%%%%%%%%%%%%%%%%%%%%%%%%%%%%%%%%%%%%%%%%%%%%%%%%%%%%%%%%%%%%%%%%%%%%%%%%

%%%%%%%%%%%%%%%%%%%%%%%%%%%%%%%%%%%%%%%%%%%%%%%%%%%%%%%%%%%%%%%%%%%%%%%%%%%%%%%%%%
\begin{table}
\begin{center}
loop shape $\#1$\\
\vspace{0.2cm}
\begin{tabular}{ c| c c c}
\hline\\
$A_{1}^{++}$  &  1  &  1  &  1  \\[1ex]
\hline\\
$E^{++}$      &  2   & -1    & -1    \\[1ex]
              &  0   &  1    & -1 \\
\hline\\
$T_{1}^{+-}$  &  0   &  0    &  1 \\[1ex]
              &  0   &  1    &  0\\
              &  1   &  0    &  0\\
\hline
\end{tabular}
\end{center}
\caption[Orthogonal wave functions of the irreducible operators that can be built from the plaquette.]{\footnotesize{Orthogonal wave functions of the irreducible operators that can be built from the plaquette.}\label{shape1_ort}}
\end{table}
%%%%%%%%%%%%%%%%%%%%%%%%%%%%%%%%%%%%%%%%%%%%%%%%%%%%%%%%%%%%%%%%%%%%%%%%%%%%%%%%%%%%%%%%%%%%%

\section{Simulation details}
For the numerical simulation, we followed essentially the method proposed by
\cite{Niedermayer:2000yx}. We performed a test simulation using the standard Wilson action for
$\beta=6.0$ on a $16^4$ lattice, for which results can be found in the
literature.\\
We decided to concentrate on $\mzpp$ and $\mtpp$ by measuring the masses in
the representations $A_{1}^{++}$, $E^{++}$ and $T_{2}^{++}$. The $\mzpp$ is
particularly interesting for our purpose to investigate the lattice artefacts
on RG actions since these turn out to be quite sizeable for the Wilson
action.\\
For our test simulation we measured all 22 loop shapes up to length 8 and
formed the wave functions corresponding to $A_{1}^{++}$, $E^{++}$ and
$T_{2}^{++}$ as explained in the previous sections. Our timeslice observable
at time $t$ is
\begin{equation}
S_{n}^{R}(t)=\frac{L^{-3/2}}{K}\sum_{{\vec x}}\sum_{i=1}^{d_{n}}c_{n}^{i
  R}\Re W_{n}^{i}({\vec x},t),\quad n=1,...,22,
\end{equation}
where the coefficient $c_{n}^{iR}$ are taken from the literature \cite{billoire,wenger}, the sum over $i$ indicates the
sum over all $d_{n}$ orientations of a given shape $n$ and $K$ is a suitable normalization constant.\\
As for the calculation of the potential, a smearing procedure is necessary in
order to improve the overlap with the ground state. We applied the smearing
operator \eq{smearing} to every space-like link; for $\beta=6.0$ we found that 
$M=4$, $n_{l}=(5,10,15,20)$ was a reasonably good choice for the smearing levels.\\
Then we build the correlation matrices \footnote{Notice that the vacuum
  subtraction is required only in the $A_{1}^{++}$ channel, since it has the
  same quantum numbers as the vacuum.}
\begin{equation}\label{corrma_glueball}
C_{kl}^{R}(t)=\langle S_{k}^{R}(t)S_{l}^{R}(0)\rangle_{c}=\langle
S_{k}^{R}(t)S_{l}^{R}(0)\rangle-\langle S_{k}^{R}(t)\rangle\langle S_{l}^{R}(0)\rangle,
\end{equation}
for $R=A_{1}^{++},E^{++},T_{2}^{++}$. The indices $k,l$ assume $(22\times M)$
values, where $M$ is the number of smearing levels.\\
We expect that 
\begin{equation}
\langle S_{k}^{R}(t)S_{l}^{R}(0)\rangle_{c}=\langle\Psi_{k}^{R}|e^{-Ht}|\Psi_{l}^{R}\rangle=
\sum_{\alpha}\langle\Psi_{k}^{R}|\Psi_{\alpha}\rangle\langle\Psi_{\alpha}|\Psi_{l}^{R} \rangle e^{-m_{\alpha}(R)t},
\end{equation}
where $|\Psi_{\alpha}\rangle$ are eigenstates of the Hamiltonian.
At large $t$ the lowest mass $m_{0}(R)$ dominates; 
it belongs to a glueball state which in the continuum limit will have 
the lowest spin contained in the representation $R$.\\

Starting from our large basis, we analyzed the signal/noise ratio of
the different operators in order to eliminate those which introduce large
noise and to reduce the correlation matrix to a set of well
measurable operators.
It is important not to include noisy operators in the analysis performed
through the variational method, because they could introduce numerical instabilities in the
generalized eigenvalue problem.\\
We considered the effective masses evaluated from the diagonal elements of the
correlation matrix \eq{corrma_glueball}
\begin{equation}
m_{eff}^{lR}=-\log\left(\frac{C_{ll}^{R}(t)}{C_{ll}^{R}(t-1)}\right)
\end{equation} 
and took the relative error as quality of the corresponding operator.\\
We found general agreement in the classification of the "bad" 
operators with \cite{wenger}. Our final choice for the operators to measure was
then
\begin{eqnarray}
A_{1}^{++}  & : & \#2,\#5,\#7,\#8,\#10,\#12,\#14   \\
E^{++}      & : & \#2,\#5,\#7,\#10,\#12,\#14 \\
T_{2}^{++}  & : & \#7,\#12,\#14,
\end{eqnarray}
where for $E^{++}$ and $T_{2}^{++}$ channels we took respectively 2 and 3
orthogonal projections for each shape. For each operator we then considered all
$M$ smearing levels. \\
One has to keep in mind that our choice contains a certain degree of arbitrariness, since the behavior of the different operators may
depend on the considered lattice spacing. We did not perform a systematic
study and we assumed that this choice is reasonable for each value of the
coupling and also for the Iwasaki and DBW2 actions.\\
The simulation parameters for our new Monte Carlo simulations are listed in
\tab{sim_par_glue}. Measurements were taken after a number of sweeps between three and five.\\

\section{Analysis details}
\begin{figure} 
\begin{center}
\includegraphics[width=7cm]{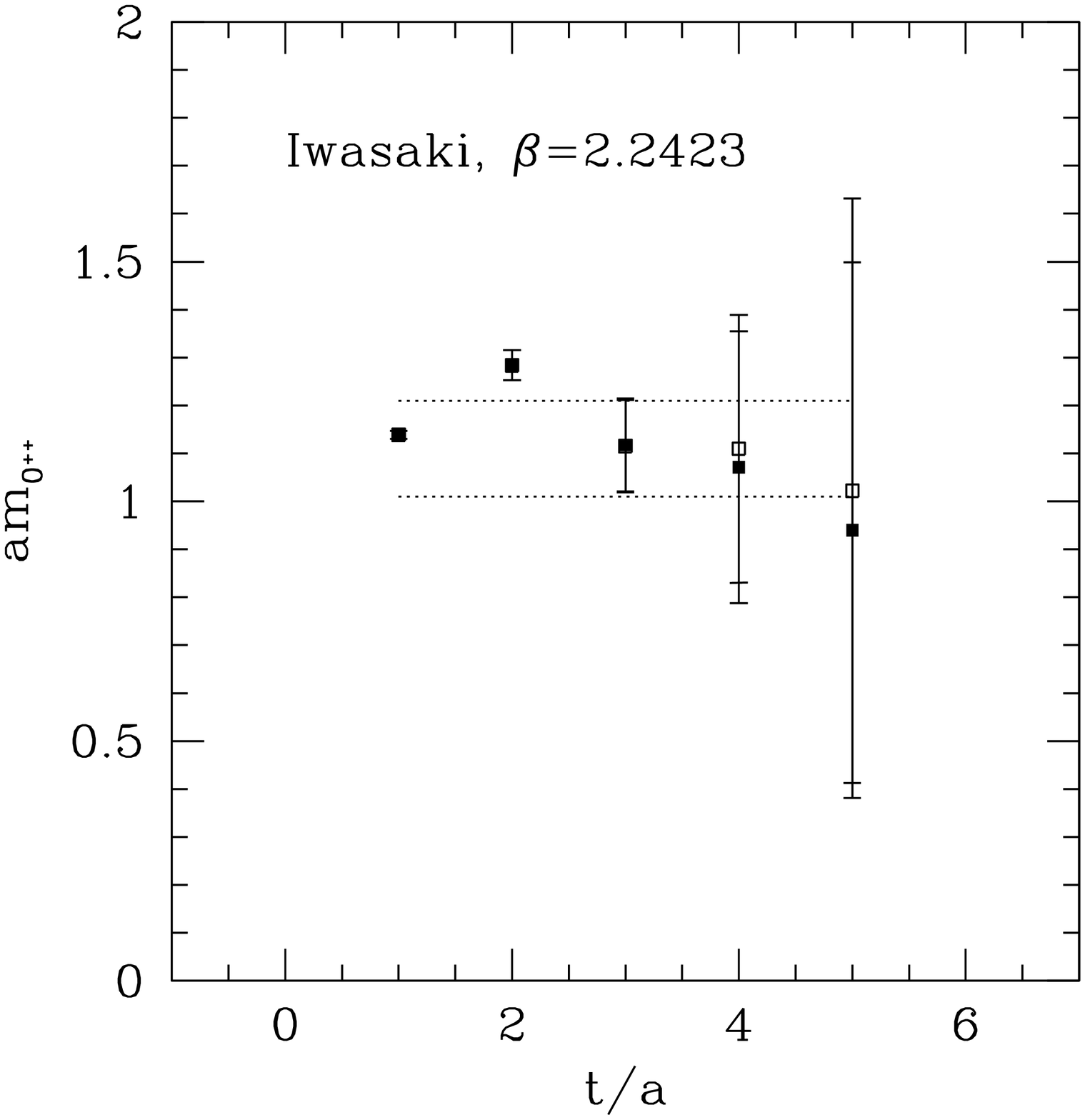} 
\includegraphics[width=7cm]{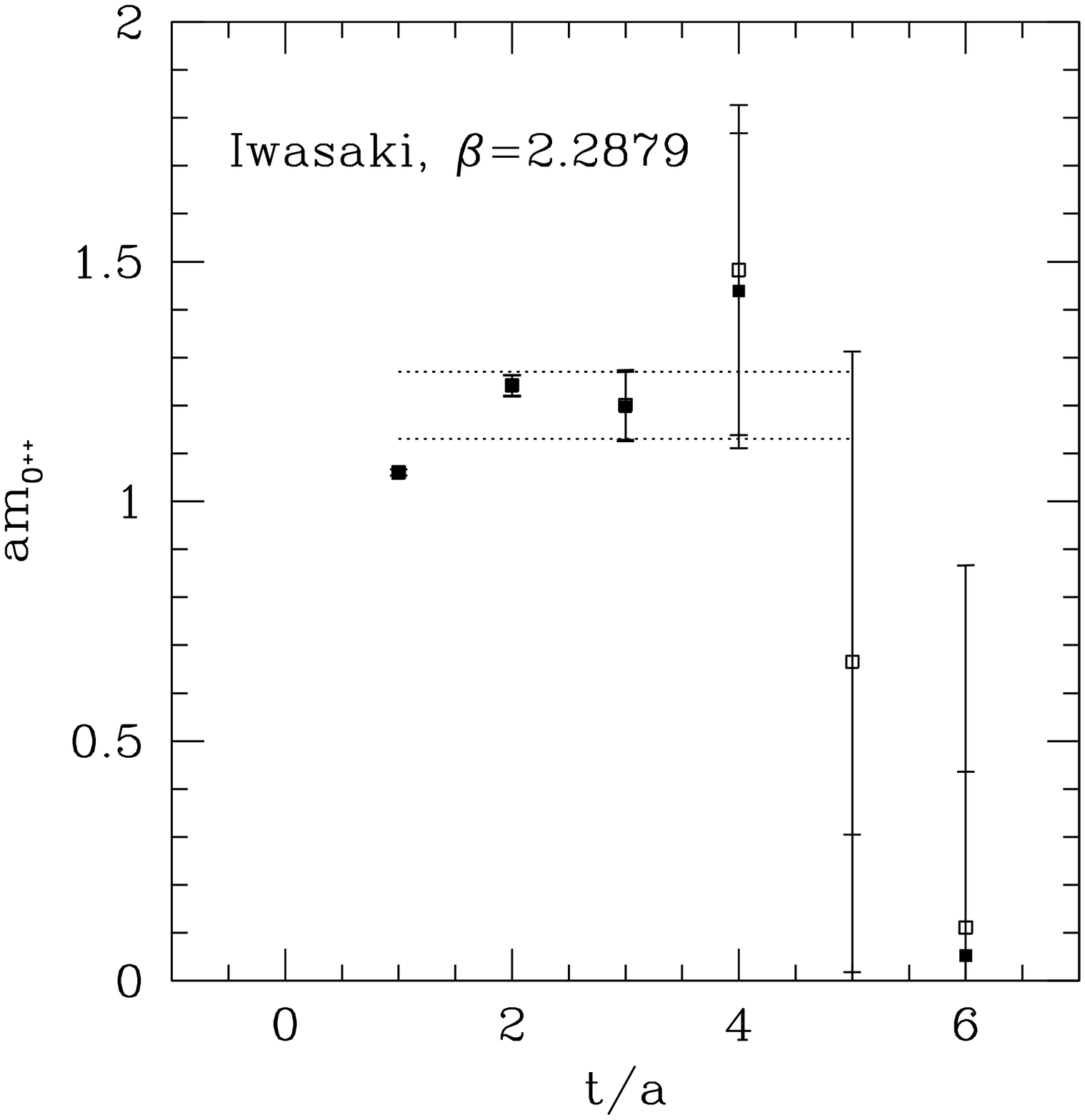} 
\includegraphics[width=7cm]{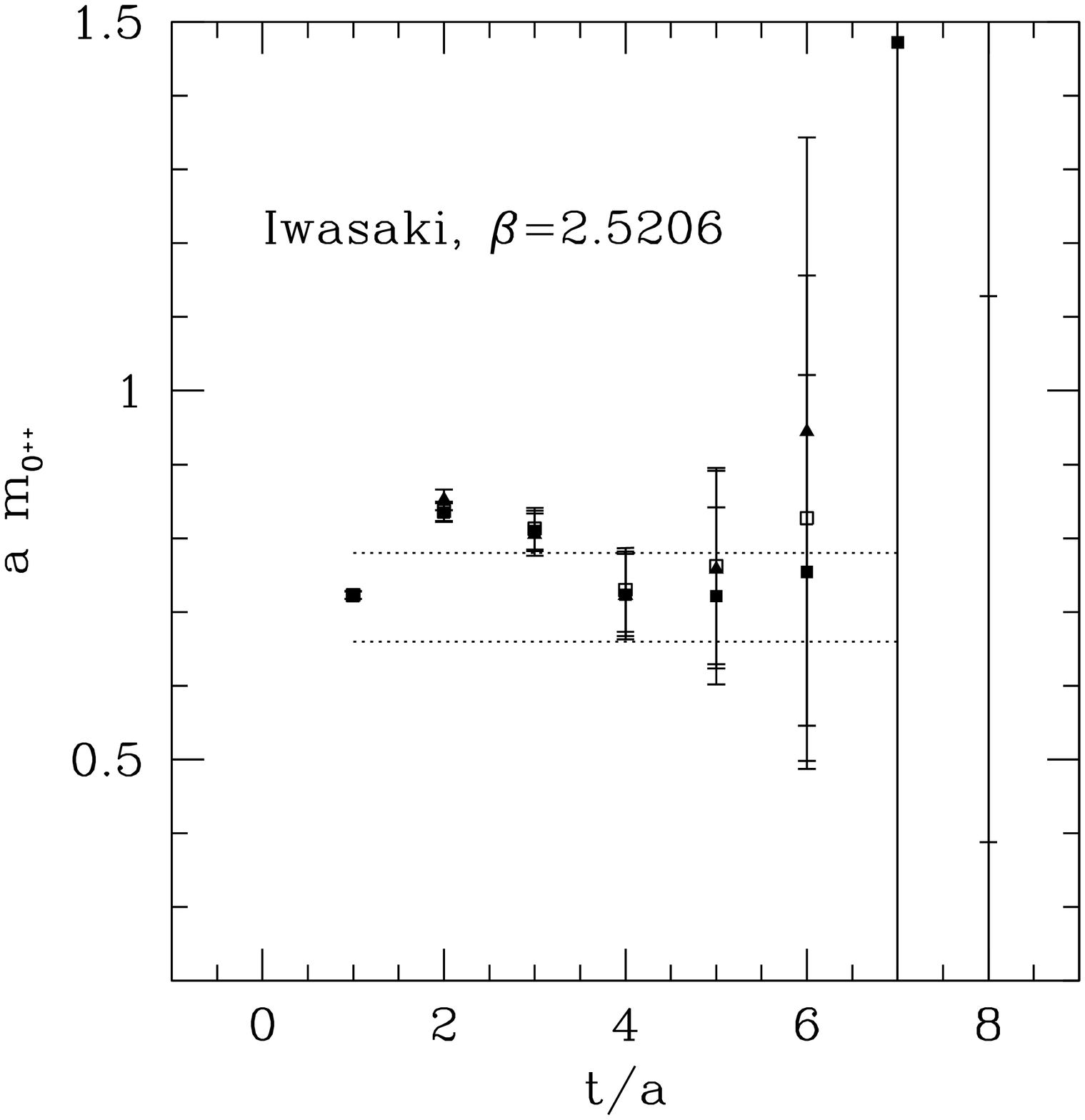} 
\end{center}
\caption[The effective masses for the $A_{1}^{++}$ channel,
evaluated with Iwasaki action, at different lattice spacings.]{\footnotesize{The effective masses for the $A_{1}^{++}$ channel,
evaluated with Iwasaki action, at different lattice spacings. The filled and empty squares corresponds to respectively to \protect\eq{meff_eigv} and \protect\eq{meff_corrma} with $t_{0}=0$. For $\beta=2.5206$, the filled triangles correspond to \protect\eq{meff_eigv} with $t_{0}=a$.}\label{fig_meff_iwasaki}}
\end{figure}

\begin{figure}
\begin{center}
\includegraphics[width=7cm]{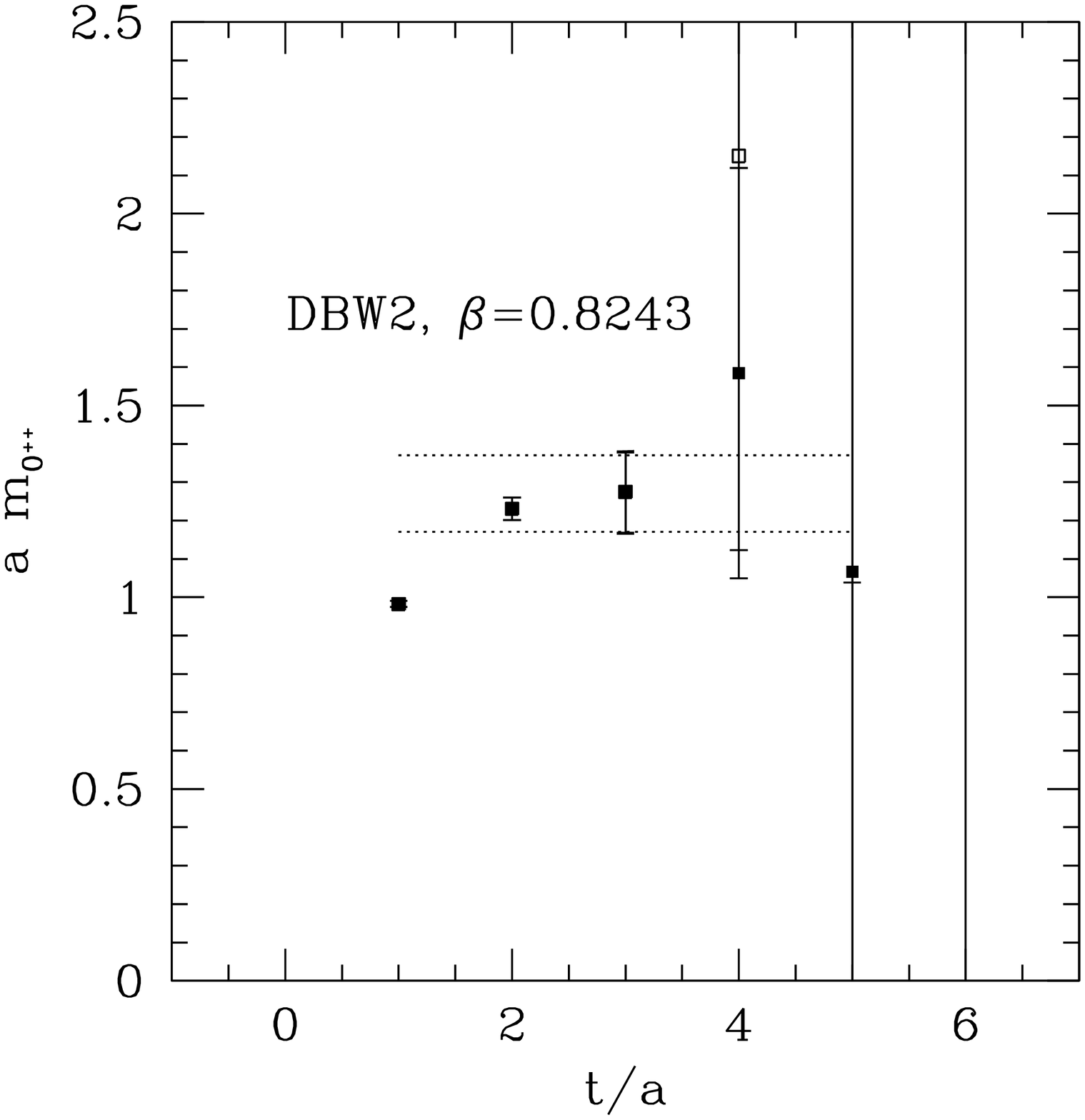} 
\includegraphics[width=7cm]{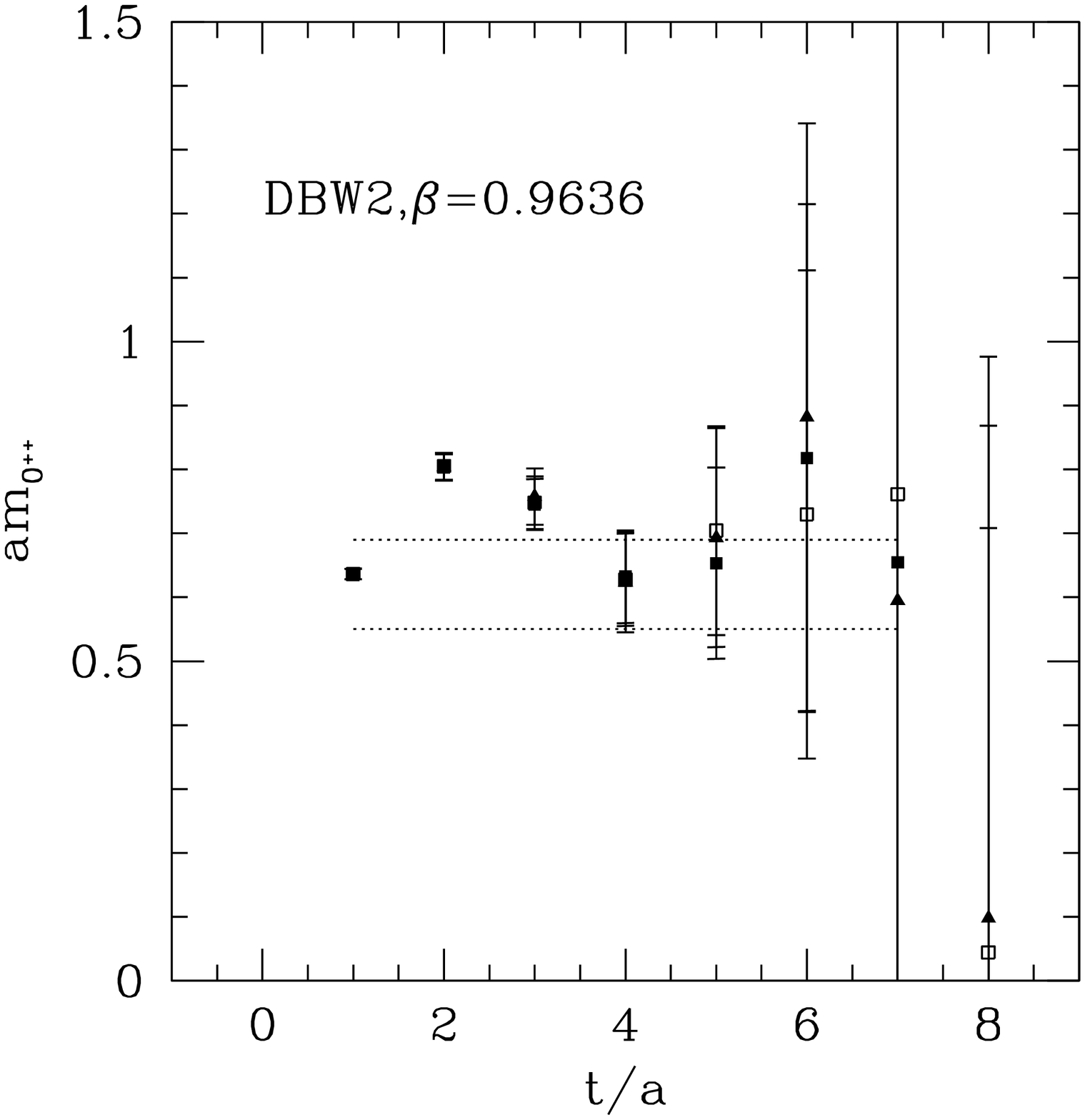} 
\end{center}
\caption[The effective masses for the $A_{1}^{++}$ channel,
evaluated with DBW2 action, at different lattice spacings.]{\footnotesize{The effective masses for the $A_{1}^{++}$ channel,
evaluated with DBW2 action, at different lattice spacings. The filled and empty squares corresponds to respectively to \protect\eq{meff_eigv} and \protect\eq{meff_corrma} with $t_{0}=0$. For $\beta=0.9636$, the filled triangles correspond to \protect\eq{meff_eigv} with $t_{0}=a$.}\label{fig_meff_dbw2}}
\end{figure}

\begin{figure} 
\begin{center}
\includegraphics[width=7cm]{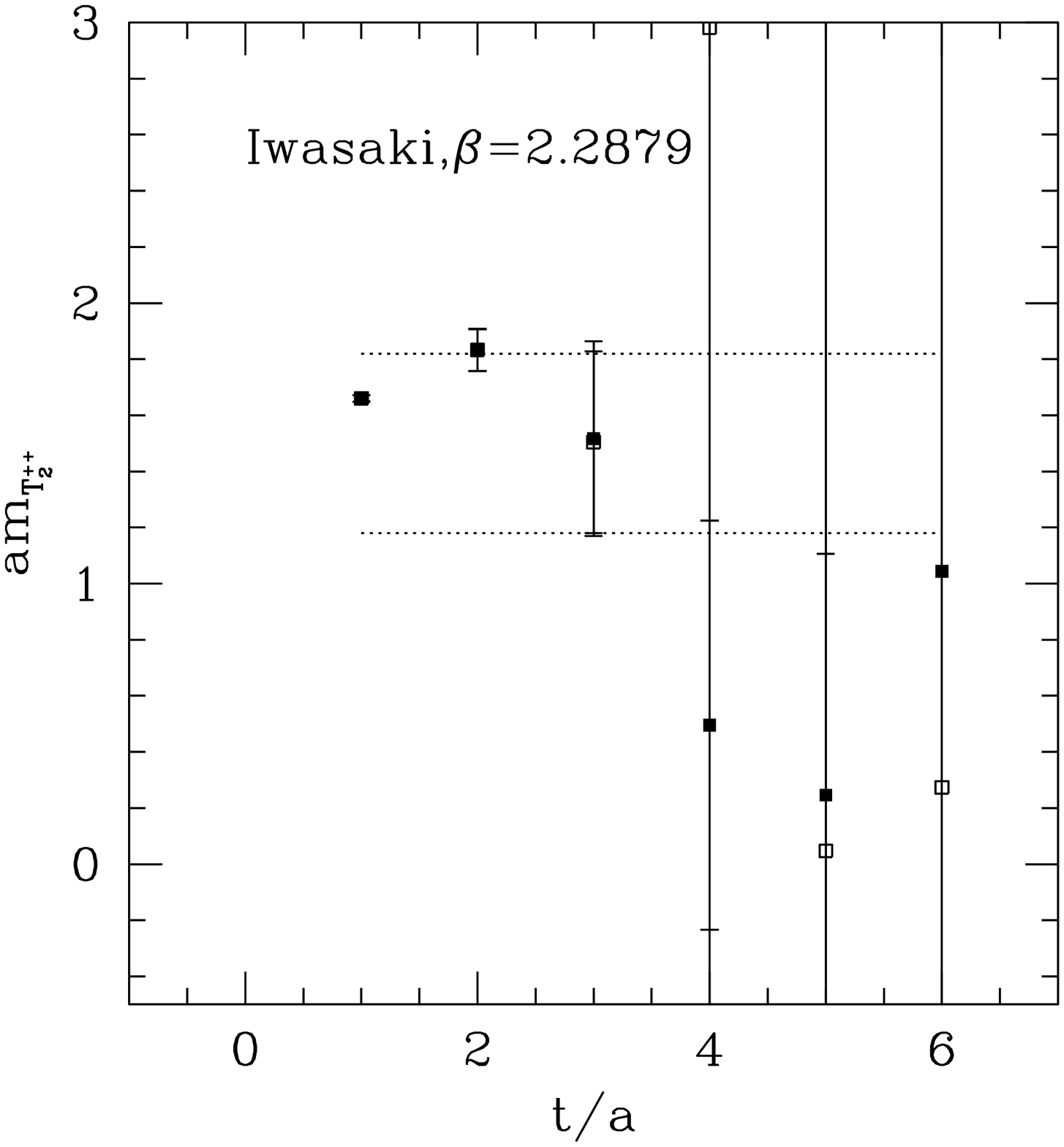} 
\includegraphics[width=7cm]{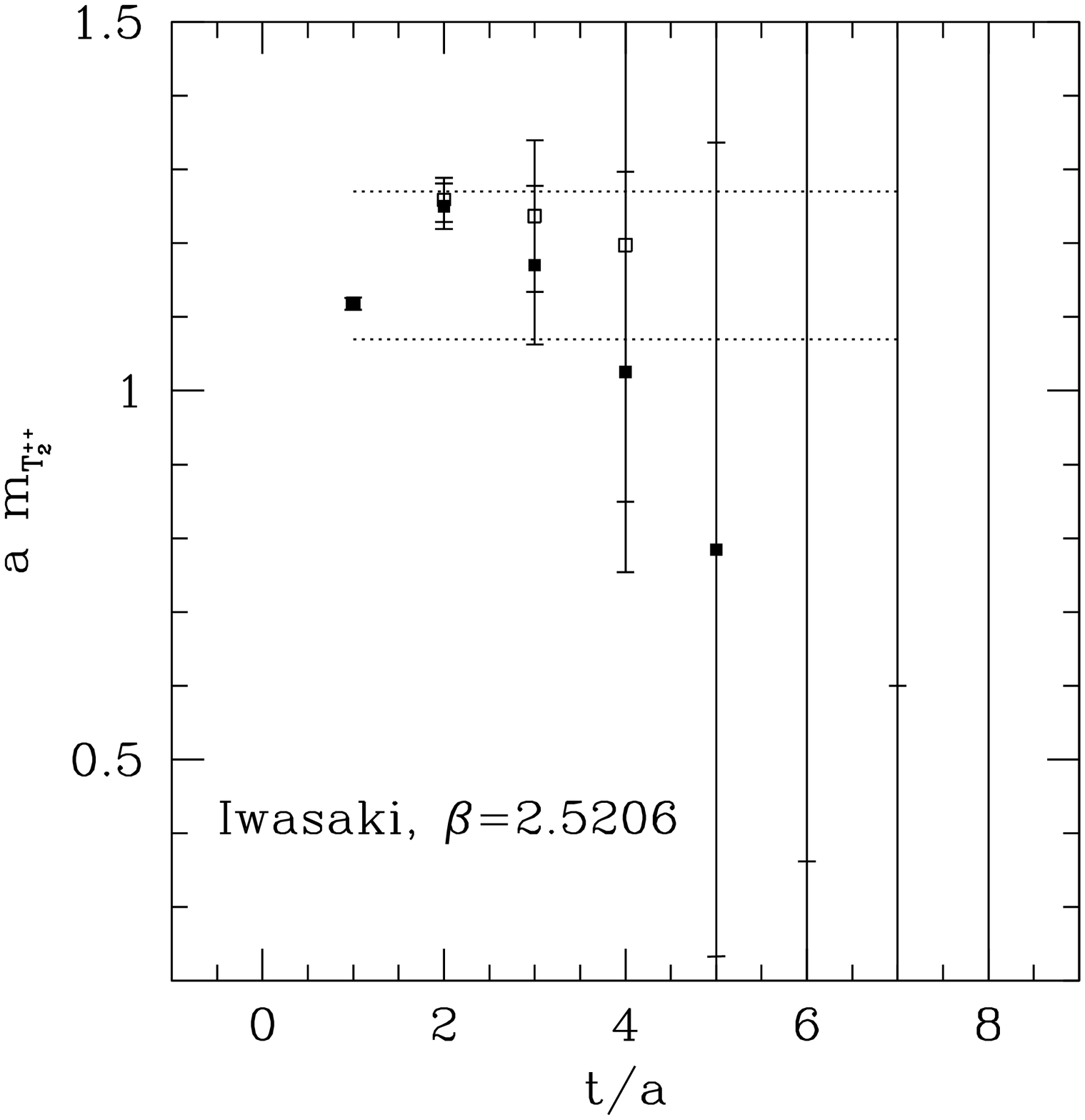} 
\end{center}
\caption[The effective masses for the $T_{2}^{++}$ channel,
evaluated with Iwasaki action, at different lattice spacings.]{\footnotesize{The effective masses for the $T_{2}^{++}$ channel,
evaluated with Iwasaki action, at different lattice spacings. The filled and empty squares corresponds to respectively to \protect\eq{meff_eigv} and \protect\eq{meff_corrma} with $t_{0}=0$.}\label{fig_meff2_iwasaki}}
\end{figure}

\begin{figure} 
\begin{center}
\includegraphics[width=7cm]{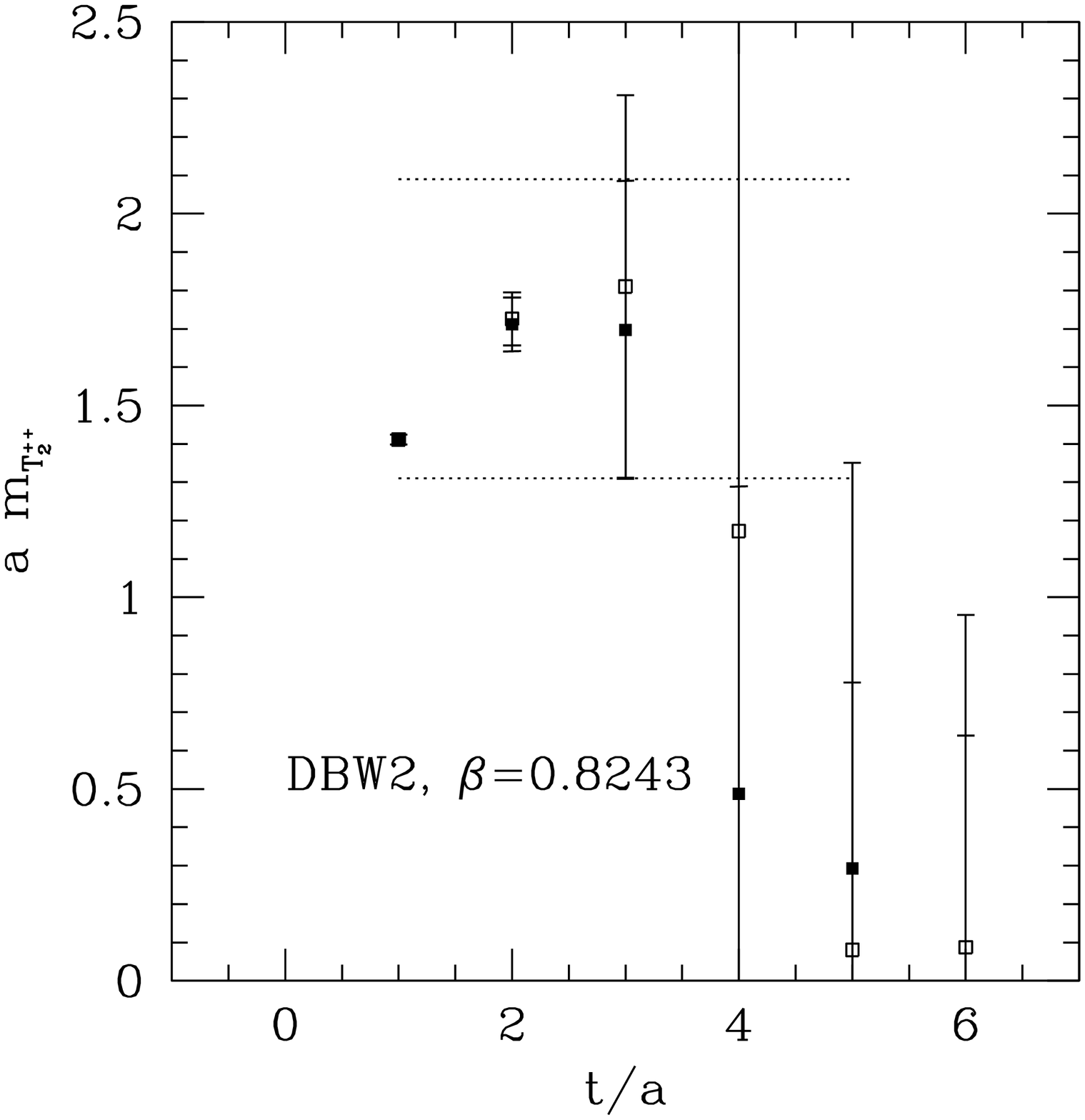} 
\includegraphics[width=7cm]{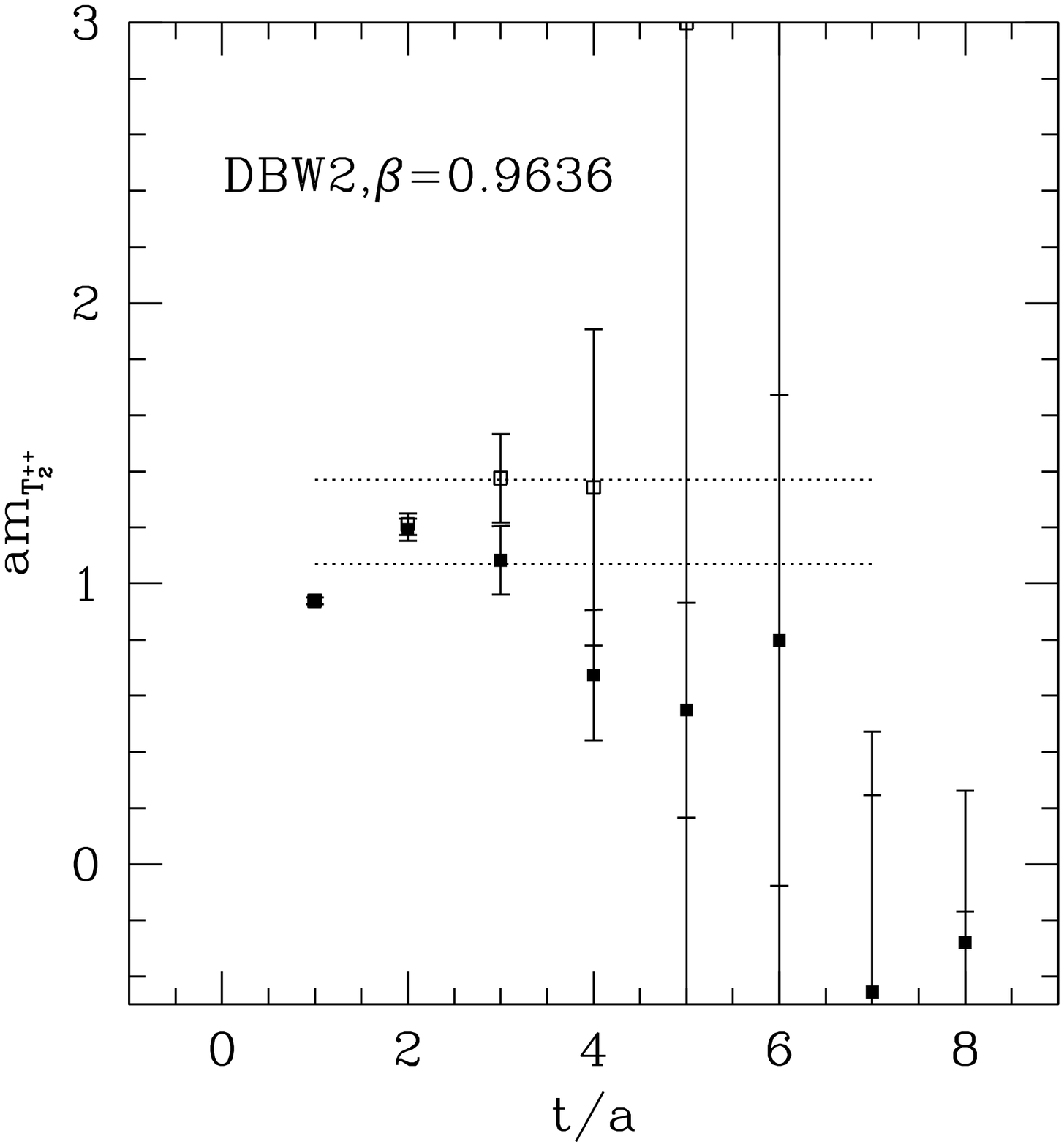} 
\end{center}
\caption[The effective masses for the $T_{2}^{++}$ channel,
evaluated with DBW2 action, at different lattice spacings.]{\footnotesize{The effective masses for the $T_{2}^{++}$ channel,
evaluated with DBW2 action, at different lattice spacings. The filled and empty squares corresponds to respectively to \protect\eq{meff_eigv} and \protect\eq{meff_corrma} with $t_{0}=0$.}\label{fig_meff2_dbw2}}
\end{figure}

For the extraction of glueball masses from the correlation matrices
\eq{corrma_glueball} variational techniques are necessary. Due to the
violation of physical positivity for the RG actions (appendix \ref{app_trans}), the
variational method is mathematically not well founded, at least not at very small
$t_{0}$, where one would like to apply it. The statistical errors are indeed
drastically increasing with $t$ and already at $t=4a$ the signal in the
correlation function \eq{corrma_glueball} is lost in noise.\\
We decided to adopt the following procedure: first we solved the generalized
eigenvalue problem 
\begin{equation}\label{gen_eigv_first}
C(t_{1})v_{\alpha}(t_1,t_0)=\lambda_{\alpha}C(t_{0})v_{\alpha}(t_1,t_0),
\end{equation}
with $t_{0}=0$, $t_{1}=a$. Then we projected the correlation matrices to the
space of eigenvectors corresponding to the $N$ eigenvalues which satisfy the
condition \footnote{in \cite{Niedermayer:2000yx} the direct eigenvalues and eigenvectors of $C(t_{0})$ are considered, instead of the generalized one. We have tried both possibilities and found consistent results.}
\begin{equation}\label{epsilon_eigv}
\lambda_{\beta}>\epsilon,\quad\beta=1,...,N,
\end{equation}
where $\epsilon$ is an adjustable (small) parameter. So the reduced matrix is
obtained by \footnote{The index $R$ for the representation is now omitted.}
\begin{equation}
C_{ij}^{N}(t)=(v_{i}(t_{1},t_0),C(t)v_{j}(t_{1},t_0)),\quad i,j=1,...,N.
\end{equation}
By choosing $\epsilon$ appropriately in \eq{epsilon_eigv} one hopes to get rid
of most of the unphysical modes caused by negative and very small eigenvalues.\\
Then one can apply the variational method to the reduced matrix
\begin{equation}
C^{N}(t)w_{\beta}(t,t_{0})=\lambda_{\beta}(t,t_{0})C^{N}(t_{0})w_{\beta}(t,t_{0}).
\end{equation}
The effective glueball mass can be read off directly from the lowest eigenvalue
\footnote{Due to the periodic boundary conditions in the time-direction, the correlation matrix \protect\eq{corrma_glueball} describes not only the correlation $C(t)$, but also $C(T-t)$, where $T$ is the lattice time-extent. Therefore one would expect
 $C(t)\sim A(e^{-mt}+e^{-m(T-t)})$. If one assumes that $T$ is much larger than the value of $t$ at which we are able to evaluate the mass, one can neglect the contribution from $(T-t)$. For our largest lattice spacing ($T=10a$) we included this corrections and check that they are irrelevant for the measurement of the glueball masses. }
\begin{equation}\label{meff_eigv}
m_{eff}(t)=-\log\left(\frac{\lambda_{0}(t,t_{0})}{\lambda_{0}(t-a,t_{0})}\right).
\end{equation}
or, alternatively one can project again $C^{N}$ to the subspace corresponding to the lowest eigenvalue
\begin{equation}
C^{1}(t)=(w_{1},C^{N}(t)w_{1}),
\end{equation}
where $w_{1}=w_{1}(t_{0}+a,t_{0})$;
then one evaluates the glueball masses by
\begin{equation}\label{meff_corrma}
m_{eff}(t)=-\log\left(\frac{C^{1}(t)}{C_{1}(t-a)} \right).
\end{equation}
We tested that these two different evaluation yield results which are
compatible within the statistical errors.\\
We chose $\epsilon$ in \eq{epsilon_eigv} so that the reduced matrix had
dimension between 2 and 5.\\
We applied the same procedure also for $t_{0}=a$, $t_{1}=2a$ in \eq{gen_eigv_first}, but due to the fact that the statistical fluctuations are already quite large one has to start from the beginning from a reduced number of operators in order to be able to solve the generalized eigenvalue problem.\\
We decided to extract the glueball masses by taking our effective mass at
$t=3a$ for the small $\beta$ regime and $t=4a$ for the large value of the
coupling at our disposal. For the RG actions we performed numerical simulations up to $L=16$, with a minimum lattice spacing $a\sim 0.1\quad\fm$; in this sense we are quite far away from the continuum limit and we will not be able to perform a continuum extrapolation of the results, but on the other hand this is the regime of lattice spacings that have been used for simulations with dynamical fermions and hence it is desirable to obtain informations about the discretization errors in this region.\\
The figures \ref{fig_meff_iwasaki},\ref{fig_meff_dbw2} show the effective masses in the $A_{1}^{++}$ channel
computed by using \eq{meff_eigv} (filled squares) and \eq{meff_corrma} (empty squares) for the Iwasaki and the DBW2 action, with
$t_{0}=0$. For the largest value of $\beta$, which are more precise, we show also the results obtained with $t_{0}=a$ (filled triangles).
As for the potential, one notices also in this case the presence
of negative contributions in the correlation functions due to unphysical
states; the situation here is somehow more drastic since one has to discard
for this reason the small $t$ region but on the other hand the errors
increase very rapidly and one is forced to extract the glueball mass at
$t=3-4a$. The results can then be affected by systematic errors that
can not be easily estimated. In the plots we have indicated 
with the dotted lines the range in which we decided to take the mass.\
The statistical errors were evaluated by using a jackknife procedure.\\
Concerning the determination of the $m_{2^{++}}$ we observed that the signal for the $E^{++}$ channel is usually worse than for the $T_{2}^{++}$ and the errors on the effective masses are very large already at $t=3a$. For this reason we decided to use  $m_{T_{2}^{++}}$ as estimate of $m_{2^{++}}$ at finite lattice spacing.\\
The figures \ref{fig_meff2_iwasaki},\ref{fig_meff2_dbw2} show the effective masses for the $T_{2}^{++}$ channel. The determination of the masses is more problematic that the $A_{1}^{++}$ case; in particular for the largest coupling one can observe that at $t=4a$ the errors are too large to have a significant measurement of the mass. We decided hence to extract the mass at $t=3a$, taking care that this value is compatible with the one at $t=4a$ within the statistical errors.

%%%%%%%%%%%%%%%%%%%%%%%%%%%%%%%%%%%%%%%%%%%%%%%%%%%%%%%%%%%%%%%%%%%%%%%%%%%%%%

%%%%%%%%%%%%%%%%%%%%%%%%%%%%%%%%%%%%%%%%%%%%%%%%%%%%%%%%%%%%%%%%%%%%%%%%%%%%%

%%%%%%%%%%%%%%%%%%%%%%%%%%%%%%%%%%%%%%%%%%%%%%%%%%%%%%%%%%%%%%%%%%%%%%%%%%%%%
\section{Results}
\begin{figure}
\begin{center}
\includegraphics[width=12cm]{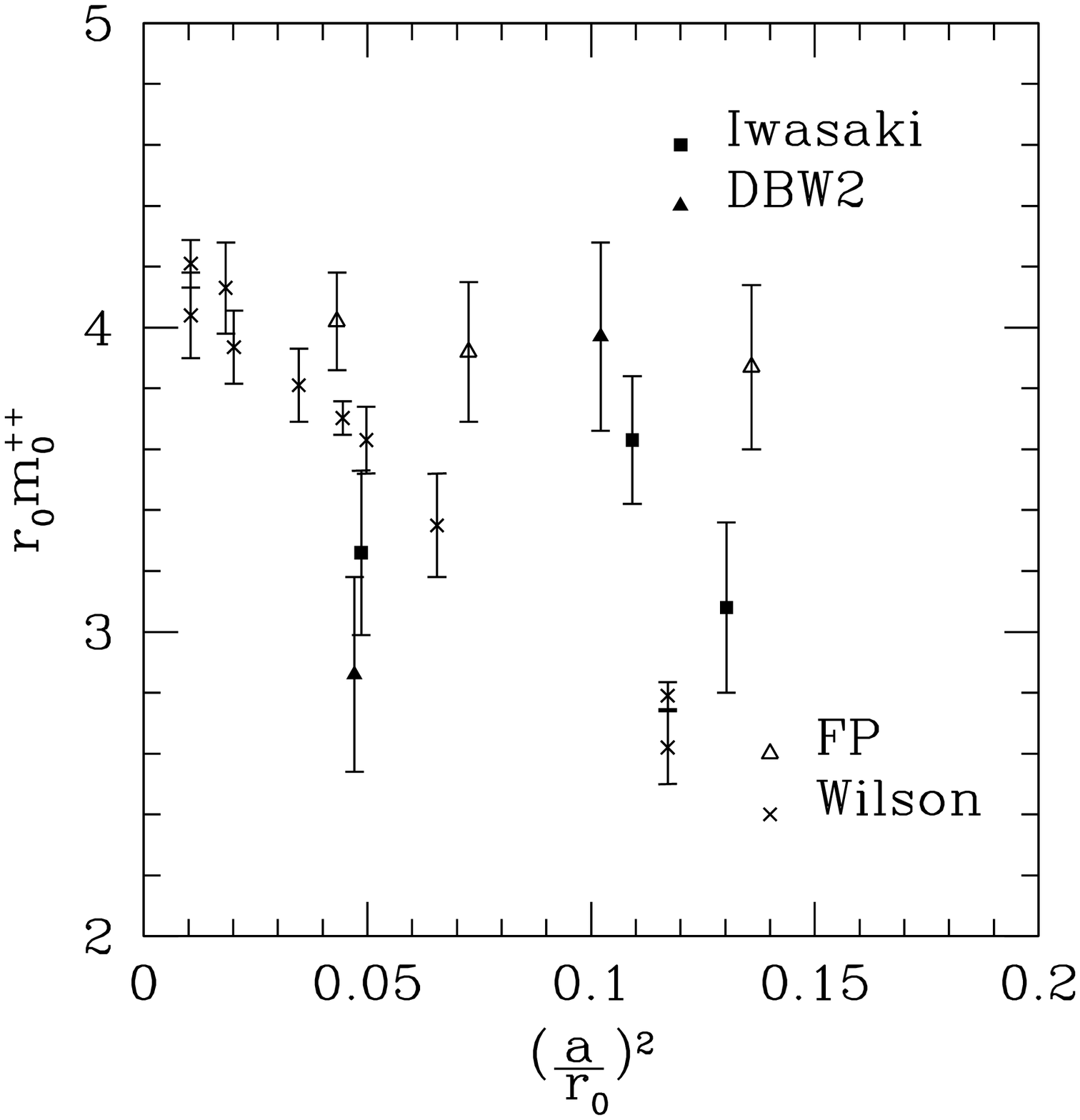} 
\end{center}
\caption[The $0^{++}$ glueball mass normalized with $\rnod$ as
    function of $(a/\rnod)^2$ for different actions.]{\footnotesize{The $0^{++}$ glueball mass normalized with $\rnod$ as
    function of $(a/\rnod)^2$ for different actions.}\label{f_0++}}
\end{figure}

In \tab{glue_res_iwasaki1}, \tab{glue_res_dbw21} the results for the masses in
lattice units are reported. For the smallest $\beta$ for the Iwasaki action,
it was not possible to obtain a reliable evaluation of the mass in the $T_{2}^{++}$ channel.\\
The renormalized quantities $\rnod\m_{G}$ are reported in
\tab{glue_res_iwasaki2}, \tab{glue_res_dbw22}; due to the fact that the errors
on $a m_{G}$ are quite large (between $8\%$ and $9\%$ for $A_{1}^{++}$ and between $10\%$ and $20\%$ for $T_{2}^{++}$), the difference which
arises by choosing $r_{I}$ or $r_{n}$ is in this case much smaller than the total
uncertainty; we have reported however both results in the table.
For $\beta=2.2423$ we had no direct measurement of $\rnod/a$ and we used the
interpolation formula \eq{fit_iwasaki} to evaluate it.\\
\Fig{f_0++} shows the results for $\rnod m_{0^{++}}$ as function of
$(a/\rnod)^2$, displaying only the results obtained with $r_{n}$.
 For the comparison we included the results for FP action \cite{Niedermayer:2000yx}
and several calculations performed with the Wilson action \cite{Ishikawa:1983xg,Vaccarino:1999ku,Bali:1993fb}.\\
The continuum values avaliable in the literature are listed in \tab{glue_cont}
and have been taken from \cite{Niedermayer:2000yx} for the FP action and from
\cite{Wittig:1999kb} for the Wilson action, where the results of \cite{Ishikawa:1983xg,Vaccarino:1999ku,Bali:1993fb} have been
expressed in units of $\rnod$.\\
The interpretation of our results is not very clear, also due to the large errors.\\
%%%%%%%%%%%%%%%NEW PART%%%%%%%%%%%%%%%%%%%%%%%%%%%%%%%%%%%%%%%%%%%%%%%%%%%
In any case, we want to stress that our determination can be seen at least as an upper limit for $m_{0^{++}}$ and $m_{2^{++}}$. We expect that at the values of $t/a$ at which we extracted the masses, the effects of positivity violations 
have already disappeared; this assumption is justified by our study based on perturbation theory and our estimation of $t_{min}$ \eq{tmin}.\\
For this reason we believe that possible systematic uncertainties on the glueball masses could only be due to the presence of excited states and hence could affect our measurement only in such a way that the real values of $m_{0^{++}}$, $m_{2^{++}}$ are lower with respect to our determination.\\
\noindent
%%%%%%%%%%%%%%%%%%%%%%%%%%%%%%%%%%%%%%%%%%%%%%%%%%%%%%%%%%%%%%%%%%%%%%%%%%%
At lattice spacings $a\sim 0.15\,\fm$ we notice a
improvement of the RG actions with respect to the Wilson action;
comparing with the continuum limit we find no significant discrepancy both for DBW2 and Iwasaki action,
while for the Wilson action one finds  $30-40\%$ deviation.\\
At lattice spacing $a\sim 0.1\,\fm$ we find on the other hand large lattice artefacts for RG actions: the result obtained with the Iwasaki
action is practically compatible with the one calculated through the plaquette action at the same lattice spacing,
while for the DBW2 action it is even further away from the continuum limit.\\
%%%%%%%%%%%%%%%%%%NEW%%%%%%%%%%%%%%%%%%%%%%%%%%%%%%%%%%%%%%%%%%%%%%%%%%%%%%%%%
If one considers our measurement as upper limit, one could conclude that the RG improved actions are not able to cure the problem of large lattice artefacts for the $0^{++}$ glueball mass. 
%%%%%%%%%%%%%%%%%%%%%%%%%%%%%%%%%%%%%%%%%%%%%%%%%%%%%%%%%%%%%%%%%%%%%%%%%%%%%%
At very small lattice spacings one
expects that the dominant lattice artefacts are of order $a^2$; for the
Wilson action this is indeed well confirmed by the numerical results in the
range $a\lesssim 0.17\fm$, as one can see in \fig{f_0++}.\\
For alternative actions there is no reason a priori to observe the same
behavior: while at small lattice spacings the ${\rm O}(a^2)$ should in any
case be the dominant one, at larger $a$ it is possible that lattice artefacts
are governed by higher orders (${\rm O}(a^4)$ and higher).\\
We have indeed already observed deviations from ${\rm O}(a^2)$ behavior in RG
actions for the quantity $T_{c}\rnod$ and even already for the force computed
at tree level (chapter \ref{improved}).\\
With our results it is not
possible to investigate how the continuum limit is approached, because one
should evaluate the glueball masses at smaller lattice spacing and this was
beyond the purpose of this work. The evident problem shown in our plot is
that a continuum extrapolation of the form
\begin{equation}
\rnod\m_{0^{++}}=\rnod\m_{0^{++}}|_{a=0}+s\times\left(\frac{a}{\rnod}\right)^2
\end{equation} 
of the data obtained with the RG action in the considered range of lattice
spacings would lead to results which are rather different from the expected
continuum limit.\\
One has however to remember that the extraction of the glueball masses is for
RG actions particularly difficult and the systematic errors 
could be underestimated.\\
%%%%%%%%%%%%%%%%%%%%%%%%%%%%%%%%%%%%%%%%%%%%%%%%%%%%%%%%%%%%%%%%%%%%%%%%%%%%%%%%%%

\begin{figure}
\begin{center}
\includegraphics[width=12cm]{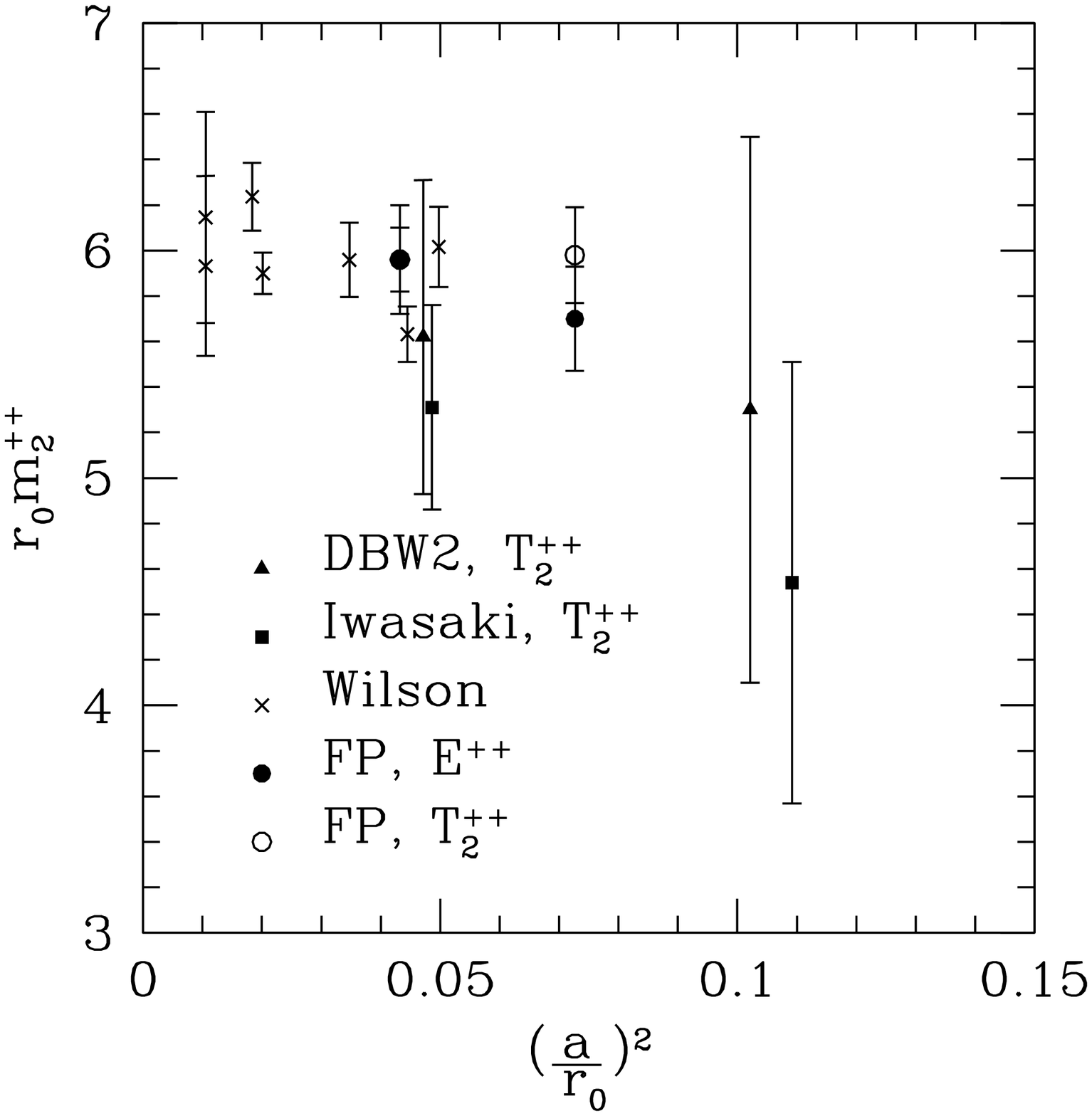} 
\end{center}
\caption[The $2^{++}$ glueball mass normalized with $\rnod$ as
    function of $(a/\rnod)^2$ for different actions.]{\footnotesize{The $2^{++}$ glueball mass normalized with $\rnod$ as
    function of $(a/\rnod)^2$ for different actions.}\label{f_2++}}
\end{figure}

In \fig{f_2++} we report our results for $\rnod m_{2^{++}}$; 
for this particular observables the calculation performed with the Wilson action do not show significant lattice artefacts.\\
At our smallest lattice spacing we do not observe a deviation from the results obtained with the Wilson action; one has however to notice that our errors are too large to make any conclusive statement.\\
For our largest lattice spacing the results with the Wilson action are not available (there are only know results with anisotropic lattices \cite{Morningstar:1999rf}) and it is indeed difficult to have a reliable estimation of the mass also in our case: we decided to show however our results, even with the very large error bars.\\
From our computation one can not deduce if for $\rnod m_{2^{++}}$ the RG actions show significative discretization errors  
 and further investigations are needed to clarify the issue.

%%%%%%%%%%%%%%%%%%%%%%%%%%%%%%%%%%%%%%%%%%%%%%%%%%%%%%%%%%%%%%%%%%%%%%%%%%%%%
\begin{table}
\begin{center}
\begin{tabular}{c c c}
\hline
Collab.  & $\rnod m_{0^{++}}$   &  $\rnod m_{2^{++}}$ \\[1ex]
\hline
M \& P \cite{Morningstar:1999rf}   & 4.21(11)(4)  & 5.85(2)(6) \\   
GF11 \cite{Vaccarino:1999ku}     & 4.33(10)     & 6.04(18)    \\
Teper\cite{glueb:teper98}     & 4.35(11)     & 6.18(21)     \\
UKQCD\cite{Bali:1993fb}     & 4.05(16)     & 5.84(18)   \\
FP  \cite{Niedermayer:2000yx}      & 4.12(21)     & [5.96(24)]      \\[1ex]
\hline
\end{tabular}
\end{center}
\caption[Continuum extrapolations of the two lowest glueball
    masses in units of $\rnod$.]{\footnotesize{Continuum extrapolations of the two lowest glueball
    masses in units of $\rnod$. For the FP action, the $2^{++}$ value is not extrapolated to the continuum but denotes the mass obtained at a lattice spacing $a=0.10\,\
fm$.}\label{glue_cont}}
\end{table}

%%%%%%%%%%%%%%%%%%%%%%%%%%%%%%%%%%%%%%%%%%%%%%%%%%%%%%%%%%%%%%%%%%%%%%%%%%%%%%%%%%%%
%\begin{table}
%\begin{center}
%Iwasaki action\\[1ex]
%\begin{tabular}{c c c c}
%\hline
%$\beta$  & $am_{A_{1}^{++}}$ &  $am_{E^{++}}$ & $am_{T_{2}^{++}}$ \\[1ex]
%\hline
%2.2423   & 1.11(10)          &               &                  \\
%2.2879   & 1.20(7)     &  1.82(20)      & 1.50(32) \\
%2.5206   & 0.72(6)      &  1.14(5)      & 1.10(9)      \\   
%\hline
%\end{tabular}
%%\end{center}
%\caption{\footnotesize{Glueball masses in lattice units, Iwasaki action.}\label{glue_res_iwasaki1}}
%\end{table}
%%%%%%%%%%%%%%%%%%%%%%%%%%%%%%%%%%%%%%%%%%%%%%%%%%%%%%%%%%%%%%%%%%%%%%%%%%%%%%%%%%%
%%%%%%%%%%%%%%%%%%%NEW RESULTS%%%%%%%%%%%%%%%%%%%%%%%%%%%%%%%%%%%%%%%%%%%%%%%%%

\begin{table}
\begin{center}
Iwasaki action\\[1ex]
\begin{tabular}{c c c}
\hline
$\beta$  & $am_{A_{1}^{++}}$ & $am_{T_{2}^{++}}$ \\[1ex]
\hline
2.2423   & 1.11(10)          &                             \\
2.2879   & 1.20(7)     &    1.50(32) \\
2.5206   & 0.72(6)      &   1.17(10)      \\   
\hline
\end{tabular}
\end{center}
\caption[Glueball masses in lattice units, Iwasaki action.]{\footnotesize{Glueball masses in lattice units, Iwasaki action.}\label{glue_res_iwasaki1}}
\end{table}
\begin{table}
\begin{center}
DBW2 action\\[1ex]
\begin{tabular}{c c c}
\hline
$\beta$  & $am_{A_{1}^{++}}$ &   $am_{T_{2}^{++}}$ \\[1ex]
\hline
0.8243   & 1.27(10)               &  1.70(39)  \\
0.9636   & 0.62(7)                &  1.22(15)   \\
\hline
\end{tabular}
\end{center}
\caption[Glueball masses in lattice units, DBW2 action.]{\footnotesize{Glueball masses in lattice units, DBW2 action.}\label{glue_res_dbw21}}
\end{table}
\begin{table}
\begin{center}
Iwasaki action, $r_{n}$\\[1ex]
\begin{tabular}{c c c }
\hline
$\beta$  & $\rnod m_{A_{1}^{++}}$ &  $\rnod m_{T_{2}^{++}}$ \\[1ex]
\hline
2.2423   &  3.08(28)  &                               \\
2.2879   &  3.63(21)        & 4.54(97)   \\
2.5206   &  3.26(27)        & 5.31(45)    \\[1ex]
\hline
\hline
\end{tabular}
\\[1ex]
Iwasaki action, $r_{I}$\\[1ex]
\begin{tabular}{c c c }
\hline
$\beta$  & $\rnod m_{A_{1}^{++}}$ & $\rnod m_{T_{2}^{++}}$ \\[1ex]
\hline
2.2423   &  3.07(28)  &                            \\
2.2879   &  3.63(21)       &   4.54(97)       \\
2.5206   &  3.25(27)       &   5.28(45)     \\
\hline
\end{tabular}

\end{center}
\caption[Results for $\rnod m_{G}$ for the channels $A_{1}^{++}$
    and $T_{2}^{++}$, using the Iwasaki action.]{\footnotesize{Results for $\rnod m_{G}$ for the channels $A_{1}^{++}$
    and $T_{2}^{++}$, using the Iwasaki action. }\label{glue_res_iwasaki2}}
\end{table}

%%%%%%%%%%%%%%%%%%%%%%%%%%%%%%%%%%%%%%%%%%%%%%%%%%%%%%%%%%%%%%%%%%%%%%%%%%%%%%%

\begin{table}
\begin{center}
DBW2 action, $r_{n}$\\[1ex]
\begin{tabular}{c c c}
\hline
$\beta$  & $\rnod m_{A_{1}^{++}}$  & $\rnod m_{T_{2}^{++}}$ \\[1ex]
\hline
0.8243   &  3.97(31)         &   5.3(1.2)               \\
0.9636   &  2.86(32)         &   5.62(69)                 \\[1ex]
\hline
\hline
\end{tabular}
\\[1ex]
DBW2 action, $r_{I}$\\[1ex]
\begin{tabular}{c c c}
\hline
$\beta$  & $\rnod m_{A_{1}^{++}}$ & $\rnod m_{T_{2}^{++}}$ \\[1ex]
\hline
0.8243   &  3.86(30)         &  5.2(1.2)               \\
0.9636   &  2.82(32)         &  5.56(68)              \\
\hline
\end{tabular}
\end{center}
\caption[Results for $\rnod m_{G}$ for the channels $A_{1}^{++}$
    and $T_{2}^{++}$, using the DBW2 action.]{\footnotesize{Results for $\rnod m_{G}$ for the channels $A_{1}^{++}$
    and $T_{2}^{++}$, using the DBW2 action.}\label{glue_res_dbw22}}
\end{table}

%%%%%%%%%%%%%%%%%%%%%%%%%%%%%%%%%%%%%%%%%%%%%%%%%%%%%%%%%%%%%%%%%%%%%%%%%%%%%%
%%%%%%%%%%%%%%%%%%%%%%%%%%%%%%%%%%%%%%%%%%%%%%%%%%%%%%%%%%%%%%%%%%%%%%%%%%%%%%

\begin{table}
\begin{center}
Iwasaki action\\[1ex]
\begin{tabular}{c c c c}
\hline
$\beta$  &  $L$   &  $n_{l}$    &   $n_{meas}$ \\[1ex]    
\hline
2.2423   &  10    &   2,4,6,8   &   12000 \\
2.2879   &  12    &   2,4,6,8   &   16000 \\
2.5206   &  16    &   3,6,9,12  &   8000  \\       
\hline
\hline
\end{tabular}
\\[1ex]
DBW2 action\\[1ex]
\begin{tabular}{c c c c}
\hline
$\beta$  &  $L$   &  $n_{l}$    &   $n_{meas}$ \\[1ex]    
\hline
0.8342   &  12    &   2,4,6,8   &   8000 \\
0.9636   &  16    &   3,6,9,12  &   2500  \\       
\hline
\end{tabular}
\end{center}
\caption[Simulation parameters for the evaluation of the glueball masses for the Iwasaki and DBW2 actions.]{\footnotesize{Simulation parameters for the evaluation of the glueball masses for the Iwasaki and DBW2 actions.}\label{sim_par_glue}}
\end{table}

%%%%%%%%%%%%%%%%%%%%%%%%%%%%%%%%%%%%%%%%%%%%%%%%%%%%%%%%%%%%%%%%%%%%%%%%%%%%%%%%%%
\newpage
\section{Open questions}
\begin{figure}
\begin{center}
\includegraphics[width=10cm]{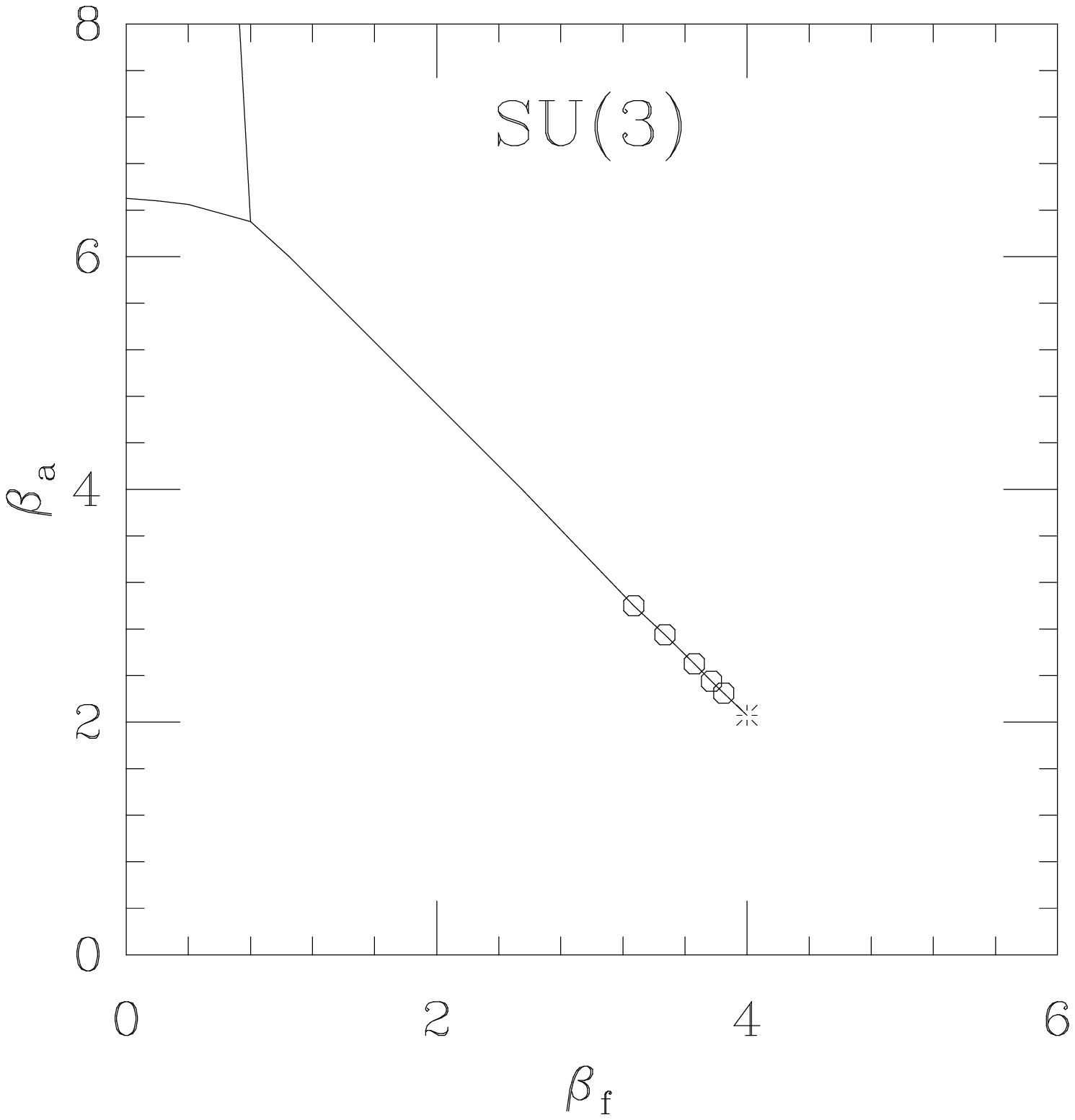} 
\end{center}
\caption[The phase diagram of $\SUthree$ gauge theories in the fundamental-adjoint coupling space.]{\footnotesize{The phase diagram of $\SUthree$ gauge theories in the fundamental-adjoint coupling space \protect\cite{Blum:1995xb}.}\label{f_phase_trans}}
\begin{center}
\includegraphics[width=10cm]{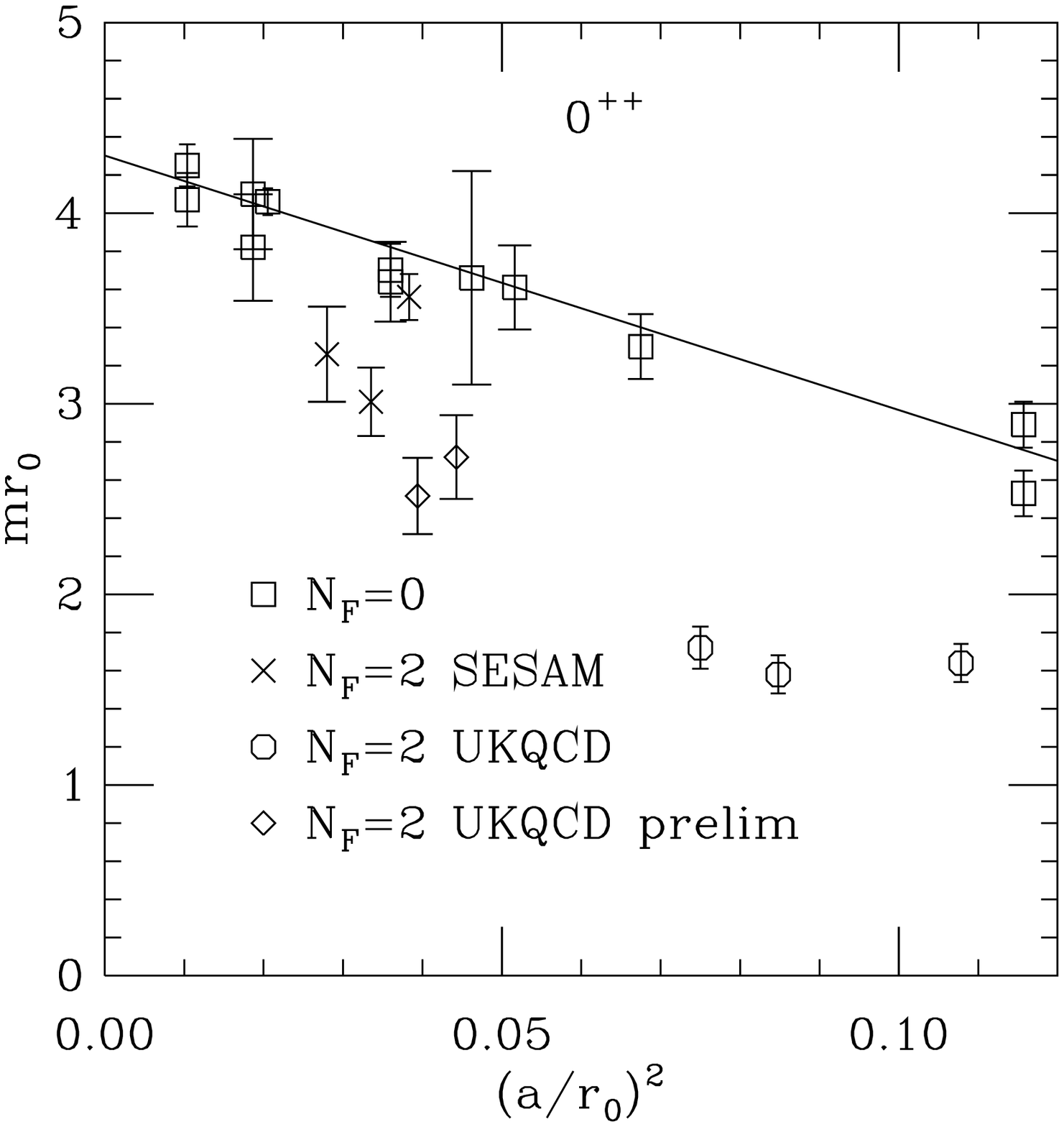} 
\end{center}
\caption[Plot for the value of mass of the $0^{++}$ glueball state from quenched data  and for the lightest scalar meson with $N_{f}=2$ flavours of dynamical quarks.]{\footnotesize{Plot from \cite{Michael:2003ai}for the value of mass of the $0^{++}$ glueball state from quenched data \cite{deForcrand:1985rs,Michael:1989jr,Bali:1993fb,Morningstar:1999rf} and for the lightest scalar meson with $N_{f}=2$ flavours of dynamical quarks \cite{Bali:2000vr,McNeile:2000xx,Hart:2002sp}.}\label{0pp_unquenched}}
\end{figure}

As already mentioned, RG action have been indicated  by some Collaborations
\cite{AliKhan:2001tx,Orginos:2001xa,Aoki:2002vt} as best candidates to be used in the simulations with dynamical fermions.\\
In view of these next simulations, it is important to have under control the properties of alternative gauge actions by testing their features, the universality and the discretization errors for several physical quantities.\\
This motivated our investigations and requires further studies on this subject.\\
An important point to stress is that in full lattice QCD it can happen
that the features of different gauge actions are modified, also
depending on which formulation of fermions on the lattice is adopted.\\
For example, the JLQCD and CP-PACS collaborations have used a ${\rm O}(a)$ formulation of
the fermionic action, the so-called Sheikholeslami-Wohlert (clover) action
\cite{impr:SW}, which is based on the Symanzik approach (see chapter
\ref{c_lattice}).
In order to achieve the improved action one has to add a counterterm such that
the order $a$ term in the action of the effective continuum theory is
canceled. For the improved action one thus obtains
\begin{equation}\label{oa_imp}
S=S_{Wilson}+a^{5} \sum_{x}c_{sw}\overline{\psi}(x)\frac{i}{4}\sigma_{\mu\nu}\hat{F}_{\mu\nu}(x)\psi(x),
\end{equation}
where $S_{Wilson}$ corresponds to the Wilson fermionic action 
\footnote{Fermions on the lattice are described by Grassmann variables
  $\psi$, $\bar{\psi}$ defined on the lattice sites.
There are several possibilities to define the lattice action
  associated to fermions, leading to different properties.
The so-called Wilson action on the lattice can be written in
  the form $S_{Wilson}=a^{4}\sum_{x}\bar{\psi}(x)(D+m_{0})\psi(x)$,
  where $m_{0}$ is the bare quark mass and
  $D=\frac{1}{2}\{\gamma_{\mu}(\nabla_{\mu}^{*}+\nabla_{\mu})-a\nabla_{\mu}^{*}\nabla_{\mu}\}$ is the lattice Dirac operator.                 }
and $\hat{F}_{\mu\nu}$ is a lattice representation of the gluon field
tensor.\\
The coefficient $c_{sw}$ in \eq{oa_imp} has to be tuned appropriately. 
%%%%%%%%%%%%%%%%NEW%%%%%%%%%%%%%%%%%%%%%%%%%%%%%%%%%%%%%%%%%%%%%%%%%%%%%%%%%
For example, for simulations with 2 flavors of dynamical fermions, the CP-PACS collaboration used the mean field ${\rm O}(a)$ improved definition for the fermionic action together with the Iwasaki gauge action \cite{AliKhan:2001tx}, while the JLQCD collaboration used non-perturbative ${\rm O}(a)$ improved Wilson fermion together with the plaquette gauge action \cite{Aoki:2002uc}.\\ 
%%%%%%%%%%%%%%%%%%%%%%%%%%%%%%%%%%%%%%%%%%%%%%%%%%%%%%%%%%%%%%%%%%%%%%%%%%%%%
%For
%example, the JLQCD Collaboration used the non-perturbative determination for $\nf=2$ \cite{Jansen:1998mx} 
It was observed \cite{Aoki:2001xq} that for $N_{f}=3$ simulations with ${\rm O}(a)$-improved
Wilson fermions and plaquette gauge action one finds an unexpected first order
phase transition at zero-temperature.\\
The investigations indicate that this
phase transition is a lattice artefact restricted to strong coupling
regions ($\beta\lesssim 5.0$). This phenomenon spoils the possibility to
perform a continuum extrapolation of physical quantities unless one
considers only couplings in the region $\beta\gtrsim 5.0$.\\
The interesting fact is that this transition disappears if one adopts
alternative gauge actions, like Iwasaki or tadpole-improved Symanzik action, or alternatively if one uses the unimproved Wilson fermionic action
($c_{sw}=0$) with the plaquette gauge action.\\
A plausible scenario is that the clover term added in \eq{oa_imp} produces a term in the adjoint representation of the gauge fields with a positive coupling.\\
It is indeed known that pure $\SUthree$ lattice gauge theory containing
fundamental and adjoint representations of the gauge fields undergoes
a first order phase transition for positive values of the adjoint
coupling \cite{Bhanot:1981eb,Greensite:1981hw,Bhanot:1982pj}. A
simple way to construct a gauge action with an adjoint coupling would be to generalize the Wilson plaquette action in the following way:
\begin{equation}
S=\beta_{F}\sum_{x}\sum_{\mu<\nu}\left\{1-\frac{1}{3}\Re W^{1\times 1}_{\mu\nu}(x)\right\}+
\beta_{A}\sum_{x}\sum_{\mu<\nu}\left\{1-\frac{1}{9}|W^{1\times 1}_{\mu\nu}(x)|^{2}\right\},     
\end{equation}
where $W^{1\times 1}$ is the plaquette defined in \eq{plaquette}.\\
The resulting line of the first order phase transition with the location of
the endpoint for the gauge group $\SUthree$ is shown in \fig{f_phase_trans}. The
Wilson plaquette action corresponds to the line $\beta_{A}=0$: it is located below the end point of the phase transition and hence it is safe to take the continuum limit along this line. If one would approach the continuum limit by crossing the transition line above this point, there would be discontinuities in the expectation values of Wilson loops and in glueball masses. In particular, at the point of intersection with the transition line the $0^{++}$ glueball mass is expected to be zero.\\
The fact that for the Wilson action the lattice artefacts for
$m_{0^{++}}$ are large and the glueball mass in lattice units becomes
small could then also be interpreted as the ``influence'' of the endpoint of the first order phase transition on the $\beta_{A}=0$ line corresponding to the Wilson action.\\ 
In order to test the validity of this scenario, one should add an
adjoint term in the action and check how does it change the behavior
of the $O^{++}$ glueball masses.\\

An important test is also the computation of scalar glueball mass in
full QCD \footnote{By inserting dynamical fermions in the theory, the mass eigenstates with $0^{++}$ quantum numbers are not distinguished as ``glueballs'' or as ``quark-antiquark'' states; in general mixing will occur.}; recent results \cite{Michael:2003ai} are shown in
\fig{0pp_unquenched}.
The SESAM \cite{Bali:2000vr} results have been obtained with Wilson
gauge and fermion action, while UKQCD \cite{McNeile:2000xx,Hart:2002sp} adopted ${\rm O}(a)$ Wilson improved fermions. The larger lattice spacing results 
show a significant reduction in the lightest scalar mass, as shown in \fig{0pp_unquenched}.\\
It is very important to understand if it implies a lower scalar mass also in the continuum limit or if an enhanced ${\rm O}(a^2)$ correction might be present.\\

A useful study would be to repeat these investigations with alternative gauge actions, starting for example by locating the position of the end point of the first order phase transition;
it is interesting to know if it lies closer or further away from
the $\beta_{A}=0$ line with respect to the plaquette action.\\
From our calculation of the glueball masses with Iwasaki and DBW2 action one can not deduce any possible scenario and a systematic investigation which includes adjoint terms would be needed.\\
%Another useful investigation would be to use analytic techniques in order to estimate the contributions induced by the clover term in \eq{oa_imp} in the adjoint representation.\\
Discussing these open issues will be helpful to understand properties
of full QCD on the lattice and choose an optimal combination of gauge and
fermionic actions to perform reliable unquenched
simulations.
\vspace{-2cm}

%%% Local Variables: 
%%% mode: latex
%%% TeX-master: t
%%% End: 

%% file: concl.tex
\chapter*{Conclusions}
        \label{conclusions}
         \addcontentsline{toc}{chapter}{Conclusions}
        \pagestyle{fancy}
        \fancyhead{}
        \fancyhead[LE,RO]{\bfseries\thepage}
        \fancyhead[LO]{\bfseries Conclusions}
        \fancyhead[RE]{\bfseries Conclusions}

The main object of the first part of this work was the potential between a
static quark and antiquark in the pure $\SUthree$ gauge theory.
In particular, we concentrated on the question whether at short distances
large non perturbative terms are present. To test this possibility we used
both the potential and the force, which define two different renormalization
schemes. The first step was to obtain a parameter-free perturbative prediction
for these quantities from the Renormalization Group equation at two and tree
loops; this was possible due to the fact that the $\Lambda$ parameter is known
from lattice studies \cite{mbar:pap1} and the coefficients of the $\beta$-function have been
calculated up to tree loops \cite{Peter:1997me,Schroder:1998vy,Fischler:1977yf,Billoire:1980ih,Larin:1993tp,pert:2loop}.\\ 
Then we performed a numerical Monte Carlo simulation to evaluate potential and
force non-perturbatively. 
In the pure gauge theory it was possible to compute these quantities over a
large range of distances with high precision by considering two sets of lattices and matching them
(in the continuum limit) in a region of overlap. 
By setting the length scales $\rnod$ and $\rc$ we
built renormalization group invariant quantities and we performed a continuum
extrapolation of our results. At this stage we investigated the scaling violations
and we were able to confirm Symanzik's picture of lattice artifacts.
By comparing these results with the perturbative prediction we did not observe
the presence of large non-perturbative terms. 
Furthermore, we remarked that
different renormalization schemes have different convergence properties and
the scheme defined through the force is in this sense more appropriate: here
perturbation theory works remarkably well up to $\alpha_{q\bar{q}}\sim 0.3$,
while for the potential only up to $\alpha_{\bar{V}}\sim 0.15$.\\
This result confirms what can already be argued from the apparent ``convergence'' by comparing the
two- and three-loops expansions for the running coupling in the two
schemes; the fact that the perturbative coupling defined through the potential fails in describing non-perturbative data is due the restricted domain of
applicability of perturbation theory for this scheme rather than a signal
of non-expected terms.\\
Although the situation in full QCD is much more complicated due to presence of different mass scales, this study may be a guideline for future investigations and applications of perturbation theory.

We then compared our results with the bosonic string model, which is
expected to describe the long-distance properties \cite{nambu} of the theory. We observed a
surprising good agreement with the next-to-leading order prediction  \cite{Luscher:1980fr,Luscher:1981ac} of the effective theory
already at $0.4\fm$, where in principle the validity of the string picture is
questionable.\\ 
By including a higher order correction estimated in another work \cite{Luscher:2002qv}
we observed that at large distances the agreement with the non-perturbative
data is improved, while at intermediate distances it is deteriorated. 
This fact is not surprising since we are dealing with asymptotic expansions 
and it could indicate that the agreement at intermediate distances is rather accidental.
However we want to remark that we evaluated the potential only up to $\sim 0.8\fm$,
but a deep understanding of the string nature of the strong interactions
requires the investigation at larger distances, where the statistical errors
are usually very large. With the help of the numerical techniques
introduced in \cite{Luscher:2001up} is now possible to reach an higher precision
and hence it will be possible to perform stringent tests, also on the spectrum of excited states.\\

The second part of this work was dedicated to the study of scaling and
universality properties of gauge actions alternative to the plaquette
formulation; in particular we concentrated our attention on the Iwasaki and
DBW2 actions, which contain also rectangular $(1\times 2)$ loops and have been formulated by using Renormalization Group arguments.\\
Our main purpose was to test the universality and the scaling behavior for
some relevant physical observable: we chose in particular the deconfinement
temperature $T_{c}$ and the mass of the glueballs $0^{++}$ and $2^{++}$.
The first issue that we faced was the setting of a suitable scale to
build renormalized dimensionless quantities. As already pointed out in
previous studies it is important to use a quantity which can be measured with
good precision and which is not subjected to uncontrolled systematic errors;
the intermediate length $\rnod$ had been introduced with this purpose and
turned out to be a good choice.\\
We performed new numerical simulations at several lattice spacings to compute 
$\rnod/a$, which was up to now not available on the literature for the Iwasaki
and DBW2 actions. 
In the analysis of our results we remarked that for these actions the physical
positivity is violated \cite{Luscher:1984is}: in particular, negative contributions are expected in
Euclidean correlation functions for small time. By investigating the location
of the poles in the propagator at tree level, we
evaluated for different actions the value of $t_{min}$ such that for $t\gg
t_{min}$ these unphysical contributions are expected to disappear. 
We discussed the implications for the extraction of physical observables, in
particular for the applicability of the variational method, which is founded
per definition on physical positivity.\\
Interesting features of these actions could also be learnt by studying the force at tree level: we noticed that RG actions are ``overcorrected'' at tree level and that beside the leading lattice artifacts of order $a^2$, also higher powers are expected to be important.\\
With our determinations of $\rnod/a$
we then formed the quantity $T_{c}\rnod$ and we compared the results at
different lattice spacings with the ones obtained with the plaquette Wilson
action. For the Iwasaki action we observed a reduction of the lattice
artifacts, while for the DBW2 action the situation is less clear since the
results depend quite strongly on which definition of the force at finite
lattice spacing is adopted. 
The important result is that universality is confirmed for this particular
observable; in previous calculations the quantity $T_{c}/\sqrt{\sigma}$ 
was used and a discrepancy in the continuum extrapolations between Iwasaki and
Wilson actions was observed \cite{Okamoto:1999hi}. Our result exclude a violation of the universality and seem to confirm that this disagreement was generated by the
problems related with the
extraction of the string tension $\sigma$ and hence by the choice of the
quantity used to set the scale.\\
With our results we were able to give a parametrization formula which describes $\rnod/a$ in the range $2.1551\leq\beta \leq 2.7124$ for the Iwasaki action and $0.75696\leq\beta\leq 1.04$ for the DBW2 action, which can be useful for future applications.\\
We checked the scaling of RG actions also on the running coupling extracted from the force by
comparing the results at our smallest lattice spacing with the
continuum results obtained in the first part of the work. As expected, at
sufficiently large distances, we found no significative deviation, while at
small $r/\rnod$ we observed a difference of $2\%$ for the Iwasaki action and $4-10\%$
for the DBW2 action, depending also in this case on the definition of the force
at finite lattice spacings.\\
Finally we computed the glueball masses for the $0^{++}$ and $2^{++}$
states. We started by selecting a basis of "good" operators from Wilson loops up to length eight, using the signal/noise ratio as criterion for the choice.\\
For the evaluation of the glueball masses the violation of physical positivity
is even more problematic that for the potential, since one can not avoid to use
variational techniques. We analyzed our data by truncating the correlation
matrices to suitable subspaces where the presence of unphysical states is
supposed to be reduced.
%%%%%%%%%%%%%%%%%%%%%%%NEW%%%%%%%%%%%%%%%%%%%%%%%%%%%%%%%%%%%%%%%%%%%%%%%
We build the quantities $\rnod m_{0^{++}}$ and $\rnod m_{2^{++}}$; our final results are however affected by large errors; additional studies and numerical techniques are necessary to deduce clear conclusions on the lattice artefacts for these observables.
However, by considering our determinations as upper limit for the glueball masses at different lattice spacings, one could conclude that the RG actions are not able to significantly reduce the discretization errors for $m_{0^{++}}$ with respect to the plaquette action.\\
%%%%%%%%%%%%%%%%%%%%%%%%%%%%%%%%%%%%%%%%%%%%%%%%%%%%%%%%%%%%%%%%%%%%%%%%%%%%%%
\noindent
Further investigations on the properties of RG actions could be very useful to clarify this question more definitively as well as further open questions, also in view of the future unquenched simulations.

%% file: app_a.tex
\chapter{Numerical results}\label{numerical_results}
In this appendix the numerical results for potential and force at finite lattice spacing are collected, for different gauge actions. 
\section{Wilson action}
The tables
\ref{t_force1a} and \ref{t_force2a} report the data for potential and force
at finite lattice spacing (chapter \ref{chapt_potential}) for the Wilson action.
For the couplings $5.7\leq\beta\leq 6.4$ we took the data from Guagnelli
et. al. \protect\cite{pot:r0_SU3}, while in the range $6.57\leq\beta\leq 6.92$ the results were obtained from our new numerical simulations. 
$\rI$ and $d_{\rm I}$ define the tree-level improved observables and are expressed in \eq{e_rI}, \eq{pot_impr}.
\input taba2.tex
\input taba1.tex
\section{RG actions}
The tables \ref{tab_iwasaki_results} and \ref{tab_dbw2_results} report the
data for the potential and the force evaluated at finite lattice spacing with
Iwasaki and DBW2 actions, which has been used in chapter \ref{improved}. Also
in this case $\rI$ defines the tree-level improved force. For the
potential only the naive definition is reported.
\input tab_iwasaki.tex
\input tab_dbw2.tex

%% file: taba2.tex
  \begin{table}[ht] 
\caption[Force and potential
from the data of Guagnelli et al.]{\label{t_force1}\footnotesize{Force and potential
from the data of Guagnelli et al. \protect\cite{pot:r0_SU3}}}  
\begin{center}
\begin{tabular}{ c  c  c  c  c }
  \hline 
 $\beta$  & $\rI/a$  & $a^2\,F(\rI)$  &  $d_{\rm I}/a$ &  $a\,V_{\rm I}(d_{\rm I})$     \\ 
  \hline 
 $ 5.7 $ & $  $ & $   $ & $ 1.855 $ & $  0.80463(94) $ \\
 $  $ & $ 2.277 $ & $ 0.2160(14) $ & $ 2.889 $ & $ 1.0206(22) $ \\
 $  $ & $ 3.312 $ & $ 0.1847(25) $ & $ 3.922 $ & $ 1.2053(44) $ \\
 $  $ & $ 4.359 $ & $ 0.1730(43) $ & $ 4.942 $ & $ 1.3784(77) $ \\[1ex]
 $ 5.95 $ & $  $ & $   $ & $ 1.855 $ & $ 0.61794 ( 16 )$ \\
 $  $ & $ 2.277 $ & $ 0.11212 ( 20 ) $ & $ 2.889 $ & $ 0.73006 ( 33 )$ \\
 $  $ & $ 3.312 $ & $ 0.08319 ( 27 ) $ & $ 3.922 $ & $ 0.81325 ( 55 )$ \\
 $  $ & $ 4.359 $ & $ 0.07164 ( 36 ) $ & $ 4.942 $ & $ 0.88489 ( 84 )$ \\
 $  $ & $ 5.393 $ & $ 0.06613 ( 48 ) $ & $ 5.954 $ & $ 0.9510 ( 12 )$ \\
 $  $ & $ 6.414 $ & $ 0.06296 ( 56 ) $ & $ 6.962 $ & $ 1.0140 ( 17 )$ \\
 $  $ & $ 7.428 $ & $ 0.0606 ( 12 ) $ & $ 7.967 $ & $ 1.0728 ( 30 )$ \\[1ex]
 $ 6.07 $ & $  $ & $   $ & $ 1.855 $ & $ 0.571729 ( 97 )$ \\
 $  $ & $ 2.277 $ & $ 0.09211 ( 11 ) $ & $ 2.889 $ & $ 0.66384 ( 19 )$ \\
 $  $ & $ 3.312 $ & $ 0.06427 ( 13 ) $ & $ 3.922 $ & $ 0.72811 ( 30 )$ \\
 $  $ & $ 4.359 $ & $ 0.05301 ( 26 ) $ & $ 4.942 $ & $ 0.78116 ( 55 )$ \\
 $  $ & $ 5.393 $ & $ 0.04771 ( 28 ) $ & $ 5.954 $ & $ 0.82887 ( 72 )$ \\
 $  $ & $ 6.414 $ & $ 0.04468 ( 27 ) $ & $ 6.962 $ & $ 0.87355 ( 89 )$ \\
 $  $ & $ 7.428 $ & $ 0.04262 ( 38 ) $ & $ 7.967 $ & $ 0.9162 ( 11 )$ \\
 $  $ & $ 8.438 $ & $ 0.04215 ( 45 ) $ & $ 8.971 $ & $ 0.9583 ( 13 )$ \\
 $  $ & $ 9.445 $ & $ 0.04087 ( 73 ) $ & $ 9.974 $ & $ 0.9992 ( 17 )$ \\[1ex]
 $ 6.2 $ & $  $ & $   $ & $ 1.855 $ & $ 0.533457 ( 82 )$ \\
 $  $ & $ 2.277 $ & $ 0.07804 ( 11 ) $ & $ 2.889 $ & $ 0.61145 ( 17 )$ \\
 $  $ & $ 3.312 $ & $ 0.05135 ( 14 ) $ & $ 3.922 $ & $ 0.66279 ( 28 )$ \\
 $  $ & $ 4.359 $ & $ 0.04054 ( 13 ) $ & $ 4.942 $ & $ 0.70333 ( 38 )$ \\
 $  $ & $ 5.393 $ & $ 0.03511 ( 20 ) $ & $ 5.954 $ & $ 0.73844 ( 54 )$ \\
 $  $ & $ 6.414 $ & $ 0.03238 ( 20 ) $ & $ 6.962 $ & $ 0.77082 ( 69 )$ \\
 $  $ & $ 7.428 $ & $ 0.03018 ( 25 ) $ & $ 7.967 $ & $ 0.80100 ( 85 )$ \\
 $  $ & $ 8.438 $ & $ 0.02884 ( 26 ) $ & $ 8.971 $ & $ 0.8298 ( 10 )$ \\
 $  $ & $ 9.445 $ & $ 0.02813 ( 27 ) $ & $ 9.974 $ & $ 0.8580 ( 12 )$ \\
 $  $ & $ 10.451 $ & $ 0.02766 ( 30 ) $ & $ 10.977 $ & $ 0.8856 ( 14 )$ \\
 $  $ & $ 11.455 $ & $ 0.02752 ( 34 ) $ & $ 11.979 $ & $ 0.9131 ( 16 )$ \\
  \hline 
  \end{tabular} 
\end{center}
  \end{table} 
  \begin{table}[ht] 
\caption[Force and potential
from the data of Guagnelli et al.]{\label{t_force1a}\footnotesize{Force and potential
from the data of Guagnelli et al. \protect\cite{pot:r0_SU3}}}  
\begin{center}
\begin{tabular}{ c  c  c  c  c }
  \hline 
 $\beta$  & $\rI/a$  & $a^2\,F(\rI)$  &  $d_{\rm I}/a$ &  $a\,V_{\rm I}(d_{\rm I})$     \\ 
  \hline 
 $ 6.4 $ & $  $ & $   $ & $ 1.855 $ & $ 0.488379 ( 48 )$ \\
 $  $ & $ 2.277 $ & $ 0.064318 ( 51 ) $ & $ 2.889 $ & $ 0.552697 ( 81 )$ \\
 $  $ & $ 3.312 $ & $ 0.039580 ( 70 ) $ & $ 3.922 $ & $ 0.59228 ( 13 )$ \\
 $  $ & $ 4.359 $ & $ 0.029360 ( 79 ) $ & $ 4.942 $ & $ 0.62164 ( 19 )$ \\
 $  $ & $ 5.393 $ & $ 0.024367 ( 84 ) $ & $ 5.954 $ & $ 0.64600 ( 24 )$ \\
 $  $ & $ 6.414 $ & $ 0.02145 ( 12 ) $ & $ 6.962 $ & $ 0.66744 ( 40 )$ \\
 $  $ & $ 7.428 $ & $ 0.01939 ( 16 ) $ & $ 7.967 $ & $ 0.68683 ( 50 )$ \\
 $  $ & $ 8.438 $ & $ 0.01819 ( 18 ) $ & $ 8.971 $ & $ 0.70502 ( 66 )$ \\
 $  $ & $ 9.445 $ & $ 0.01757 ( 17 ) $ & $ 9.974 $ & $ 0.72258 ( 79 )$ \\
 $  $ & $ 10.451 $ & $ 0.01677 ( 17 ) $ & $ 10.977 $ & $ 0.73936 ( 84 )$ \\
 $  $ & $ 11.455 $ & $ 0.01651 ( 15 ) $ & $ 11.979 $ & $ 0.75586 ( 94 )$ \\
 $  $ & $ 12.459 $ & $ 0.01609 ( 17 ) $ & $ 12.980 $ & $ 0.7720 ( 11 )$ \\
 $  $ & $ 13.462 $ & $ 0.01616 ( 29 ) $ & $ 13.982 $ & $ 0.7881 ( 11 )$ \\
 $  $ & $ 14.465 $ & $ 0.01564 ( 18 ) $ & $ 14.983 $ & $ 0.8038 ( 12 )$ \\
 $  $ & $ 15.467 $ & $ 0.01513 ( 39 ) $ & $ 15.984 $ & $ 0.8189 ( 14 )$ \\
  \hline 
  \end{tabular} 
\end{center}
  \end{table} 

%%% Local Variables: 
%%% mode: latex
%%% TeX-master: "pap1"
%%% End: 

%% file: taba1.tex
  \begin{table}[ht] 
\caption[Force and potential in the
short distance region.]{\label{t_force2}\footnotesize{Force and potential in the
short distance region. }}  
\begin{center}
  \begin{tabular}{ c  c  c  c  c }
  \hline 
 $\beta$  & $\rI/a$  & $a^2\,F(\rI)$  &  $d_{\rm I}/a$ &  $a\,V_{\rm I}(d_{\rm I})$     \\  
  \hline 
 $ 6.57 $ & $  $ & $   $ & $ 1.855 $ & $ 0.457898 ( 71 )$ \\
 $  $ & $ 2.277 $ & $ 0.056623 ( 65 ) $ & $ 2.889 $ & $ 0.51452 ( 12 )$ \\
 $  $ & $ 3.312 $ & $ 0.033391 ( 97 ) $ & $ 3.922 $ & $ 0.54795 ( 21 )$ \\
 $  $ & $ 4.359 $ & $ 0.023752 ( 94 ) $ & $ 4.942 $ & $ 0.57170 ( 28 )$ \\
 $  $ & $ 5.393 $ & $ 0.01904 ( 12 ) $ & $ 5.954  $ & $ 0.59074 ( 35 )$ \\
 $  $ & $ 6.414 $ & $ 0.01629 ( 13 ) $ & $ 6.962  $ & $ 0.60703 ( 41 )$ \\[1ex]
 $ 6.69 $ & $  $ & $   $ & $ 1.855 $ & $ 0.43918 ( 41 )$ \\
 $  $ & $ 2.277 $ & $ 0.052308 ( 38 ) $ & $ 2.889 $ & $ 0.491487 ( 72 )$ \\
 $  $ & $ 3.312 $ & $ 0.030174 ( 66 ) $ & $ 3.922 $ & $ 0.52162 ( 13 )$ \\
 $  $ & $ 4.359 $ & $ 0.021054 ( 75 ) $ & $ 4.942 $ & $ 0.54267 ( 18 )$ \\
 $  $ & $ 5.393 $ & $ 0.016439 ( 72 ) $ & $ 5.954 $ & $ 0.55911 ( 23 )$ \\
 $  $ & $ 6.414 $ & $ 0.013728 ( 82 ) $ & $ 6.962 $ & $ 0.57284 ( 28 )$ \\
 $  $ & $ 7.428 $ & $ 0.012036 ( 96 ) $ & $ 7.967 $ & $ 0.58487 ( 35 )$ \\
 $  $ & $ 8.438 $ & $ 0.010869 ( 82 ) $ & $ 8.971  $ & $ 0.59574 ( 41 )$ \\
 $  $ & $ 9.445 $ & $ 0.010123 ( 97 ) $ & $ 9.974 $ & $ 0.60587 ( 46 )$ \\[1ex]
 $ 6.81 $ & $  $ & $   $ & $ 1.855 $ & $ 0.422617 ( 22 )$ \\
 $  $ & $ 2.277 $ & $ 0.048833 ( 26 ) $ & $ 2.889 $ & $ 0.471411 ( 47 )$ \\
 $  $ & $ 3.312 $ & $ 0.027650 ( 32 ) $ & $ 3.922  $ & $ 0.499062 ( 66 )$ \\
 $  $ & $ 4.359 $ & $ 0.018860 ( 31 ) $ & $ 4.942 $ & $ 0.517921 ( 90 )$ \\
 $  $ & $ 5.393 $ & $ 0.014471 ( 32 ) $ & $ 5.954 $ & $ 0.53239 ( 11 )$ \\
 $  $ & $ 6.414 $ & $ 0.011870 ( 40 ) $ & $ 6.962 $ & $ 0.54426 ( 13 )$ \\
 $  $ & $ 7.428 $ & $ 0.010163 ( 54 ) $ & $ 7.967 $ & $ 0.55437 ( 18 )$ \\
 $  $ & $ 8.438 $ & $ 0.009072 ( 56 ) $ & $ 8.971 $ & $ 0.56344 ( 21 )$ \\
 $  $ & $ 9.445 $ & $ 0.008267 ( 52 ) $ & $ 9.974 $ & $ 0.57171 ( 24 )$ \\
 $  $ & $ 10.451$ & $ 0.007701 ( 58 ) $ & $ 10.977 $ & $ 0.57941 ( 27 )$ \\
 $  $ & $ 11.455$ & $ 0.007232 ( 62 ) $ & $ 11.979 $ & $ 0.58664 ( 30 )$ \\
  \hline 
  \end{tabular} 
\end{center}
  \end{table} 
  \begin{table}[ht] 
\caption[Force and potential in the
short distance region.]{\label{t_force2a}\footnotesize{Force and potential in the
short distance region. }}  
\begin{center}
  \begin{tabular}{ c  c  c  c  c }
  \hline 
 $\beta$  & $\rI/a$  & $a^2\,F(\rI)$  &  $d_{\rm I}/a$ &  $a\,V_{\rm I}(d_{\rm I})$     \\  
  \hline 
 $ 6.92 $ & $  $ & $   $ & $ 1.855 $ & $ 0.408642 ( 19 )$ \\
 $  $ & $ 2.277 $ & $ 0.046081 ( 19 ) $ & $ 2.889 $ & $ 0.454723 ( 34 )$ \\
 $  $ & $ 3.312 $ & $ 0.025696 ( 24 ) $ & $ 3.922 $ & $ 0.480418 ( 47 )$ \\
 $  $ & $ 4.359 $ & $ 0.017266 ( 29 ) $ & $ 4.942 $ & $ 0.497662 ( 75 )$ \\
 $  $ & $ 5.393 $ & $ 0.012969 ( 33 ) $ & $ 5.954 $ & $ 0.510631 ( 88 )$ \\
 $  $ & $ 6.414 $ & $ 0.010412 ( 39 ) $ & $ 6.962 $ & $ 0.52104 ( 11 )$ \\
 $  $ & $ 7.428 $ & $ 0.008855 ( 37 ) $ & $ 7.967 $ & $ 0.52990 ( 13 )$ \\
 $  $ & $ 8.438 $ & $ 0.007755 ( 43 ) $ & $ 8.971  $ & $ 0.53765 ( 16 )$ \\
 $  $ & $ 9.445 $ & $ 0.006974 ( 51 ) $ & $ 9.974 $ & $ 0.54463 ( 19 )$ \\
 $  $ & $ 10.451$ & $ 0.006402 ( 53 ) $ & $ 10.977 $ & $ 0.55103 ( 23 )$ \\
 $  $ & $ 11.455$ & $ 0.006022 ( 53 ) $ & $ 11.979 $ & $ 0.55705 ( 26 )$ \\
  \hline 
  \end{tabular} 
\end{center}
  \end{table} 

%%% Local Variables: 
%%% mode: latex
%%% TeX-master: "pap1"
%%% End: 

%% file: tab_iwasaki.tex
\begin{table}
\caption[The potential and the force in lattice units for the Iwasaki action.]{\footnotesize{The potential and the force in lattice units for the Iwasaki action.}\label{tab_iwasaki_results}}
\begin{center}
\begin{tabular}{c c c c c}
\hline
$\beta$   &  $r_{I}/a$ & $a^{2}F(r_{I})$ & $r/a$   &  $aV(r)$  \\
\hline
2.1551    &   1.4858 & 0.4059(21)  &  2        &  0.9519(20) \\  
          &   2.4626 & 0.2965(75)  &  3        &  1.2487(61)\\
          &   3.5663 & 0.2585(98)   &  4              &  1.515(12)\\
          &   4.6059 & 0.249(29)    &  5        &  1.783(12)\\
\hline
2.2879   & 1.4858 & 0.6055(29)   &  2   &     0.7730(12 \\
         & 2.4626 & 0.19796(76)  &  3   &     0.9707(33)\\
         & 3.5663 & 0.1689(70)    &  4  &     1.1418(41)\\
         & 4.6059 & 0.1622(64)   &  5   &     1.3047(95)\\
         & 5.6082 & 0.155(10)   &  6    &     1.460(16)\\
         & 6.6036 & 0.148(16)  &  7     &     1.608(33)\\
\hline
2.5208    & 1.4858 & 0.21301(25) &  2  &   0.60342(29) \\
          & 2.4626 & 0.11734(39) &  3  &   0.72099(46)\\
          & 3.5663 & 0.09011(70)  &  4  &   0.8107(14) \\
          & 4.6059 & 0.0803(10) &  5    &   0.8902(20)\\
          & 5.6082 & 0.07575(61)  &  6  &   0.9658(36)\\
          & 6.6036 & 0.0744(32)  &  7   &   1.0378(37)\\
          & 7.5977 & 0.0711(15)  &  8   &   1.1098(74) \\
\hline
2.7124    & 1.4858  &  0.17512(12)  &  2  &  0.52165(14) \\
          & 2.4626  & 0.08691(14)  &  3  &  0.60856(20) \\
          & 3.5663  & 0.06138(53) &  4  &  0.66994(56) \\
          & 4.6059   & 0.05194(81) &  5  &  0.7219(13) \\
          & 5.6082   & 0.04693(85)  &  6  &  0.7688(20) \\
          & 6.6036   &  0.0451(23) &  7  &  0.8139(11)  \\
          & 7.5977   & 0.0425(10)   &  8  &  0.8564(10)\\
          & 8.5915   & 0.0420(15) &  9  &  0.8984(23) \\
          & 9.5854   & 0.0414(19)  &  10 &  0.9398(25) \\
\hline
\end{tabular}
\end{center}
\end{table}

%%% Local Variables: 
%%% mode: latex
%%% TeX-master: t
%%% End: 

%% file: tab_dbw2.tex
\begin{table}
\caption[The potential and the force in lattice units for the DBW2 action.]{\footnotesize{The potential and the force in lattice units for the DBW2 action.}\label{tab_dbw2_results}}
\begin{center}
\begin{tabular}{c c c c c}
\hline
$\beta$   &  $r_{I}$ & $a^{2}F(r_{I})$ &  $r/a$   &  $aV(r)$  \\
\hline 
0.75696   &  1.9233  & 0.3885(13)  & 2 & 0.9004(17) \\ 
	  &  2.7793  &  0.2723(59) & 3 & 1.1726(33) \\ 
          &  3.7435  & 0.241(10)  & 4 & 1.410(11)  \\ 
          &  4.7364  & 0.2424(83)  & 5 & 1.658(20) \\ 
          &  5.7535  &  0.236(20)  & 6 & 1.894(35) \\ 
\hline
0.8243     & 1.9233 &  0.29697(61) & 2	&    0.7357(11) \\
           & 2.7793 & 0.1887(12)  & 3	&    0.9244(11)\\
	   & 3.7435 &  0.1613(12)  &  4	&    1.0857(17)  \\
           & 4.7364 &  0.1532(62)   &  5	&    1.2389(70)\\
	   &  5.7535 & 0.1476(91)  &  6	&    1.386(16)\\
	   & 6.7843 & 0.1494(61)  &  7	&    1.536(20)\\
\hline
0.9636    & 1.9233 & 0.20810(27)  &  2  &    0.55693(64)   \\     
          & 2.7793 &  0.1136(16) &  3  &    0.6703(20)  \\
          & 3.7435 & 0.08680(85) &  4	&    0.7569(20) \\
          & 4.7364 &  0.0783(13)  &  5	&    0.8340(29) \\
          & 5.7535 & 0.0745(37)  &  6	&    0.9076(57)\\
          & 6.7843 & 0.0700(57)  &  7	&    0.977(10)\\
          & 7.8120 & 0.0640(47)  &  8	&    1.038(10)\\
\hline
1.04      & 1.9233 & 0.18182(26) &  2  & 0.49710(24)  \\
          & 2.7793 & 0.09392(54)  &  3  & 0.59094(58) \\
          & 3.7435 & 0.06847(84)   &  4  & 0.65914(76) \\
          & 4.7364 & 0.0575(18)   &  5  & 0.7180(26)\\
          & 5.7535 & 0.0544(12)  &  6  & 0.7724(21) \\
          & 6.7843 & 0.0527(22)  &  7  & 0.8254(42) \\
          & 7.8120 & 0.0498(23)    &  8  & 0.8753(65) \\
          & 8.8296 & 0.0481(42)  &  9  & 0.9230(51)\\
          & 9.8369 & 0.0485(44)  &  10 & 0.9713(56) \\
          & 10.8365  & 0.0464(34) &  11 & 1.0175(95)  \\
	  & 11.8310  &  0.048(13) &  12 & 1.067(20)  \\
\hline
\end{tabular}
\end{center}
\end{table}

%% file: app_trans.tex
\chapter{Transfer matrix and Hamiltonian formalism}\label{app_trans}
The Hamiltonian formalism constitutes a very useful framework in order to
extract physical observables from the exponential decay of Euclidean correlation functions.
In this appendix the transfer matrix formulation will be reviewed, with a
particular attention to the improved actions, where the violation of
physical positivity at cutoff-energies can be a problem for the extraction of physical observables.\\
In the formulation of \eq{expect_value}, lattice gauge theories are represented as a statistical mechanical system with partition function $Z$ and thermal averages $\langle\mathcal{O}\rangle$. The transfer matrix provides an equivalent representation of the model as a quantum mechanical system with a Hilbert space  $\mathcal{H}$, a Hamiltonian operator $\mathbb{H}$ and linear operators $\hat{\mathcal{O}}$ corresponding to the observables $\mathcal{O}$.\\
We will now assume that the Euclidean space-time lattice $\Lambda$ is finite, with $T\times L\times L\times L$ points and we take periodic boundary conditions for the gauge fields
\begin{eqnarray}
U(x+aT\hat{\nu},\mu) & = & U(x,\mu),\quad \nu =  0\\
U(x+aL\hat{\nu},\mu) & = & U(x,\mu),\quad \nu > 0.
\end{eqnarray}
In this appendix we will set $a=1$ for simplicity.

\section{Wilson action}
We denote the set of links that are completely contained in the time slice $x_{0}=t$ by $B_{t}$; those links which connect the time slices $x_{0}=t+1$ and $x_{0}=t$ form the set $B_{t+1,t}$. Let $b$ be a link connecting the lattice points $x$ and $(x+\hat{\mu})$ and $U(b)\equiv U(x,\mu)$; then the links belonging  to $B_{t}$ and $B_{t+1,t}$ are denoted by
\begin{eqnarray}\label{time_slices}
U_{t} & \equiv & U(b),\quad b \in B_{t},\\
U_{t+1,t} & \equiv & U(b),\quad b\in B_{t+1,t} .
\end{eqnarray}

The Wilson action can be written as
\begin{equation}\label{tr_wilson}
S=\sum_{t}L\left[U_{t+1},U_{t+1,t},U_{t}\right],
\end{equation}
where
\begin{equation}\label{action_split}
L\left[U_{t+1},U_{t+1,t},U_{t}\right]=\frac{1}{2}L_{1}\left[U_{t+1}\right]+\frac{1}{2}L_{1}\left[U_{t}\right]+L_{2}\left[U_{t+1},U_{t+1,t},U_{t}\right].
\end{equation}
The first two terms in \eq{action_split}
\begin{equation}
L_{1}\left[U_{t}\right]=-\frac{\beta}{N}\sum_{\vec{x}}\sum_{k<l}\Re W_{kl}^{1\times 1}(x)|_{x_{0}=t}
\end{equation}
are the contribution of space-like plaquettes on a time slice, while
\begin{equation}
L_{2}\left[U_{t+1},U_{t+1,t},U_{t}\right]=
-\frac{\beta}{N}\sum_{\vec{x}}\sum_{k}\Re W_{k0}^{1\times 1}(x)|_{x_{0}=t}
\end{equation}
represents the contribution of time-like plaquettes between two time slices.
The plaquette $W_{\mu\nu}^{1\times 1}(x)$ is defined in \eq{plaquette}.\\
We then consider wave functions $\Psi[U_{t}]$ depending on the link variables
on a certain time slice; these describe states belonging to different charged
sectors depending on how they transform under gauge transformations.
In particular, we will consider gauge invariant wave functions, which
correspond to the vacuum sector
\begin{equation}
\Psi[U'_{t}]=\Psi[U_{t}],
\end{equation}
where $U'_{t}$ and  $U_{t}$ are gauge-equivalent.\\
A convenient way to define a scalar product of two given wave functions 
$\Psi$ and $\Phi$ is given by 
\begin{equation}
(\Phi,\Psi)=\int D[U_{t}]\Phi[U_{t}]^{*}\Psi[U_{t}],
\end{equation}
with
\begin{equation}
D[U_{t}]=\prod_{b\in B_{t}}dU(b).
\end{equation}
The Hilbert space $\mathcal{H}$ corresponds to the space of all gauge
invariant functions with finite norm,
\begin{equation} 
(\Psi,\Psi) < \infty.
\end{equation}

The transfer matrix $\mathbb{T}$ \cite{Luscher:TM} is an integral operator acting on $\mathcal{H}$ via 
\begin{equation}\label{transfer_matrix_wilson}
(\trans\Psi)[U_{t+1}]=\int D[U_{t}]K[U_{t+1},U_{t}]\Psi[U_{t}],
\end{equation}
where $K$ is an integral kernel
\begin{equation}\label{trans_wilson_kernel}
K[U_{t+1},U_{t}]=\int D[U_{t+1,t}]\exp\{-L\left[U_{t+1},U_{t+1,t},U_{t}\right]\}.
\end{equation}
From these considerations follows that 
\begin{equation}\label{z_t}
Z=\int D[U]e^{-S[U]}=\tr\{\trans ^{T}\},
\end{equation}
where ``$\tr$'' indicates the trace in the Hilbert space.\\
The Euclidean correlation functions $\langle A(x)B(y)...\rangle$ of local, gauge invariant fields $A(x), B(y)...$ can be given in a quantum mechanical interpretation. In particular, if these fields are polynomials of the gauge field variables $U_{t}$ at a fixed time $t$, we may associate an operator $\hat{A}(\vec{x})$ through
\begin{equation}\label{2points}
(\hat{A}(\vec{x})\Psi)[U_{t}]=A(0,\vec{x})\Psi[U_{t}].
\end{equation}
For $x_{0}=t$, $0\leq x_{0}\leq T$ we have
\begin{equation}\label{two-point-f}
\langle A(x)A(0)\rangle =\frac{1}{Z}\tr\left\{\mathbb{T}^{T-t}\hat{A}(\vec{x})\mathbb{T}^{t}\hat{A}(\vec{0})\right\}.
\end{equation}
Similar formulas hold for higher correlation functions. The transfer matrix
$\mathbb{T}$ turns out to be a bounded, self-adjoint positive operator with a completely discrete spectrum (for finite $L$). \\
%For pure gauge theories with Wilson action, one can show that reflection positivity (both \emph{site} and \emph{link}) holds \cite{Osterwalder:1978pc,Seiler:1982pw,} :
%this amounts to the positivity of scalar products on $\mathcal{H}$:
%\begin{equation}
%\langle \psi|\psi\rangle\geq 0.
%\end{equation}
All eigenvalues of $\trans$ are then strictly positive and the maximal eigenvalue $\lambda_{0}$ is not degenerate. Thus one can define the Hamiltonian
\begin{equation}\label{hamiltonian}
\mathbb{H}=-\ln\frac{\trans}{\lambda_{0}},
\end{equation}
which is a self-adjoint operator in $\mathcal{H}$. 
For $T\rightarrow\infty$ the two-point function \eq{two-point-f} then assumes the form
\begin{equation}\label{spectral_repr}
\langle A(x)A(0)\rangle =\langle 0|\hat{A}(\vec{x})e^{-t\mathbb{H}}\hat{A}(\vec{0})|0\rangle,
\end{equation}
which is also called \emph{spectral representation} and establishes the exponential decay of Euclidean two-point functions, which is the basis of the extraction of physical observables from lattice QCD.
%%%%%%%%%%%%%%%%%%%%%%%%%%%%%%%%%%%%%%%%%%%%%%%%%%%%%%%%%%%%%%%%%%%%%%%%%%%%%%

\section{Improved actions}
We now consider actions containing not only the plaquette term, but also other classes of loops $\mathcal{C}$ included in a real gauge invariant and orientation independent function of the ordered product of link variables along $\mathcal{C}$.
These include both
L\"uscher-Weisz and RG improved actions as particular cases.
One can prove that for those actions it is still possible to construct a transfer matrix \cite{Luscher:1984is}. 
Restricting to actions of kind \eq{lw_action}, one can notice that in the temporal gauge, $U(x,0)=1$, the associated field equations assume the form of a fourth-order difference equation in the time coordinate $x_{0}$. One has then to define a transition amplitude to go from a given gauge configuration on a pair of consecutive equal time hyperplanes to some other gauge fields on a later pair of hyperplanes. For the wave function $\Psi$ one then makes the ansatz
\begin{equation}
\Psi=\Psi[U_{t+1},U_{t+1,t},U_{t}].
\end{equation}
A gauge invariant scalar product is now\\
\begin{equation}
(\Phi,\Psi)\int D[U_{t+1}]D[U_{t+1,t}]D[U_{t}]
\Phi[U_{t+1},U_{t+1,t},U_{t}]^{*}\Psi[U_{t+1},U_{t+1,t},U_{t}].
\end{equation}
The transfer matrix can be defined in analogy to the Wilson case as transition amplitude between fields configurations defined on two consecutive pairs of hyperplanes $(B_{t},B_{t+1})$ and $(B_{t+1},B_{t+2})$, and one can argue that the properties \eq{z_t} and \eq{2points} are satisfied.

%The improved action can be then written as
%\begin{equation}\label{tr_improved}
%S=\sum_{t}L\left[U_{t+2},U_{t+2,t+1},U_{t+1},U_{t+1,t},U_{t}\right],
%\end{equation}
%where
%\begin{equation}
%L\left[U_{t+2},U_{t+2,t+1},\mathcal,U_{t+1},U_{t+1,t},U_{t}\right] =
%\end{equation}
%$$
%=\frac{1}{3}L_{1}\left[U_{t+2}\right]
%+\frac{1}{3}L_{1}\left[U_{t+1}\right]
%+\frac{1}{3}L_{1}\left[U_{t}\right]
%$$
%$$
%+\frac{1}{2}L_{2}\left[U_{t+2},U_{t+2,t+1},U_{t+1}\right]
%$$
%$$
%+\frac{1}{2}L_{2}\left[U_{t+1},U_{t+1,t},U_{t}\right]
%$$
%$$
%+L_{3}\left[U_{t+2},U_{t+2,t+1},U_{t+1},U_{t+1,t},U_{t}\right].
%$$
%In analogy to \eq{action_split}, $L_{1}$, $L_{2}$, $L_{3}$ represent terms in the action \eq{lw_action} with time extend 0,1,2 respectively. 

%The transfer matrix can be then written as\\

%\begin{equation}\label{transfer_matrix_improved}
%(\trans\Psi)[U_{t+2},U_{t+2,t+1},U_{t+1}]=
%\end{equation}
%$$
%\int D[U_{t+1}]D[U_{t+1,t}]D[U_{t}]
%K[U_{t+2},U_{t+2,t+1},U_{t+1}]
%\Psi[U_{t+1},U_{t+1,t},U_{t}],
%$$
%where the kernel is now defined as
%\begin{equation}\label{trans_impr_kernel}
%K[U_{t+2},U_{t+1},U_{t}]=
%\end{equation}
%$$
%\int D[U]_{t+2,t+1}D[U_{t+1,t}]
%\exp\{-L\left[U_{t+2},U_{t+2,t+1},U_{t+1},U_{t+1,t},U_{t}\right]  \}.
%$$

But, unlike the Wilson case, one finds that $\trans$ is no longer hermitian; rather, one obtains
\begin{equation}
\trans^{\dagger}=\theta\trans\theta,
\end{equation}
where 
\begin{equation}
\theta\Psi[U_{t+1},U_{t+1,t},U_{t}]=\Psi[U_{t},U^{-1}_{t+1,t},U_{t+1}],\quad\theta^{\dagger}=\theta,\quad\theta^{2}=1.
\end{equation}
The spectrum $\sigma(\trans)$ of $\trans$ can be defined as the set of complex numbers $\lambda$, for which the operator $\lambda-\trans$ has no bounded inverse. Further considerations \cite{Luscher:1984is} lead to the conclusion that spectral values always occur in pairs of complex conjugated numbers and there is a positive (non-degenerate) eigenvalue $\lambda_{0}$ such that
\begin{equation}
\lambda_{0}>|\lambda|,\quad\forall\lambda\in\sigma(\trans),
\quad\lambda\neq\lambda_{0}.
\end{equation}
The eigenfunction $\Psi_{\lambda_{0}}$ may be interpreted as ground state wave function.
Due to the violation of physical positivity, it is no longer possible to define an Hamiltonian operator and hence to write down the spectral representation \eq{spectral_repr}. For $T\rightarrow\infty$ the two-point function \eq{two-point-f} takes 
the form
\begin{equation}
\langle A(x)A(0)\rangle=\lambda_{0}^{-t}\langle 0|\theta\hat{A}(\vec{x})\trans^{t}\hat{A}(\vec{0})|0\rangle,
\end{equation}
where $|0\rangle\equiv|\Psi_{\lambda_{0}}\rangle$, and in the limit $t\rightarrow\infty$ one ends up with an asymptotic expansion
\begin{equation}\label{two-point-impr}
\langle A(x)A(0)\rangle\sim_{t\rightarrow\infty}\lambda_{0}^{-t}\sum_{\lambda\in\sigma(\trans)\backslash\lambda_{0}}\lambda^{t}p_{\lambda}(t),
\end{equation}
where $p_{\lambda}(t)$ is a polynomial of $t$ with degree strictly less than a certain $n_{\lambda}$: usually $n_{\lambda}=1$ as in the hermitian case, but one can find examples where $n_{\lambda}>1$ for some $\lambda$.
The \eq{two-point-impr} establishes that connected two-point functions always decay exponentially at large times, but if the leading spectral value is not real and positive, the exponential factor is multiplied by an oscillating amplitude and contributions in the spectral decomposition with negative weight appear, as remnant of the positivity violation.
This fact has been observed in the evaluation of the effective potential and
effective masses for improved actions \ref{improved}, \ref{chapter_glueball} and can be a serious problem for
the applicability of the variational method.\\
Nevertheless, one expects that physical positivity is recovered in the
continuum limit; restricting ourself to the subspace $\mathcal{H}_{phys}$ of the states with small energy with respect to the cutoff ($\lambda/\lambda_{0}\sim 1$), then
the non-hermiticity of the transfer matrix can be eliminated by choosing a new scalar product relative to which $\trans$ becomes hermitian
and physical positivity is then completely restored.
For example
\begin{equation}
(\Phi,\Psi)_{new}=(\Phi,\theta\Psi)_{old}
\end{equation}
is not well defined on the full Hilbert space, but one expects that there exists a $0<\epsilon<1$  such that independently of the cutoff the following properties hold:\\
\noindent
(i) all spectral values $\lambda\in\sigma(\trans)$ with $|\lambda|\geq\epsilon\lambda_{0}$ are real and positive; \\
\noindent
(ii) $(\Psi,\Psi)_{new}>0,\quad\forall\Psi\in\mathcal{H}_{phys},\quad\Psi\neq 0$.\\
The existence of such $\epsilon$ has not been rigorously proved, but is
supported by perturbative calculations, which can be used to estimate a value
of $\epsilon$ by determining the location of unphysical poles in the propagator, as shown in the next section.

\subsection{Unphysical poles in the propagator}

\begin{figure}
\begin{center}
\includegraphics[width=9cm]{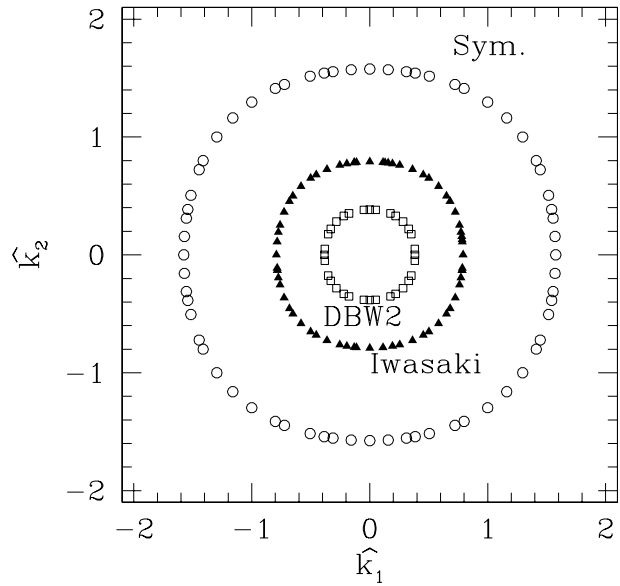}
\end{center}
\caption[The curves defined by the condition $\mathfrak{Im}(w)=0$ in the $\hat{k}_{3}=0$ plane. Inside the curve $\mathfrak{Im}(w)=0$, while outside $\mathfrak{Im}(w)\neq 0$.]{\footnotesize{\label{brill}The curves defined by the condition $\mathfrak{Im}(w)=0$ in the $\hat{k}_{3}=0$ plane. Inside the curve $\mathfrak{Im}(w)=0$, while outside $\mathfrak{Im}(w)\neq 0$.}}
\end{figure}

In \cite{Weisz:1983zw} the propagator $D_{\mu\nu}$ is evaluated  to lowest order in $g_{0}$ with covariant gauge fixing; it is defined by
\begin{equation}\label{pert_propagator}
\langle A_{\mu}^{i}(x) A_{\nu}^{j}(y)\rangle=\delta^{ij}\int_{k}e^{ik(x-y)}e^{i(k_{\mu}-k_{\nu})/2}D_{\mu\nu}(k),
\end{equation}
where $A_{\mu}(x)$ is related to the link variables by 
$$
U(x,\mu)=e^{-A_{\mu}(x)}
$$
and the momenta integration corresponds to
$$
\int_{k}=\prod_{\mu=0}^{3}\int_{-\pi}^{\pi}\frac{dk_{\mu}}{2\pi}.
$$ 
We are now referring to the case $L=\infty$, so that the momenta take continuous values in the Brillouin zone.\\
\Eq{pert_propagator} can be rewritten in the form 
\begin{equation}\label{pert_propagator2}
D_{\nu\tau}(k)=(\hat{k}^{2})^{-2}\left[\alpha\hat{k}_{\nu}\hat{k}_{\tau}+\sum_{\sigma}(\hat{k}_{\sigma}\delta_{\tau\nu}-\hat{k}_{\tau}\delta_{\sigma\nu})A_{\tau\sigma}(k)\hat{k}_{\sigma}\right],
\end{equation}
where $\alpha$ is the gauge parameter, $\hat{k}_{\mu}=2\sin(k_{\mu}/2)$ and 
$A_{\mu\nu}$  is independent of $\alpha$ and satisfies the following properties:
\begin{enumerate}
\item $A_{\mu\mu}=0,\quad\forall \mu$;
\item $A_{\mu\nu}=A_{\nu\mu}$;
\item $A_{\mu\nu}(k)=A_{\mu\nu}(-k)$;
\item $A_{\mu\nu}(0)=1,\quad\mu\neq\nu$.
\end{enumerate}
$A_{\mu\nu}$ has the general form
\begin{equation}
A_{\mu\nu}=\frac{f(\hat{k})}{D},
\end{equation}
where 
\begin{equation}
D=\sum_{\mu}\hat{k}_{\mu}^{4}\prod_{\nu\neq\mu}q_{\mu\nu}+\sum_{\mu>\nu,\rho >\tau,\{\rho,\tau\}\cup\{\mu,\nu\}=\emptyset}\hat{k}_{\mu}^{2}\hat{k}_{\nu}^{2}q_{\mu\nu}(q_{\mu\rho}q_{\nu\tau}+q_{\mu\tau}q_{\nu\rho}),
\end{equation} 
and the explicit form of $f(\hat{k})$ can be found in \cite{Weisz:1983zw} and will not reported here.
The denominator of \eq{pert_propagator2} is then given by
\begin{equation}\label{delta}
\Delta=D(\hat{k}^2)^2=
\end{equation}
$$
(\hat{k}^2)^2\left(\sum_{\mu}\hat{k}_{\mu}^{4}\prod_{\nu\neq\mu}q_{\mu\nu}+\sum_{\mu>\nu,\rho >\tau,\{\rho,\tau\}\cup\{\mu,\nu\}=\emptyset}\hat{k}_{\mu}^{2}\hat{k}_{\nu}^{2}q_{\mu\nu}(q_{\mu\rho}q_{\nu\tau}+q_{\mu\tau}q_{\nu\rho})\right),
$$
where $q_{\mu\nu}$ satisfies the properties (1)-(4)
and in our specific form of the action takes the form
\begin{equation}
q_{\mu\nu}=(1-\delta_{\mu\nu})[1-c_{1}(\hat{k}_{\mu}^{2}+\hat{k}_{\nu}^{2})],
\end{equation}
$c_{1}$ being the coefficient of the $1\times 2$ planar loops term in the action \eq{c1_impr}.
Then \eq{delta} becomes
\begin{equation}
\Delta=\left(\hat{k}^{2}-c_{1}\sum_{\mu}\hat{k}_{\mu}^{4}\right)\Bigg[\hat{k}^{2}-c_{1}\left((\hat{k}^{2})^{2}+\sum_{\mu}\hat{k}_{\mu}^{4}\right)+
\end{equation}
$$
\frac{1}{2}c_{1}^{2}\left((\hat{k}^{2})^{3}+2\sum_{\mu}\hat{k}_{\mu}^{6}-\hat{k}^{2}\sum_{\mu}\hat{k}_{\mu}^{4}\right)\Bigg]
-4c_{1}^{3}\sum_{\mu}\hat{k}_{\mu}^{4}\prod_{\nu\neq\mu}\hat{k}_{\nu}^{2}.
$$
In order to search for the poles of the propagator,
one substitutes\\
$$
k=(\hat{k}_{1},\hat{k}_{2},\hat{k}_{3},iw)                                     $$
and looks for solutions of the equation $\Delta=0$ scanning the whole Brillouin zone. For the plaquette action $c_{1}=0$ one finds the usual result
$$
w=\pm\sqrt{\hat{k}^{2}_1+\hat{k}^{2}_2+\hat{k}^{2}_3}
$$
while for general $c_{1}$ one expects complex conjugated solutions
$$                                       
w=\mathfrak{Re}(w) \pm i\mathfrak{Im}(w)=w(\hat{k}_{1},\hat{k}_{2},\hat{k}_{3}).
$$
Numerical investigations showed that the condition
$\mathfrak{Im}(w)=0$ (for \emph{all} solutions at a given momentum) defines 
the equation $f(\hat{k}_{1},\hat{k}_{2},\hat{k}_{3})=0$ which
represents a compact 3-dimensional object. In \fig{brill} the
2-dimensional intersection with the plane $\hat{k}_{3}=0$ is plotted
for several values of $c_{1}$ corresponding to different actions.\\ 
For our numerical studies, we discretized the Brillouin zone in finite  
intervals that could be made arbitrarily small. 

\begin{figure}
\begin{center}
\includegraphics[width=9cm]{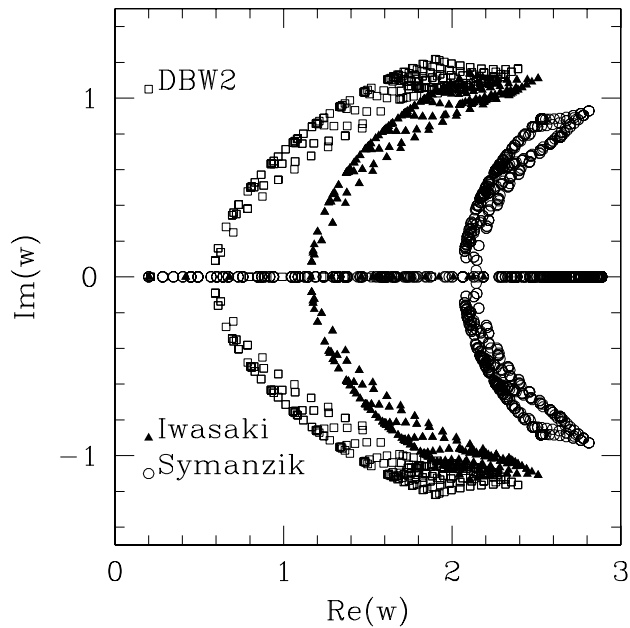}
\end{center}
\caption[Distribution of the poles for different actions, scanning the Brillouin zone.]{\footnotesize{\label{distr_poli}Distribution of the poles for different actions, scanning the Brillouin zone.}}
\end{figure}
In the region enclosed by the curve one has $\mathfrak{Im}(w)=0$, while outside $\mathfrak{Im}(w)\neq 0$.\\ 
The \fig{distr_poli} represents the distribution of the poles; by scanning the Brillouin zone and varying the intervals of the momenta,
one reaches the conclusion that
there exists a maximal value of $\mathfrak{Re}(w)$, which is related to $c_{1}$ by                                    
$$
w_{max}=2\textrm{arcsinh} \left(\frac{1}{2\sqrt{-2c_{1}}}\right)=
\left\{\begin{array}{ll}
 2.063 & \textrm{Symanzik, tree level}\\
 1.162 & \textrm{Iwasaki}\\
 0.588 & \textrm{DBW2}
\end{array}\right.
$$
such that for $w<w_{max}$
the imaginary part $\mathfrak{Im}(w)$ vanishes.\\
The value of $w_{max}$ yields an estimate of $\epsilon$ defined above through the simple relation\\
$$
\epsilon=e^{-w_{max}}=\left\{\begin{array}{ll}
0.127  & \textrm{Symanzik, tree level}\\
0.313  & \textrm{Iwasaki}\\
0.555  & \textrm{DBW2}
\end{array}\right.
$$
Evaluating observables from exponential decays of correlation
functions, one expects the presence of unphysical states unless $t\gg t_{min}$, where $t_{min}$ in lattice units is given by
\begin{equation}\label{tmin}
t_{min}=\frac{1}{w_{max}}=\left\{\begin{array}{ll}
 0.484 & \textrm{Symanzik, tree level}\\
 0.860 & \textrm{Iwasaki}\\
 1.702 & \textrm{DBW2}
\end{array}\right.
\end{equation}
This fact has to be taken into account in the extraction of physical observables from correlation functions evaluated with RG actions, and in general with every action which contains not only plaquette terms. In particular this restricts the applicability of the variational method.\\
One has to remember that our estimation of the unphysical poles has been performed in perturbation theory and one can not exclude a priori that at large coupling $g_{0}$ the numerical value of $\epsilon$ can be larger than what we found.

%%% Local Variables: 
%%% mode: latex
%%% TeX-master: t
%%% End: 

%% file: app1.tex
\chapter[3-dimensional lattice propagator in coordinate space]{Evaluation of the 3-dimensional lattice propagator in coordinate space
\label{app1}}
An efficient method to calculate the lattice propagator in 
coordinate space is proposed in \cite{pert:LW95b}.  
It is based on a recursion relation which allows to express the propagator 
as linear function of its values near the origin. In that paper, 
the 4-dimensional case is discussed; it is straitghforward to apply 
the same method in 3 dimensions. 
We use lattice units, $a=1$, in this appendix and restrict the attention to
the plaquette Wilson action.

\section{Lattice 3-dimensional propagator.}
The 3-dimensional scalar propagator in the coordinate space 
satisfies the Laplace equation 
\begin{equation}\label{green}
-\triangle G({\vec x})=\left\{\begin{array}{ll}
1 & \textrm{if ${\vec x}=0$}\\
0 & \textrm{otherwise}.
\end{array}\right.
\end{equation}
where the lattice Laplacian is given by
\begin{equation}
\triangle=\sum_{j=1}^{3}\nabstar j\nab j,
\end{equation}
and $\vec{x}=(x_{1},x_{2},x_{3})$.\\
The forward and backward lattice derivatives are defined through
\begin{equation}
\nab j f({\vec x})=f({\vec x}+\hat{j})-f({\vec x}),\quad \nabstar j f({\vec x})=f({\vec x})-f({\vec x}-\hat{j}),
\end{equation}
where $\hat {j}$ denotes the unit vector in direction $j$.\\
In the continuum limit, which here coincides with the limit $|{\vec x}|\rightarrow\infty$, we expect that the propagator converges to $4\pi|x|$, which is the Green function of the continuum Laplacian on $\mathbb{R}^{3}$.\\

\section{Recursion relation}
The first step to obtain a recursion relation is the observation that
\begin{equation}\label{acca}
(\nabstar j +\nab j)G({\vec x})=x_{j}H({\vec x}),
\end{equation}
where
\begin{equation}
H({\vec x})=\int_{-\pi}^{+\pi}\frac{d^{3}k}{(2\pi)^3}e^{i{\vec k}\cdot{\vec x}}
\ln(\hat{k}^{2})
\end{equation}
is independent of the direction $j$ and
\begin{equation}
\hat{k}^2=\sum_{j=1}^{3}\hat{k_j}^2,\quad\hat{k}_{j}=2\sin(k_{j}/2).
\end{equation}
Summing over $k$ and using \eq{green} one obtains
\begin{equation}
H({\vec x})=\frac{2}{\rho}\sum_{j=1}^{3}[G({\vec x})-G({\vec x}-\hat{j})].
\end{equation}
with $\rho=\sum_{j=1}^{3}x_{j}$.\\
This expression can be used to eliminate $H({\vec x})$ in \eq{acca}, giving\\
\begin{equation}\label{recursion}
G({\vec x}+\hat{j})=G({\vec x}-\hat{j})+2\frac{x_{j}}{\rho}\sum_{i=1}^{3}[G({\vec x})-G({\vec x}-\hat{i})]
\end{equation}
which is valid only for the points with $\rho\neq 0$.\\
Because of isotropy, we can restrict our attention to the points ${\vec x}$ with $x_{1}\geq x_{2} \geq x_{3} \geq 0$. In this region \eq{recursion} is a recursion relation which can be used to express $G({\vec x})$ as a linear combination of the values of the Green function at the corners of the unit cube,\\
\be\label{cubic}
G(0,0,0),\quad G(1,0,0),\quad G(1,1,0),\quad G(1,1,1).
\ee
This four initial values are not independent; from \eq{green} at ${\vec x}=0$ 
\begin{equation}\label{relation1}
G(0,0,0)-G(1,0,0)=\frac{1}{6}
\end{equation}
follows directly.
Another relation can be deduced in the following way: first one can observe that \eq{recursion} becomes one-dimensional along the lattice axes. Defining
$$
g_{1}(n)=G(n,0,0), \quad g_{2}(n)=G(n,1,0), \quad g_{3}(n)=G(n,1,1),
$$
and using the lattice symmetries, one finds
\begin{eqnarray}
g_{1}(n+1) & = & 6g_{1}(n)-4g_{2}(n)-g_{1}(n-1), \\
g_{2}(n+1) & = & \frac{2n}{n+1}\left[3g_{2}(n)-g_{1}(n)-g_{3}(n)\right]-\frac{n-1}{n+1}g_{2}(n-1),\\
g_{3}(n+1) &  =  &\frac{2n}{n+2}\left[3g_{3}(n)-2g_{2}(n)\right]-\frac{n-2}{n+2}g_{3}(n-1).
\end{eqnarray}
Next one notices that  
\begin{equation}\label{invariant}
I(n)=(n-1)g_{1}(n)+2ng_{2}(n)+(n+1)g_{3}(n)-ng_{1}(n-1)-
\end{equation}
$$
2(n-1)g_{2}(n-1)-(n-2)g_{3}(n-1)
$$
is an invariant of the recursion relation, that is\\
$$
I(n+1)=I(n)
$$
for $n\geq 1$.\\
The value of $I$ can be worked out in the limit $n\rightarrow\infty$, where\\
$$
g_{j}(n)=\frac{1}{4\pi n}+O(1/n^{2}), \quad \textrm{for} \quad j=1,2,3,
$$
yielding $I=0$.
Setting $n=1$, \eq{invariant} gives
\begin{equation}\label{relation2}
3G(1,1,0)+2G(1,1,1)-G(0,0,0)=0.
\end{equation}
Thus from the four initial values two can be eliminated using \eq{relation1} and \eq{relation2} and the propagator can be written in the form\\
\begin{equation}\label{principal}
G({\vec x})=r_{1}({\vec x})G(0,0,0)+r_{2}({\vec x})G(1,1,0)+r_{3}({\vec x}),
\end{equation}
where the coefficients $r_{1},r_{2},r_{3}$ are rational numbers that can be computed recursively from \eq{recursion}.

\section{Numerical computation of $G({\vec x})$}
From \eq{principal} one obtains the initial conditions for $r_{1},r_{2},r_{3}$ :
\begin{eqnarray}
r_{1}(0,0,0)=1, & r_{1}(1,1,0)=0 \\     
r_{2}(0,0,0)=0, & r_{2}(1,1,0)=1 \\
r_{3}(0,0,0)=0, & r_{3}(1,1,0)=0.
\end{eqnarray}
\Eq{relation1}, \eq{relation2} give
\begin{eqnarray}
r_{1}(1,0,0)=1, & r_{1}(1,1,1)=\frac{1}{2} \\     
r_{2}(1,0,0)=0, & r_{2}(1,1,1)=-\frac{3}{2} \\
r_{3}(1,0,0)=-\frac{1}{6}, & r_{3}(1,1,1)=0.
\end{eqnarray}
The other coefficients $r_{1}({\vec x}),r_{2}({\vec x}),r_{3}({\vec x})$ can be
calculated exactly using for example an algebraic programming language. 
We are
then left with the task to compute numerically $G(0,0,0)$ and $G(1,1,0)$ with
high accuracy. For this purpose we consider the lattice points $x=(n,0,0)$ and
$y=(n,1,0)$, where $n$ is an adjustable integer. The numerical investigations
show that the associated coefficients $r_{k}(x)$ and $r_{k}(y)$ are rapidly
growing, roughly proportional to $10^{n-1}$ (see \tab{coeffi}). Solving \eq{principal} for $G(x)$ and $G(y)$, we obtain
\be
G(1,0,0)=\frac{1}{C}\left\{G(y)-G(x)\frac{r_{1}(y)}{r_{1}(x)}+\frac{r_{3}(x)r_{1}(y)}{r_{1}(x)}-r_{3}(y)\right\}, 
\ee
with
\be
\quad C=r_{2}(y)-\frac{r_{2}(x)r_{1}(y)}{r_{1}(x)},
\ee
and
\be
G(0,0,0)=-\frac{r_{2}(x)}{r_{1}(x)}G(1,1,0)-\frac{r_{3}(x)}{r_{1}(x)}.
\ee
The coefficients multiplying $G(x)$ and $G(y)$ are of order $10^{-(n-1)}$ and
can be neglected, obtaining 
\begin{equation}
G(1,0,0)=\frac{1}{C}\left\{\frac{r_{3}(x)r_{1}(y)}{r_{1}(x)}-r_{3}(y)\right\}+
{\rm O}(10^{-(n-1)}).
\end{equation}
This method is exponentially convergent and the error can be estimated well since the neglected terms are known.\\
By applying this technique one obtains\\
$$
G(0,0,0)=0.2527310098586630030260020266135701299...,
$$
$$
G(1,0,0)=0.0551914336877373170165449460300639378...
$$
\begin{table}
\caption[Numerical values of the coefficients $r_{k}(x),r_{k}(y)$ for $n=1,..,7$.]{\label{coeffi}\footnotesize{Numerical values of the coefficients $r_{k}(x),r_{k}(y)$ for $n=1,..,7$.}}
\begin{center}
\begin{tabular}{ccccccc}
\hline
$n$ & $r_{1}(x)$ & $r_{1}(y)$ & $r_{2}(x)$ & $r_{2}(y)$ & $r_{3}(x)$ & $r_{3}(y)$   \\
\hline
1 & 1 &  0 & 0 & 1 & -0.17 & 0 \\
2 & 5 &  -1.5 & -4 & 4.5 & -1 & 0.17 \\
3 & 35 &  -14 &  -42 & 28.3 & -6.5 &  2 \\
4 & 261 & -123.75 & -361.3 & 219.75 & -46 &  19.17\\
5 & 2026 & -1091.2 & -3005 & 1844.413 & -346.17 &  174\\
6 & 16259.8 & -9663.5 & -25046.32 & 16000.9 & -2727 & 1559.17\\
7 & 134186.8 &-86122.9 & -211276.52 & 141288.85 & -22252.5 & 13968 \\
\hline
\end{tabular}
\end{center}
\end{table}

%%% Local Variables: 
%%% mode: latex
%%% TeX-master: "app2"
%%% End: 

%% file: akno.tex
\chapter*{Acknowledgments}
        \label{akno}
         \addcontentsline{toc}{chapter}{ Acknowledgments}
        \pagestyle{fancy}
        \fancyhead{}
        \fancyhead[LE,RO]{\bfseries\thepage}
        \fancyhead[LO]{\bfseries }
        \fancyhead[RE]{\bfseries }
I would like to thank first of all Rainer Sommer for suggesting the argument of my thesis, for the 
competence and also for encouragement and constant support.\\

Thanks also to all people of DESY Zeuthen, in particular the theory/NIC 
group who created a stimulating and enjoyable working atmoshpere and made these three years very fruitful.\\

I would like to thank Ulli Wolff, Peter Weisz and Urs Wenger for useful discussions and suggestions.\\

Special thanks to Francesco Knechtli for his help and the $SU(2)$ code which was the starting point for our programm, and to Marco Guagnelli and Hartmut Wittig for discussions and for providing the data generated in an earlier project.\\

Thank to Konrad-Zuse-Zentrum f\"ur Informationstechnik Berlin (ZIB) -in particular Hinnerk St\"uben - and Forschungszentrum J\"uelich for providing CPU-resources on CRAY T3E machines and assistance.\\

Special thanks to Michele Della Morte, Chris Ford, Hiro Kawamura and 
Stephan D\"urr for suggestions and comments related to the thesis.\\

I am grateful to every PhD student and postdoc with whom I shared not only the working time but also very nice moments of my life.\\

Finally I would like to thank my family and special friends: Elisa, Floriana, Lucia, Davide, Stefano, Marcella, Mauro, Vincenzo, Andrea, Alberto.